\documentclass{article}

\usepackage[preprint,nonatbib]{neurips_2025}
\usepackage[backend=biber,style=authoryear,natbib=true,maxcitenames=2,
            maxbibnames=99,giveninits=true,uniquename=false,uniquelist=false,
            urldate=comp]{biblatex}
\DeclareBibliographyAlias{legislation}{report}

\AtEveryBibitem{%
  \iffieldundef{doi}{}{\clearfield{url}\clearfield{urlyear}%
    \clearfield{urlmonth}\clearfield{urlday}}%
}

\usepackage{fontspec}
\usepackage[table]{xcolor}
\usepackage{tabularx}
\usepackage{ltablex}
\keepXColumns
\usepackage{booktabs}
\usepackage{caption}
\usepackage{graphicx}
\graphicspath{{images/}{./}}
\usepackage{ragged2e}
\usepackage{enumitem}
\usepackage{placeins}
\usepackage{needspace}
\usepackage{amsmath}
\usepackage{amssymb}
\usepackage{url}

\newcolumntype{Y}{>{\RaggedRight\arraybackslash}X}
\newcolumntype{C}[1]{>{\Centering\arraybackslash}p{#1}}
\newcolumntype{L}[1]{>{\RaggedRight\arraybackslash}p{#1}}

\newenvironment{indentedpara}[2]
  {\par\penalty-100\begingroup\leftskip=#1\relax\rightskip=#2\relax
   \interlinepenalty=10000\noindent\ignorespaces}
  {\par\endgroup}

\setlist[itemize]{leftmargin=2.5em}

\title{Mapping General-Purpose AI Governance in Twenty AI Middle-Power
  Jurisdictions}

\author{%
  Josephine Schwab\textsuperscript{1}\thanks{Working paper. This is a draft; comments are welcome and may be sent to \texttt{josephine.schwab@arcadiaimpact.org}. All figures reported here reflect the dataset as at 30 July 2026. The live base continues to be revised, and versioned snapshots are deposited at \texttt{doi:10.5281/zenodo.21978946}. Instruments, governance actors, and definitions are openly deposited; provision records available on request.} \And
  Nathan Naidoo\textsuperscript{1} \And
  Ferruccio Barazzutti\textsuperscript{1} \And
  Sheryn Lee\textsuperscript{1} \And
  Caio Machado\textsuperscript{2} \AND
  {\mdseries\textsuperscript{1}\,Arcadia Impact AI Governance Taskforce \qquad
   \textsuperscript{2}\,The Future Society} \AND
  {\mdseries August 2026}
}

\DeclareBibliographyCategory{statutes}
\DeclareBibliographyCategory{intlpolicy}
\DeclareBibliographyCategory{scholarly}
\DeclareBibliographyCategory{nonscholarly}
\addtocategory{statutes}{AREAE-DU_243}
\addtocategory{statutes}{BRA_032}
\addtocategory{statutes}{CHL_054}
\addtocategory{statutes}{EUFRADEU_144}
\addtocategory{statutes}{EUFRADEU_162}
\addtocategory{statutes}{GBR_153}
\addtocategory{statutes}{GBR_159}
\addtocategory{statutes}{IND_164}
\addtocategory{statutes}{KEN_065}
\addtocategory{statutes}{KOR_084}
\addtocategory{statutes}{KOR_085}
\addtocategory{statutes}{NGA_069}
\addtocategory{statutes}{doraDelegatedRegulation2024}
\addtocategory{statutes}{sb53TFAIA2025}
\addtocategory{intlpolicy}{AREAE-DU_284}
\addtocategory{intlpolicy}{ARE_185}
\addtocategory{intlpolicy}{ARE_266}
\addtocategory{intlpolicy}{AUS_272}
\addtocategory{intlpolicy}{BRA_209}
\addtocategory{intlpolicy}{DefiningAIIncidents2024}
\addtocategory{intlpolicy}{EUFRADEU_171}
\addtocategory{intlpolicy}{GBR_022}
\addtocategory{intlpolicy}{GBR_103}
\addtocategory{intlpolicy}{GBR_105}
\addtocategory{intlpolicy}{GBR_188}
\addtocategory{intlpolicy}{GBR_196}
\addtocategory{intlpolicy}{GBR_197}
\addtocategory{intlpolicy}{IND_166}
\addtocategory{intlpolicy}{IND_167}
\addtocategory{intlpolicy}{ISR_160}
\addtocategory{intlpolicy}{JPN_189}
\addtocategory{intlpolicy}{KOR_083}
\addtocategory{intlpolicy}{OECD_AIM}
\addtocategory{intlpolicy}{OECD_ExplanatoryMemorandum2024}
\addtocategory{intlpolicy}{SGP_178}
\addtocategory{intlpolicy}{SGP_278}
\addtocategory{intlpolicy}{TWN_219}
\addtocategory{intlpolicy}{g72023HiroshimaCodeOfConduct}
\addtocategory{intlpolicy}{lorenz2023initial}
\addtocategory{intlpolicy}{praOperationalResilience2022}
\addtocategory{scholarly}{agarwalIncorporatingAIIncidentReporting2025}
\addtocategory{scholarly}{aiindex2026ch8}
\addtocategory{scholarly}{externalAccessDangerousCapability2026}
\addtocategory{scholarly}{gomez2026}
\addtocategory{scholarly}{hardwareEnabledMechanisms2025}
\addtocategory{scholarly}{heim2024}
\addtocategory{scholarly}{kulothunganTowardsAdaptiveAIGovernance2025}
\addtocategory{scholarly}{laurenceMiddlePowers}
\addtocategory{scholarly}{mazziniConsiderationsRegulationAI2023}
\addtocategory{scholarly}{okolo2026}
\addtocategory{scholarly}{roberts2026}
\addtocategory{scholarly}{verifyInternationalAgreements2024}
\addtocategory{nonscholarly}{2025}
\addtocategory{nonscholarly}{ECstandardisation2026}
\addtocategory{nonscholarly}{aiActNewsletterAIOfficeTender2025}
\addtocategory{nonscholarly}{aiindex2026}
\addtocategory{nonscholarly}{aipolicybulletinWindowClosingBrusselsEffect}
\addtocategory{nonscholarly}{aisiIncident2026}
\addtocategory{nonscholarly}{anthropicCyberEvalIncidents2026}
\addtocategory{nonscholarly}{apolloResearchEvaluations2026}
\addtocategory{nonscholarly}{apolloWhiteBoxAccess2026}
\addtocategory{nonscholarly}{caisStatementAIRisk2023}
\addtocategory{nonscholarly}{caruana2010SystemicRisk}
\addtocategory{nonscholarly}{cetasAPacAIGovernanceMiddlePower}
\addtocategory{nonscholarly}{cltcIntolerableRiskThreshold2025}
\addtocategory{nonscholarly}{csetAddingStructureAIHarm}
\addtocategory{nonscholarly}{csisJapanAgileAIGovernance}
\addtocategory{nonscholarly}{hodesWhoGovernsAI}
\addtocategory{nonscholarly}{iasr2026}
\addtocategory{nonscholarly}{idaisBeijingConsensusRedLines2024}
\addtocategory{nonscholarly}{iradukundaLeveragingEUAICode2025}
\addtocategory{nonscholarly}{itifOneLawSetsSouthKoreaAIPolicy2025}
\addtocategory{nonscholarly}{leicht2025roadmap}
\addtocategory{nonscholarly}{leicht2026}
\addtocategory{nonscholarly}{leicht2026periphery}
\addtocategory{nonscholarly}{malonzaAutomationFirstAdoption2026}
\addtocategory{nonscholarly}{metrAbout2026}
\addtocategory{nonscholarly}{metrEvaluations2026}
\addtocategory{nonscholarly}{metrFrontierRisk2026}
\addtocategory{nonscholarly}{millernguyen2025globalcall}
\addtocategory{nonscholarly}{nist2023airmf}
\addtocategory{nonscholarly}{oecdCommonReportingFrameworkAIIncidents2025}
\addtocategory{nonscholarly}{openaiHuggingFace2026}
\addtocategory{nonscholarly}{pacingFrontierOpenLetter2026}
\addtocategory{nonscholarly}{redLinesCampaignFAQ2026}
\addtocategory{nonscholarly}{tfs2023HeavyHead}
\addtocategory{nonscholarly}{tfsAthensRoundtableRecap2025}
\addtocategory{nonscholarly}{tfsFantasticBeasts2022}
\addtocategory{nonscholarly}{tfsUnacceptableRisks2026}
\addtocategory{nonscholarly}{tsmc2026Fundamentals}
\addtocategory{nonscholarly}{wilkinson2026}
\addtocategory{nonscholarly}{zoumpalovaWhereWeDraw2026}

\begin{document}

\maketitle

\enlargethispage{3\baselineskip}
\section*{Abstract}

The most capable general-purpose AI (GPAI) models are mostly built in two jurisdictions, the United States and China, but the risks they carry land globally. Regionally advanced economies hosting no frontier developer, which we call AI middle-powers, are writing their own rules to govern these technologies. This paper investigates which GPAI-relevant provisions these AI middle-powers have enacted, mapping twenty jurisdictions including the European Union. The research focuses on the individual provision level, analyzing primary texts across four governance areas chosen to trace the accountability chain for the model layer: systemic risk assessment, evaluation and verification, prohibitions with breach monitoring and detection, and serious incident reporting. Confirmed absence is recorded as data alongside positive provision. We find that jurisdictions converge on form, but diverge on force. Sixteen engage in at least three of the four governance areas, yet only 22 percent of provisions sit in binding law, while three-quarters of the instruments that do bind do so without defining GPAI at all, and definitions cluster in guidance and strategy documents, where precision carries the least consequence. The institutional infrastructure shows the same shape: four in five of the mapped governance actors hold mandates that predate GPAI; reporting duties travel down channels borrowed from other regimes, with obligations attaching wherever the inherited regime reached in the application layer; and prohibitions address what an actor does with a system nearly four times more often than what a system can do, so enforcement waits on harm. Where these states do engage the model layer, they build capacity to observe it rather than impose duties on those who build it, and almost every evaluation body in the sample was constituted without power to act on what it finds. Nominal coverage of the full accountability chain reaches eleven jurisdictions, but only five hold more than a single provision in every area and, outside the EU, no jurisdiction imposes a binding evaluation duty on a model developer. We argue that domestic gaps can be closed before international agreement on which GPAI capabilities to prohibit, and draw policy implications from the patterns the mapping exposes. The dataset gives researchers and policymakers a provision-level basis for identifying where regimes could align, and also where coordination would have to start from scratch.

\textbf{Keywords:} general-purpose AI; middle-powers; AI governance; comparative law; AI red lines; regulatory convergence; soft law; frontier AI

\clearpage

\section*{Contribution statement}

Author Contributions (CRediT taxonomy) \\
\\
\textbf{Conceptualization, Methodology:} Josephine Schwab, Caio Machado\\
\textbf{Investigation:} Josephine Schwab, Nathan Naidoo, Ferruccio Barazzutti, Sheryn Lee\\
\textbf{Writing – Original Draft: \\}
\begin{tabular}{@{}l@{\hspace{1.5em}}l@{}}
Josephine Schwab: & 1-10 \\
Nathan Naidoo: & 5.2, 5.5 \\
Ferruccio Barazzutti: & 5.6, 5.8, 6.2, 6.5 \\
Sheryn Lee: & 5.4, 6.7 \\
\end{tabular}\\[8pt]
\textbf{Writing – LaTeX:} Josephine Schwab\\
\textbf{Writing – Review \& Editing, Supervision:} Josephine Schwab (Lead Editor), Caio Machado \\
\textbf{Project Administration:} Josephine Schwab

\section*{Acknowledgements}

We wish to thank Niki Iliadis, Kasia Jakimowicz, and Kevin Koehler for their insightful and helpful feedback, Francesca Gomez for her assistance in co-designing our data codebook, and Patrick Peel who assisted with initial research and legal mapping. We would also like to thank Charbel-Raphaël Segerie, Kathrin Gardhouse, Michael Harré, Clarissa Koh, Dongyoun Cho, Renato Leite Monteiro, Bruria Friedman, Tolulope Adebayo, Erick Iriarte Ahon, Ketana Krishna, Diana Weilguny, and David Sánchez García for their assistance in verifying our governance findings per jurisdiction, as well as others who kindly contributed to our data verification. Any remaining errors are our own.

This research was conducted as part of the AI Governance Taskforce cohort at Arcadia Impact in Summer 2026.

\section*{1. Introduction}

“General-purpose AI”, as a regulatory concept, owes its provenance to the 2024 EU Artificial Intelligence Act (henceforth EU AI Act), which defines general-purpose AI at both the model and system level, including as a model “trained with a large amount of data using self-supervision at scale, that displays significant generality and is capable of competently performing a wide range of distinct tasks regardless of the way the model is placed on the market and that can be integrated into a variety of downstream systems or applications” (Art 3(63)). \citep{EUFRADEU_144} Successive generations of general-purpose AI (GPAI) models have now arrived at capabilities their predecessors lacked, and with risks that narrow and task-specific systems do not carry, including meaningful uplift toward chemical, biological, radiological and nuclear weapons; offensive cyber capability; and behavior that resists correction or shutdown \citep{iasr2026}, \citep{tfsAthensRoundtableRecap2025}.

Incidents of this kind are now being disclosed during model evaluation: in July 2026, the UK AI Security Institute (UK AISI) and two frontier developers each published accounts of agents taking unsanctioned action against real systems and people during capability testing; UK AISI recorded nineteen such events across ten evaluation runs \citep{aisiIncident2026}, \citep{openaiHuggingFace2026}, \citep{anthropicCyberEvalIncidents2026}. These capabilities are established when a model is trained and inherited by every system built atop it, and because one model serves many purposes they are not visible from any single application. This raises the question of whether separate legal obligations should attach to the GPAI model layer.

One regulatory response names GPAI as a distinct class: the EU AI Act legislates for "general-purpose AI models" and imposes a separate obligation tier on those carrying systemic risk \citep{EUFRADEU_144}, and the G7 Hiroshima Process drew its perimeter around the most advanced AI systems \citep{g72023HiroshimaCodeOfConduct}. This is the GPAI-specific governance approach; governing a class of models by what they are capable of, rather than a class of applications by what they are used for.

In the last several years, domestic GPAI governance has begun to be drafted and enacted across a wide range of jurisdictions, in instruments of very different legal form. While the most consequential decisions about GPAI model development are taken by a small number of developers, the risks land globally. Whether a model can generate sexual imagery of children, for instance, is settled at the model layer, in its training data and safety tuning, yet the material may be generated wherever the model reaches. Moreover, risk is not the only thing distributed unevenly. Much domestic AI policy across these jurisdictions is directed at securing a fair share of the benefits, through compute access, skills, local language capability and domestic industrial strategy, and for many governments that is the more pressing half of the question. This paper addresses the risk-mitigation half only. States governing GPAI from outside the frontier of model development – AI middle powers – therefore face the problem of designing legal frameworks to prevent serious risk before it materializes as harm. We use the term for jurisdictions that currently host no frontier developer, but that carry weight in how those models are deployed, procured and regulated: as substantial markets, as regional rule-makers whose instruments are copied by neighbours, and as participants in the multilateral processes where GPAI norms are being negotiated. What they share is a common position – each must govern models built elsewhere, by companies it cannot license or tax.

Anton Leicht argues that the AI governance community has misread what these states can realistically achieve: on his account, the 2024 EU Artificial Intelligence Act – the leading instance of framework design – has faltered, with its implementation repeatedly delayed, and enforcement against frontier developers politically fragile; that international instruments cannot bind a United States federal government that treats sovereignty over domestic AI development as a strategic imperative; and that the low-salience conditions under which extraterritorial technology regulation once worked have gone \citep{leicht2026}. Their structural position, he argues, is also deteriorating: without a domestic frontier developer, they absorb AI's disruptive effects while capturing few of its gains \citep{leicht2026periphery}. The tractable work, Leicht concludes, lies in governing how AI models are deployed.

This paper considers that critique in its strategic framing. Legislative activity on AI is real and accelerating – one widely cited measure, the Stanford AI Index, tracks AI-related bills passed only by G20 states, a frame that reaches eleven of the twenty jurisdictions mapped here \citep{aiindex2026ch8}. Where broader inventories do exist, such as the OECD AI Policy Navigator – hosting 89 state profiles – these often record instruments a jurisdiction holds and their status, but not which duties those instruments impose, on whom, or with what legal force \citep{2025}. It may therefore be said that the sceptical case is not yet answerable from the published record. This paper asks the guiding question:

\begin{indentedpara}{36pt}{36pt}\textit{Across AI middle-power jurisdictions, what commonalities and divergences characterize domestic governance of general-purpose AI, as reflected in enacted legislation, regulatory guidance and codes of practice, national strategies, and pre-legislative proposals?}\end{indentedpara}

AI middle-powers have shaped international AI processes individually, through the summit series, the Hiroshima Process, the Council of Europe Framework Convention, and the International Network for Advanced AI Measurement, Evaluation and Science. Recent work has argued that the standing of the global majority in international AI governance is understated \citep{okolo2026}. Acting individually, AI middle-powers are not positioned to author binding treaties that constrain frontier laboratories, but they are positioned to build the operational machinery around those instruments. Meanwhile, documented AI incidents continue to rise: the AI Incident Database recorded 362 in 2025, against 233 in 2024 \citep{aiindex2026}, and the OECD AI Incidents and Hazards Monitor (AIM) shows a sharper increase in reports since January 2025 \citep{OECD_AIM}.\footnote{These monitors are not a complete picture but an indicator of a rising pattern.} It is this gap – between what is known about AI middle-power coordination, and what is unknown about the GPAI instruments AI middle-powers already hold – that our research addresses.

This paper takes, as its main object, four key governance areas that straddle the GPAI development and deployment sequencing divide (§3.3). Obligations imposed at either end of this divide cannot straightforwardly substitute for obligations at the other. Therefore, we look at provisions that cover systemic risk assessment duties, and evaluation and verification mechanisms (development) as well as AI prohibitions ("red lines"), serious incident monitoring and detection, and serious incident reporting infrastructure (deployment). Read in sequence these trace the accountability chain for the model layer, from anticipation of harm before deployment, to notification once it has occurred.

Four secondary research questions also structure the analysis:

\begin{itemize}[leftmargin=36pt,rightmargin=36pt,labelsep=0.6em,topsep=4pt,itemsep=2pt,parsep=0pt]
  \item \textit{How is general-purpose AI being legally defined, and what thresholds or scoping criteria do those definitions employ?}
  \item \textit{How do different legal architectures distribute obligations across the four key governance areas?}
  \item \textit{How do governance instruments interact across tiers within a jurisdiction, and which architectures recur?}
  \item \textit{Where sectoral instruments carry the governance load, which risk domains do they cover, and which governance mechanisms do they employ?}
\end{itemize}

This research proceeds in two stages. The first is descriptive: a comparative mapping that offers a baseline, rather than recommendations, to any single jurisdiction. The second draws policy implications from the patterns and gaps that mapping exposes, and directs them to multilateral coordination, rather than to individual legislatures. We distinguish throughout between what the dataset establishes and what we infer from it. The research proceeds by document analysis of primary legal and regulatory sources, under a scope test and coding procedure in which confirmed absence is coded as data on the same footing as positive provision (§3, 4). It then identifies the cross-jurisdictional patterns that a provision-level view makes visible and an instrument-level view does not.

The argument of this paper is that AI middle-powers have converged on the form of GPAI governance, but have diverged on its force. They address the same four governance areas, but they attach almost none to a binding duty on the party that builds the GPAI model. Where duties do apply, they attach at the application layer, to deployers, purchasers, platforms, and supervised firms – and the institutions carrying them were already regulating those actors for other reasons. This suggests a governance chain that is nearly complete in vocabulary, but incomplete at specific, identifiable joints. The paper develops that argument in four steps. Section 2 sets out the literature on middle-power AI governance and the sceptical case this paper tests. Section 3 builds the analytical framework: the jurisdiction sample, a six-dimension account of general-purpose AI definitions, and the four key governance areas with their severity thresholds. Section 4 states the instrument scope test, the coding procedure, and the treatment of confirmed absence. Section 5 reports the findings area by area, beginning with how GPAI is defined and then following the accountability chain from systemic risk assessment through evaluation, prohibition and incident reporting, before turning to the sectoral instruments and cross-tier architectures through which much of this governance is carried. Section 6 draws out the cross-jurisdictional patterns a provision-level view makes visible, and Section 7 states where the jurisdictions converge, setting out four axes on which they diverge, and identifies what GPAI governance gaps, on our reading, a jurisdiction can close on its own.

\section*{2. Literature review}

The term "middle power" is contested, and the international-relations literature has not settled on a definition: the term sits within a broader hierarchy of similarly contested terms, from hegemons and superpowers to regional, emerging and small powers, and the basis for placing states within any of them is disputed \citep{laurenceMiddlePowers}, though categorizations typically draw on material and military capability, regional influence, conduct within multilateral institutions, and diplomatic self-description. AI governance literature has nonetheless adopted the term in a narrower sense – it is used to refer to the "most advanced economies" outside the two jurisdictions where frontier AI models are developed, i.e. beyond the United States, and China. \citep{leicht2025roadmap} We adapt that usage. Defining the group by economic advancement excludes jurisdictions whose GPAI governance is consequential for other reasons – Kenya and Nigeria are drafting some of the most demanding provider-facing obligations in this sample, and Brazil and Chile are legislating ahead of several G7 members. We therefore define an AI middle-power by position in the AI political economy, rather than by development status: a jurisdiction that currently hosts no frontier developer, and so cannot regulate at the point of model creation, but whose market, regional influence or regulatory activity gives it purchase over how models are deployed and what their developers must demonstrate.

We treat a jurisdiction as hosting no frontier developer where no developer headquartered there has – as at 30 July 2026, and on published third-party estimates of training compute – trained a GPAI model at or above the capability frontier defined at §3.2.2 under California’s Senate Bill 53. France and Canada are boundary cases: Mistral AI and Cohere both train and release GPAI models, but both fall below that line on estimated compute. The sample criterion turns on whether a jurisdiction can regulate at the point where a GPAI model's capabilities are set, which §6.8 sets out. The sample is constructed on that basis at §3.1.

Two bodies of research bear on this paper, however they do not often intersect. The first concerns middle-powers in AI governance, which has grown quickly, though it has done so around AI governance generally, not general-purpose AI as a regulatory category. The second concerns GPAI governance as a standalone research category – definitions, compute thresholds, systemic-risk obligations, frontier safety frameworks. This body of literature remains anchored almost entirely in the EU AI Act's Chapter V regime, in US federal action, the UK institute ecosystem, and the international summit and standards processes, \citep{heim2024} \citep{roberts2026} nonetheless these remain the leading initiatives addressing model capabilities rather than applications. Research on how AI middle-powers govern GPAI specifically, at the level of the instruments they have actually adopted, sits in the gap between the two.

AI middle-power governance literature clusters around three questions adjacent to, rather than coextensive with this paper. The first concerns coordination potential. Chatham House argues that meaningful global coordination is structurally blocked as long as model capability data sits with private firms, geopolitical rivalry forecloses binding commitments, and institutions lack enforcement power, becoming feasible only after a cross-border AI crisis and only where existing institutions can be built upon \citep{wilkinson2026}. Practitioner accounts reach a similar conclusion: coordination failures are structural, but the most realistic trajectory is "convergence through practice", as voluntary commitments harden into shared assessment, convergent standards, and market expectations that function like regulation \citep{hodesWhoGovernsAI}.

A further argument holds that the 2024 EU AI Act's extraterritorial pull cannot be assumed, because AI compliance is divisible in a way that data-protection compliance was not \citep{aipolicybulletinWindowClosingBrusselsEffect}. The GDPR's Brussels Effect operated at the level of infrastructure: firms rebuilt data-processing pipelines, consent management, and storage architectures to meet EU standards, and once rebuilt there was no economic case for maintaining a weaker parallel system elsewhere. AI compliance works differently, because the trained model can remain identical across markets, but what changes is the documentation layer around it, including risk assessments, and disclosure. A provider can supply full EU-grade documentation to European customers while offering only the minimal disclosures required elsewhere, and the cost of running two tiers is low because the expensive part, building the model, is already done. The exception is frontier general-purpose AI, where mitigation requirements may eventually require changes to the model itself, and where EU requirements could shape practice only if specified early enough. Against this, it has been argued that the safety measures in the EU's GPAI Code of Practice operate at model level rather than system level throughout, so that the cost of customising models per market makes the Code the global baseline for signatories \citep{iradukundaLeveragingEUAICode2025}.

The second is a critique of the current AI middle-power governance theory of change, when treated as a coherent field. Leicht's account, set out at §1, holds that using AI middle-power domestic regulation to constrain frontier development was always marginal and is now counterproductive, and that the sovereignty impulse is itself a miscalibrated reflex which leaves these states on a "permanent periphery" \citep{leicht2026}, \citep{leicht2026periphery}. However, a comparative analysis of what AI middle-power jurisdictions have enacted in legislature is set apart from competing prescriptions about whether they should be legislating.

A distinct critique concerns who sets the rules: Chinasa Okolo and Mubarak Raji document the marginalisation of Global Majority priorities in international AI governance, and identify national and regional strategies as the most tractable countertrend \citep{okolo2026}. Take-up of the 2024 Council of Europe Framework Convention on Artificial Intelligence among non-member states is regionally uneven. Of the eleven states that negotiated the treaty, five were Latin American, and the remainder high-income OECD states, with no African participant and none from South, Southeast or continental East Asia. Cameroon and Ghana joined the Committee on Artificial Intelligence as observers only in 2025, after the text had been adopted and opened for signature.

Current GPAI-specific literature focuses on compute thresholds and their regulatory functions, framed around US and EU rule-making \citep{heim2024}. Gomez et al. (2026) compare the incident architecture of California's 2025 Senate Bill 53 (SB 53), enacting the Transparency in Frontier Artificial Intelligence Act, against the EU AI Act, and find the two diverge at the level of the harm categories themselves. SB 53 builds a two-tier structure in which a frontier model is a foundation model trained above 10\textsuperscript{26} operations, and attaches its duties to "catastrophic risk" and "critical safety incidents", neither of which has a direct counterpart in the EU regime \citep{gomez2026}. They also identify gaps that this paper finds again across the AI middle-power corpus: no mechanism for aggregating related incidents that are individually below threshold, no duty to report near misses, and no route to escalate on the ground that a harm would be irreversible. United States instruments sit outside our scope, but the comparison matters nonetheless, because it indicates that the reporting gaps documented at §5.6 are not an artefact of governing without a frontier developer in the jurisdiction.

Furthermore, evaluative frameworks applied to global GPAI initiatives mostly assess international institutions rather than domestic instruments \citep{roberts2026}. Comparative surveys compare regions or great powers, the US, EU and China \citep{kulothunganTowardsAdaptiveAIGovernance2025} or map frameworks broadly: Egypt's thirty-five-framework study identifies a "compliance divide" between Global North and Global South orientations, but works at whole-framework level and with an AI-general scope \citep{2025}. Regulatory trackers such as the OECD AI Policy Navigator, IAPP, and other commercial scanners are jurisdiction-level inventories, recording which instruments exist and their status, rather than which obligations they impose, on whom, with what legal force or enforcement consequence – and being inventories of AI policy generally, they do not separate GPAI-directed provisions from technology-neutral ones.

No existing work maps GPAI-directed governance at the level of individual provisions across a comparable AI middle-power set, at which GPAI-specific duties can be told apart from AI-general coverage, or establishes what shared domestic starting points exist. Nor does existing work record confirmed absences, so a jurisdiction that examined a governance function and declined to impose a duty is indistinguishable from one never examined.

\section*{3. Framework}

\subsection*{3.1 The twenty jurisdictions}

This paper's dataset comes from a sample of twenty AI middle-power jurisdictions: Australia, Brazil, Canada, Chile, France, Germany, India, Israel, Japan, Kenya, Nigeria, Peru, Singapore, South Africa, the Republic of Korea, Switzerland, Taiwan, the United Arab Emirates, and the United Kingdom – with European Union instruments additionally mapped, making the twentieth. Taiwan is included as a legal order that enacts and enforces its own AI governance instruments, and its semiconductor industry's market capitalization is the largest of any jurisdiction mapped \citep{tsmc2026Fundamentals}. The mapped subnational units are: the Emirate of Dubai, the Dubai International Financial Centre (DIFC), and New South Wales, Australia. The term "jurisdiction" is used in a functional sense and carries no implication as to political or diplomatic status.

Regional spread was treated as a selection constraint, and covers nine groupings: East and South East Asia, Latin America, the Middle East and North Africa, Sub-Saharan Africa, EU Europe, non-EU Europe, South Asia, North America, and the Pacific. Fourteen of the twenty are members of the OECD Global Partnership on AI, and sixteen are Friends of the Hiroshima AI Process. The sample contains both jurisdictions that routinely shape multilateral text (UN Security Council members) and jurisdictions that receive it. Jurisdictions were also chosen based on the language coverage among the research team.

The selection was further validated by a composite scoring exercise across six equally weighted clusters: general economic weight; defense; resource wealth and compute infrastructure; diplomatic agility; democracy and AI-governance infrastructure; and participation in international AI governance. Each cluster aggregates from published indicators which can be reviewed in Appendix A. The purpose of the composite score is to demonstrate that the jurisdictions span a meaningful range. Scores run from 3.42 (Kenya) to 8.15 (France) with no gap wider than 0.55 anywhere in the distribution.

\subsection*{3.2 GPAI definitions}

\subsubsection*{3.2.1 GPAI as a regulatory concept}

Our research into GPAI definitions draws heavily from The Future Society's (TFS) work, which, as of 2022, identified that there was “no common definition of GPAIS across the scientific literature,” \citep{tfsFantasticBeasts2022} and whose 2023 review of 28 academic and policy definitions found a lack of convergence, with definitions conflating foundation models, generative AI and general-purpose AI systems \citep{tfs2023HeavyHead}. TFS continues, “GPAI models are characterised by their scale (measured in number of parameters), as well as their reliance on self-supervised learning methods [...and] by its widespread use as pre-trained models for other AI systems.” (The Future Society, 2022)

As set out at §1, GPAI as a regulatory concept owes its provenance to the 2024 EU AI Act, which defines GPAI at the model level (Art. 3(63)) and at the system level, the latter as a system based on a general-purpose AI model with the capability to serve a variety of purposes, both for direct use and for integration in other AI systems (Art. 3(66)). It classifies a GPAI model as carrying systemic risk where it has high-impact capabilities, evaluated on appropriate technical tools and methodologies including indicators and benchmarks, presumed above 10\textsuperscript{25} floating-point operations of training compute (Art. 51).

Earlier fora, such as the 2019 G20 Osaka AI Principles, and UNESCO instruments like the 2021 Recommendation on the Ethics of Artificial Intelligence, take "AI systems" as their object and differentiate (if at all) by application context, but the 2023 G7 Hiroshima AI Process scoped by capability, drawing its perimeter around “the most advanced AI systems, including the most advanced foundation models and generative AI systems”, and is the first multilateral instrument to have done that. \citep{g72023HiroshimaCodeOfConduct}

\subsubsection*{3.2.2 A six-dimension framework}

Drawing on the EU AI Act, the TFS reviews of definition convergence, and the multilateral context of the 2023 G7 Hiroshima AI Process, this study defines GPAI using six dimensions. We define GPAI as:

\begin{indentedpara}{36pt}{36pt}\textit{AI models or systems that are characterized by their large scale (typically measured by the number of parameters they contain, or by the volume of compute used to train them) and their ability to perform a variety of tasks, rather than being specialized for one specific function or domain; by the way they are trained on vast amounts of unlabeled data, such as through self-supervised learning, and by their use as pre-trained foundations on which other AI systems are built}. \textit{These AI models or systems may be considered advanced (demonstrating high performance), and may also be characterized by their potentially high-impact capabilities due to their market reach, and associated risk.}\end{indentedpara}

The resulting six dimensions: 1) large scale; 2) performs a variety of tasks; 3) trained on vast unlabeled data; 4) foundation for other AI systems; 5) advanced; and 6) high impact potential due to market reach and risk profile. Dimension five scopes by model capability relative to the state of the art, while dimension six scopes by the consequence arising from the breadth of a system's deployment. Legal definitions were tested against these six dimensions. A sub-class definition, such as \textit{generative AI} or \textit{foundation model,} was eligible provided the instrument defined it distinctly and for governance.

The framework also fixes the relationship to frontier AI, the most common adjacent term in policy debate. Frontier AI is not a separate class but a subset picked out by dimension five: models at or near the state of the art. A GPAI model need not be frontier, since generality describes the range of tasks performed rather than capability relative to contemporaries.

California's Senate Bill 53 (2025) sets out that relationship in statute, defining a frontier model as a foundation model trained above 10\textsuperscript{26} operations – the parent class, plus a capability threshold, and sits outside our scope for the reasons given at §2 \citep{sb53TFAIA2025}. Where this paper uses "frontier AI" it refers to that most-capable subset, and reserves general-purpose AI for the parent class.

\subsubsection*{3.2.3 Eligibility test}

An instrument that merely used a term such as “generative AI” or “foundation model” without defining it, or gave such terms as illustrative examples inside a broader definition of AI generally, was not treated as carrying a GPAI definition. Definitions of AI in general were excluded, however capable of covering GPAI incidentally. An instrument passes the filter where its text identifies a GPAI class or recognized sub-class as the systems it governs or evaluates, and that same finding creates the definition record. An instrument that names a GPAI-equivalent term without defining it does not pass and generates no definition record, though it may still bear on whether the instrument's governance-area provisions respond to GPAI-specific risks.

Each GPAI definition was coded by how it brings systems into scope, under five mechanisms: \textit{definition-based}, where any system matching a written description is in scope; \textit{capability-based}, by crossing a quantitative threshold such as training compute or parameter count; \textit{use-context-based}, depending on how or where a model is deployed rather than on the model itself; \textit{designation-based}, where a public authority has formally designated the system; and \textit{output-modality-based}, turning on the kind of content produced rather than the breadth of tasks performed. Controlled vocabularies for each are in Appendix C.

Where a definition brings GPAI into scope through more than one route, each mechanism was recorded and the definition classed as a combination, with any numeric threshold recorded verbatim and its metric. Only instruments containing a definitional moment, whether a definitional clause, a glossary entry or an operative scoping sentence, were considered. The record-granularity rules are given in Appendix C.

\subsection*{3.3 Key governance areas}

The analytical core of the framework is built around four governance areas, against which every jurisdiction was mapped in full. The areas were operationalized as coding categories with explicit severity thresholds, so that inclusion depended on a stated test, and absence of coverage was recorded as an explicit negative finding, meaning every inclusion and exclusion traces back to a threshold. The four key governance areas are as follows:

\subsubsection*{3.3.1 Systemic risk assessment}

This area captures duties to assess GPAI risk at the system, sector or society level, pitched above any individual instance, deployment, decision or affected person. The point at which the duty attaches, whether upstream, downstream, whole-lifecycle or performed by the state itself, is recorded on the provision record. A duty falls within the area whether it binds a developer before release, a deployer in operation, or an authority monitoring the market from outside the supply chain.

Taking its origin from the finance sector, the Bank for International Settlements defines systemic risk as “a risk of disruption to financial services that is caused by an impairment of all or parts of the financial system and has the potential to have serious negative consequences for the real economy.” \citep{caruana2010SystemicRisk} Secondly, the OECD offers the risk of a serious AI hazard as equivalent, defined as “an event, circumstance or series of events where the development, use or malfunction of one or more AI systems could plausibly lead to a serious AI incident or AI disaster.” \citep{DefiningAIIncidents2024} By this criterion, the risk assessed must be one that could plausibly reach either the threshold of a serious AI hazard, or sector disruption. Consequences falling below this severity threshold do not satisfy this governance area.

Risk arising from the structure of a market – concentration of frontier capability, dependence on a few compute or model providers, tipping, entrenchment or lock-in – satisfies this area where three conditions hold. First, severity: the risk must be capable of impairing all or part of a sector, with potential for serious negative consequences for the real economy. Second, the risk must be GPAI-driven, arising from the systems governed rather than from market structure at large. Third, the provision must direct an identified actor to assess or keep under review that risk, including under a standing institutional mandate where the issuing body commits to performing the function itself.

\subsubsection*{3.3.2 Evaluation and verification}

This governance area captures mechanisms for measuring what a model does (evaluation), or checking whether claims about it hold (verification): capability, safety, red-teaming and human uplift testing are evaluative, while conformity assessment, certification, audit and attestation are verificatory. The two are treated as one area because the instruments often treat them together, though the distinction is recoverable from the recorded mechanism type. Unlike the other three areas, membership turns on a functional test not a severity threshold. The provision must direct an identified actor to measure or check, and identify what is measured, so a duty to audit at a stated frequency that is silent on what an audit examines does not qualify. Whether an instrument reaches the model or a system built on it is recorded rather than assumed.

Evaluation and verification provisions are coded on five dimensions, from mechanism type through to the object of evaluation, with controlled vocabularies given in Appendix C. Provisions carrying mechanisms with differing timings are split so that no mechanism lacks the exact lifecycle point on the record (§4.3).

\subsubsection*{3.3.3 Prohibitions, serious-incident monitoring and detection}

This area captures prohibitions on specific GPAI uses or capabilities, together with the monitoring and detection machinery surrounding them. These two elements are dimensions of one area, not a compound requirement; an enacted prohibition silent on detection has that absence recorded, identifying prohibitions that lack enforcement infrastructure.

Prohibitions were classified by target in a three-part scheme: a \textit{capability prohibition} targets what a system can do regardless of use; a \textit{use prohibition} targets a specific application or deployment context regardless of capability; an \textit{outcome prohibition} targets a specified harmful result regardless of design or intent. This is finer-grained than a behaviors-and-uses pair, under which almost all prohibitions would resolve to a single value because legislatures overwhelmingly regulate the conduct of legal persons. The distinction bears directly on serious incident detection: capability-targeting prohibitions imply \textit{ex ante} evaluation, use-targeting prohibitions imply conduct enforcement, and outcome-targeting prohibitions imply serious incident reporting and liability.

Prohibitions carried in any legal vehicle, including criminal law, are recorded. The structural form of each prohibition (absolute ban, conditional ban, moratorium, or restriction) was coded from the prohibition text itself. This paper treats GPAI red lines as the subset of prohibitions on “AI uses or behaviors deemed too dangerous to permit under any circumstances”, following the Global Call for AI Red Lines. \citep{millernguyen2025globalcall}

We define a \textit{serious incident} using the OECD, which sets the threshold at death or serious harm to health, serious and irreversible disruption to critical infrastructure, serious violation of human rights or breach of laws protecting fundamental, labor and intellectual property rights, or serious harm to property, communities or the environment \citep{DefiningAIIncidents2024}. The definition is reproduced in full at Appendix C. It is from this definition that we coded a "fundamental rights" risk domain.

\subsubsection*{3.3.4 Serious incident reporting}

Where §3.3.3 concerns catching a breach, this area concerns notification once a serious harm has occurred. The OECD definition above supplies the test, so a provision satisfies where it creates a duty to notify and the trigger engages one or more of the OECD harms. Duties triggered by events falling below that threshold, including AI hazards and near-misses where no harm materialized, will not satisfy the area. The fields recorded for each reporting duty, from bearer and addressee through to consequence and channel, are listed in Appendix C.

\section*{4. Methodology}

\subsection*{4.1 Source acquisition}

Research began from the OECD AI Policy Navigator and the UNESCO AI Ethics and Governance country profiles as starting orienting inventories, and proceeded to independent analysis of each jurisdiction’s primary sources: official gazettes, consolidated statute databases, implementing decrees, and ministerial and regulatory guidance. For each jurisdiction, the search covered the websites of the ministries and regulators identified as holding an AI mandate extensively. Summaries, secondary commentary and news reporting were not accepted as sources; every instrument was verified against its authoritative text before entry into the dataset. The source acquisition and coding were conducted between 8 June 2026 and 30 July 2026. Instruments adopted or published on or before 30 July 2026 were eligible for inclusion; developments after that date were not otherwise coded. Each jurisdiction was assigned to one of the five researchers, with searches conducted in languages native to the researcher, CEFR B2 proficiency or higher, or in English (see §4.5). Searches used a constant set of terms across all jurisdictions, applied in the language of the source: the GPAI family of terms set out at §3.2, together with terms for the four governance areas, including (but not limited to): systemic risk assessment, evaluation, verification, testing, audit, prohibition, serious incident, monitoring, and reporting. The same set was applied whether or not a jurisdiction was expected to hold relevant material. A search was treated as complete when the researcher had examined every source class above for that jurisdiction and the resulting per-jurisdiction report had been put to an external expert without a further instrument being identified.

A particular governance area was recorded as a confirmed absence for a jurisdiction only where that search had been completed and returned no positive provision for that governance area across all mapped instruments, the assessment being made at Step 5 of the decision flow (§4.2.1). Where the search could not be completed, or where the team could not satisfy itself that the relevant material had been examined, the area was recorded as “coverage uncertain”, requiring resolution by further investigation, or two-researcher validation. Confirmed absence is therefore a documented negative finding, but under a specified procedure, not a claim that no such provision exists anywhere in the jurisdiction's law.

\subsection*{4.2 Scope testing}

\subsubsection*{4.2.1 The decision flow, and GPAI anchoring}

Every candidate instrument was tested against an eight-step decision flow to assist systematic scope testing (see Appendix B for more detailed information). The figure below is a working tool – its purpose was to make scope determinations reproducible across five researchers and twenty jurisdictions. The flow is read top to bottom, following the labeled branch at each step. Blue nodes are tests, amber nodes are branches that route an instrument past a subsequent step.

The disjunctive relationship between Steps 2 and 5, and the effect of the strategy-scaffolding branch at Step 3, are set out in Appendix B. Two further features matter here. Step 5 was assessed for every in-scope instrument, including those already admitted at Step 2, because it determines which provisions are extracted and where confirmed absences are created: a Step 2 pass settles scope, not governance coverage. The brightline scan within Step 5 runs before the general test of GPAI-specificity, because an instrument drafted in AI-general terms may nonetheless carry a loss-of-control or CBRN-uplift provision, and running the general test first tends to exclude it before that provision is reached. The sequence concludes at Step 8, where the scope decision and tier classification follow.

\addvspace{10pt}

\textbf{Figure 4.2.1 The decision flow}

\begin{figure}[!ht]
\centering
\includegraphics[width=1.0\textwidth]{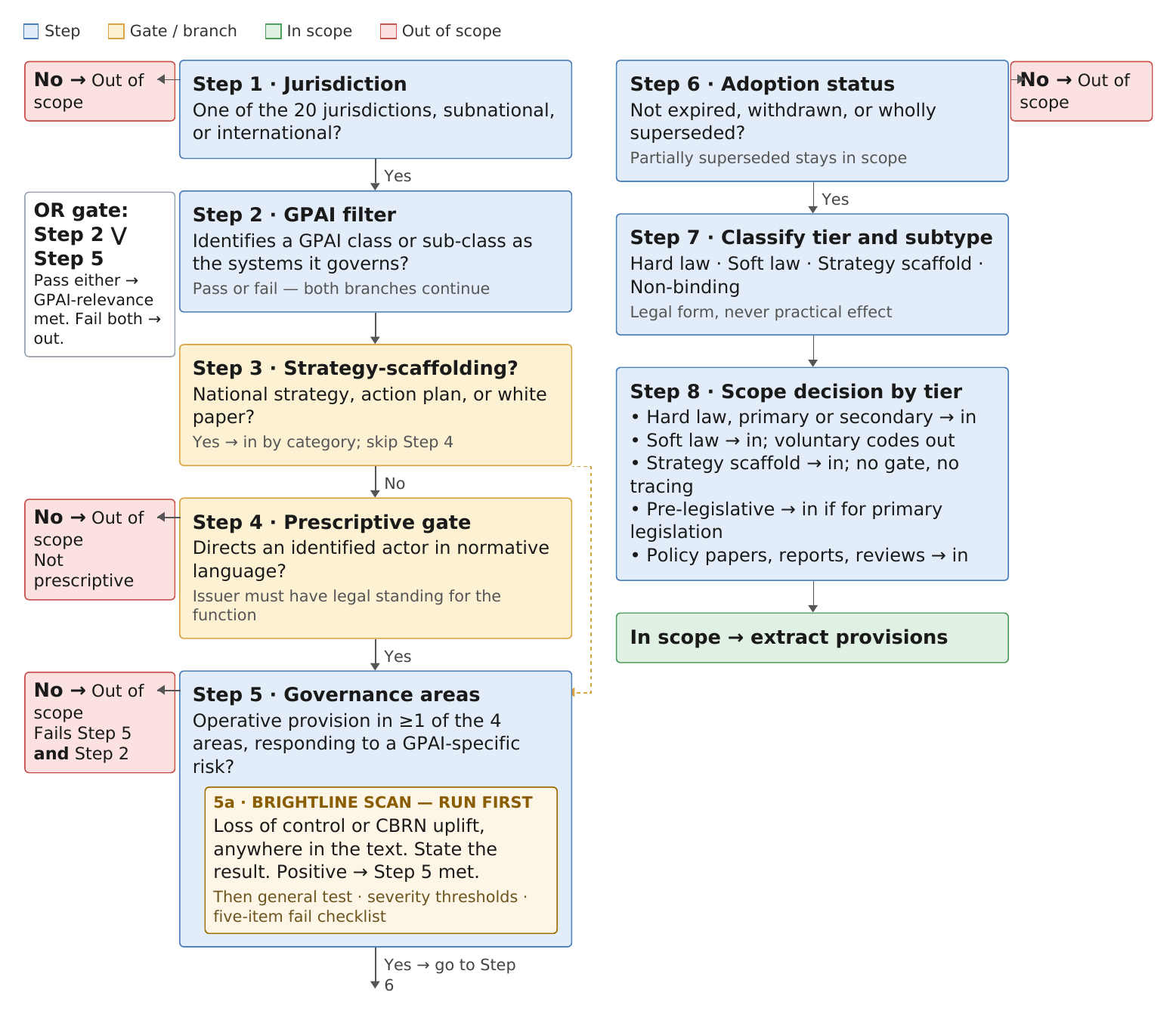}
\label{fig:figure-4-2-1-the-decision-flow}
\end{figure}
\FloatBarrier

{\footnotesize \textit{Figure 4.2.1. The scope-testing diagram used throughout mapping.}}

\subsubsection*{4.2.2 GPAI-specific risk}

The second branch in the decision flow above is engaged where the GPAI definition test was not satisfied, and required that provisions respond to GPAI-specific risks, operationalized through three rules. First, two particular GPAI risk classes were treated as brightlines because these risks are inherent to advanced general-purpose capabilities and do not attach to narrow AI systems: loss of control and CBRN uplift.

Following the 2025 Athens Roundtable and the International AI Safety Report, we define loss of control as covering systems that cannot be reliably controlled, corrected, interrupted or shut down, that resist or circumvent human oversight, or that recursively self-improve \citep{tfsAthensRoundtableRecap2025}, \citep{iasr2026}. CBRN uplift covers advanced AI systems significantly increasing the capability of state or non-state actors to develop or use chemical, biological, radiological or nuclear weapons. A provision addressing either class satisfies the brightlines, and no other risk class was treated this way.

Second, a functional distinction was applied between provisions that operate on GPAI-specific risk and provisions merely motivated by GPAI developments, such as those addressing AI risks generally, while citing a GPAI development as context. Only the former – provisions that operate on GPAI-specific risk – satisfy the test. Third, an instrument drafted in technology-neutral language was treated as GPAI-relevant where two conditions were met together: its own text identifies a GPAI-specific harm (such as synthetic-media generation or AI voice cloning) as its regulatory target; and its operative text explicitly extends its prohibitions or obligations to cover AI-generated or AI-altered material, outputs, or conduct. A sample of twelve instruments that were out-scoped can be downloaded at Appendix E.

\subsubsection*{4.2.3 Excluded governance levers and domains}

Two governance levers fall outside this study. Voluntary industry commitments are excluded because the issuer holds no governance mandate, subject to the presumption-of-conformity exception above. Judicial interpretations are excluded because a court construing an instrument does not itself create the obligation the instrument imposes; where a judicial body issues an administrative instrument governing AI use within its own institution, that is an instrument like any other and is in scope, which is why Brazil's 2025 Resolution CNJ No. 615 appears in the corpus, while the case law of any jurisdiction does not.

Instruments in the defense domain, and those pertaining to national security, were excluded from scope regardless of how a GPAI system figures within them. The exclusion is deliberate rather than incidental. These domains are frequently subject to statutory carve-outs from the AI frameworks mapped here – the EU AI Act's exclusion of systems placed on the market exclusively for military, defense or national-security purposes is the clearest instance, and comparable carve-outs recur across the sample – with the consequence that AI governance in these domains proceeds under a separate legal regime, and frequently through unpublished instruments, whose absence from the public record cannot be distinguished from non-existence. Any defense counts would therefore be systematically incomplete, and the findings describe civilian GPAI governance only.

\subsection*{4.3 Extraction and coding}

The data that informs this paper was held in a relational database linking jurisdictions, governance actors, instruments, GPAI definitions, and provisions, with automated validation formulas enforcing structural rules (Appendix D). Extraction was model-assisted for much of the corpus which can be read in detail at §8.

For every in-scope instrument, provisions falling within the four key governance areas were extracted from the full primary text, never from summaries. Provisions were extracted at a default granularity of one record per section of the instrument, with all applicable governance areas tagged to that single record and the area-specific fields completed for each. The two circumstances requiring a split beyond that default are set out in Appendix C.

Provision text was recorded verbatim, in both the original language and in English where the source was non-English, and every coded dimension follows the controlled vocabularies of the project codebook (Appendix C). Risk domains were coded only where a specific passage of the instrument's text grounds the code, never inferred from the instrument's sector or the obligation bearer's industry. Where an instrument mixed GPAI-specific and technology-neutral provisions, all governance-area provisions were extracted, and those using only technology-neutral language were then tested for a traceable connection to the instrument's GPAI-specific content under the rule at Appendix B. Where no such trace existed, the provision was removed. Verbatim provision text and citations were finalised only against enacted, adopted or in-force text. Legislative debate records, government factsheets and professional summaries were treated only as corroborating sources. A separate rule governed institutional architecture, which can also be read under Appendix B.

\subsection*{4.4 Data quality assurance}

Expert interviews were conducted for Canada, Chile and Japan in order to verify findings. The other seventeen jurisdictions received asynchronous expert verification against per-jurisdiction reports. All interviewees elected the Chatham House Rule, so contributions inform the analysis but are not attributed by name.

\subsection*{4.5 Limitations}

This study codes what AI middle-power governance instruments require, of whom, and with what stated consequence, not what happens in practice. We do not observe whether reports are made, whether authorities act on them, or whether a weak-looking obligation is treated as binding by those it addresses. Only expert interview evidence recovers that, and we hold it for few jurisdictions.

Language coverage among the research team varied. For instruments published in Arabic, German, Hebrew, and Korean the team lacked native or CEFR B2 proficiency and worked under a documented-limitation protocol combining AI-assisted translation, with round-trip verification and expert resolution of divergences (§8). Every AI-assisted record carries a confidence flag, and low-confidence records require two-researcher validation before the coding is accepted. There were no such flagged records at the time of publication. This reduces, but cannot eliminate, the risk of translation-borne coding error.

The dataset is a snapshot taken when the mapping period closed on 30 July 2026. GPAI governance is currently moving quickly enough that the findings will date within months. The relative strength or weakness of our convergence analysis also depends on the size and composition of the jurisdiction sample set. Our sample is purposive rather than random: the composite scoring at §3.1 shows that the twenty jurisdictions span a meaningful range of the AI middle-power space, but does not make them representative of it, so the findings should not be generalised. The research is also confined to four governance areas, so where we report that a jurisdiction's governance is thin, that finding is confined to those four areas.

The definitional dimensions (§3.2.2), and the governance areas with their severity thresholds (§3.3), are built from the EU AI Act, the OECD and the G7 Hiroshima Process, and the first is also an instrument in the sample. Findings that the EU covers the framework most completely therefore reflect the framework's provenance, as much as the instrument's content. We mitigate this by setting the systemic risk threshold below the EU's construction (§5.3) and by admitting sub-class definitions on the same footing as the parent class (§3.2.3), but the framework remains built, in part, from a particular regulatory tradition.

Five narrower caveats follow. Because the framework captures duties triggered by realised harm at the OECD threshold (§3.3.4), an instrument requiring notification of a near-miss does not register, and the dataset cannot distinguish a jurisdiction with no near-miss duty from one whose duty fell outside the frame. Secondly, governance at the international tier is only partially mapped: binding treaties are represented through our testing of the Council of Europe Framework Convention, but soft-law and non-binding multilateral initiatives are not. Third, cross-referencing is recorded only where an instrument names another in-scope instrument in its operative text, so an instrument that operates on another without naming it leaves no trace. Fourth, because the coding rule creates one record per section (§4.3), provision counts show where governance content is dense, rather than where obligations are strong. Fifth, the governance actor coding records the powers a body holds as constituted, not whether it has the staff, budget or technical capability to exercise them (§7.3), and a governance actor is recorded only where an instrument or provision names it, so bodies active in practice but named nowhere in the mapped instruments do not appear.

In addition, the scope rule at §3.2.2 admits an instrument whose only GPAI anchor is a defined sub-class such as generative AI, even where that definition engages none of the six dimensions of generality. Because the findings then report that governance attaches to outputs and to deployment contexts, part of that pattern could, in principle, be produced by the inclusion rule rather than observed in the instruments. We tested this by rerunning the main findings on a restricted sample, excluding the twelve instruments anchored only that way, which carry twelve percent of positive provisions. However, the pattern still holds, and sharpens: the excluded instruments are disproportionately non-binding, and are disproportionately concerned with deploying or using generative tools, so excluding them shifts the corpus toward the model layer, rather than away from it. Finally, the dataset records only instruments that entered scope.

\section*{5. Findings}

{\small
\hyphenpenalty=50\exhyphenpenalty=50\emergencystretch=1em
\setlength{\tabcolsep}{4pt}
\renewcommand{\arraystretch}{1.15}
\begin{tabularx}{\textwidth}{YC{0.145\textwidth}C{0.145\textwidth}C{0.145\textwidth}C{0.145\textwidth}}
\multicolumn{5}{@{}>{\raggedright\arraybackslash}p{\dimexpr\textwidth-2\tabcolsep}@{}}{\normalsize \textbf{Table 5 Findings}} \\[5pt]
\toprule
\cellcolor[HTML]{EFEFEF}\textbf{Jurisdiction} & \cellcolor[HTML]{EFEFEF}\textbf{Systemic risk assessment} & \cellcolor[HTML]{EFEFEF}\textbf{Evaluation \& verification} & \cellcolor[HTML]{EFEFEF}\textbf{Prohibitions \& incident monitoring} & \cellcolor[HTML]{EFEFEF}\textbf{Serious incident reporting} \\
\midrule
\endfirsthead
\toprule
\cellcolor[HTML]{EFEFEF}\textbf{Jurisdiction} & \cellcolor[HTML]{EFEFEF}\textbf{Systemic risk assessment} & \cellcolor[HTML]{EFEFEF}\textbf{Evaluation \& verification} & \cellcolor[HTML]{EFEFEF}\textbf{Prohibitions \& incident monitoring} & \cellcolor[HTML]{EFEFEF}\textbf{Serious incident reporting} \\
\midrule
\endhead
\midrule\multicolumn{5}{r}{\textit{continued on next page}}\\
\endfoot
\bottomrule
\endlastfoot
European Union & \cellcolor[HTML]{1F5C8B}\textcolor{white}{\textbf{H}} & \cellcolor[HTML]{1F5C8B}\textcolor{white}{\textbf{H}} & \cellcolor[HTML]{1F5C8B}\textcolor{white}{\textbf{H}} & \cellcolor[HTML]{1F5C8B}\textcolor{white}{\textbf{H}} \\
France (domestic) & \cellcolor[HTML]{A8D5BA}\textbf{H} & \cellcolor[HTML]{55A196}\textcolor{white}{\textbf{S}} & \cellcolor[HTML]{55A196}\textcolor{white}{\textbf{S}} & \cellcolor[HTML]{F2F2F2}\textcolor[HTML]{999999}{\textbf{—}} \\
Germany (domestic) & \cellcolor[HTML]{F2F2F2}\textcolor[HTML]{999999}{\textbf{—}} & \cellcolor[HTML]{55A196}\textcolor{white}{\textbf{H}} & \cellcolor[HTML]{55A196}\textcolor{white}{\textbf{H}} & \cellcolor[HTML]{F2F2F2}\textcolor[HTML]{999999}{\textbf{—}} \\
Brazil & \cellcolor[HTML]{1F5C8B}\textcolor{white}{\textbf{S}} & \cellcolor[HTML]{1F5C8B}\textcolor{white}{\textbf{S}} & \cellcolor[HTML]{1F5C8B}\textcolor{white}{\textbf{H}} & \cellcolor[HTML]{1F5C8B}\textcolor{white}{\textbf{S}} \\
Republic of Korea & \cellcolor[HTML]{1F5C8B}\textcolor{white}{\textbf{H}} & \cellcolor[HTML]{1F5C8B}\textcolor{white}{\textbf{H}} & \cellcolor[HTML]{1F5C8B}\textcolor{white}{\textbf{H}} & \cellcolor[HTML]{55A196}\textcolor{white}{\textbf{S}} \\
Australia & \cellcolor[HTML]{1F5C8B}\textcolor{white}{\textbf{S}} & \cellcolor[HTML]{55A196}\textcolor{white}{\textbf{H}} & \cellcolor[HTML]{1F5C8B}\textcolor{white}{\textbf{S}} & \cellcolor[HTML]{55A196}\textcolor{white}{\textbf{H}} \\
\textit{New South Wales} & \cellcolor[HTML]{F2F2F2}\textcolor[HTML]{999999}{\textbf{—}} & \cellcolor[HTML]{55A196}\textcolor{white}{\textbf{H}} & \cellcolor[HTML]{55A196}\textcolor{white}{\textbf{H}} & \cellcolor[HTML]{55A196}\textcolor{white}{\textbf{H}} \\
United Kingdom & \cellcolor[HTML]{55A196}\textcolor{white}{\textbf{S}} & \cellcolor[HTML]{55A196}\textcolor{white}{\textbf{H}} & \cellcolor[HTML]{1F5C8B}\textcolor{white}{\textbf{S}} & \cellcolor[HTML]{1F5C8B}\textcolor{white}{\textbf{S}} \\
India & \cellcolor[HTML]{55A196}\textcolor{white}{\textbf{S}} & \cellcolor[HTML]{55A196}\textcolor{white}{\textbf{H}} & \cellcolor[HTML]{55A196}\textcolor{white}{\textbf{H}} & \cellcolor[HTML]{1F5C8B}\textcolor{white}{\textbf{H}} \\
Singapore & \cellcolor[HTML]{DDE6C4}\textbf{H} & \cellcolor[HTML]{55A196}\textcolor{white}{\textbf{H}} & \cellcolor[HTML]{1F5C8B}\textcolor{white}{\textbf{H}} & \cellcolor[HTML]{55A196}\textcolor{white}{\textbf{H}} \\
Japan & \cellcolor[HTML]{55A196}\textcolor{white}{\textbf{H}} & \cellcolor[HTML]{55A196}\textcolor{white}{\textbf{H}} & \cellcolor[HTML]{55A196}\textcolor{white}{\textbf{H}} & \cellcolor[HTML]{55A196}\textcolor{white}{\textbf{H}} \\
Kenya & \cellcolor[HTML]{55A196}\textcolor{white}{\textbf{H}} & \cellcolor[HTML]{55A196}\textcolor{white}{\textbf{H}} & \cellcolor[HTML]{55A196}\textcolor{white}{\textbf{H}} & \cellcolor[HTML]{A8D5BA}\textbf{H} \\
Taiwan & \cellcolor[HTML]{55A196}\textcolor{white}{\textbf{H}} & \cellcolor[HTML]{55A196}\textcolor{white}{\textbf{H}} & \cellcolor[HTML]{1F5C8B}\textcolor{white}{\textbf{S}} & \cellcolor[HTML]{F2F2F2}\textcolor[HTML]{999999}{\textbf{—}} \\
Canada & \cellcolor[HTML]{A8D5BA}\textbf{H} & \cellcolor[HTML]{55A196}\textcolor{white}{\textbf{H}} & \cellcolor[HTML]{1F5C8B}\textcolor{white}{\textbf{H}} & \cellcolor[HTML]{F2F2F2}\textcolor[HTML]{999999}{\textbf{—}} \\
Israel & \cellcolor[HTML]{F2F2F2}\textcolor[HTML]{999999}{\textbf{—}} & \cellcolor[HTML]{55A196}\textcolor{white}{\textbf{H}} & \cellcolor[HTML]{55A196}\textcolor{white}{\textbf{H}} & \cellcolor[HTML]{55A196}\textcolor{white}{\textbf{H}} \\
United Arab Emirates & \cellcolor[HTML]{DDE6C4}\textbf{H} & \cellcolor[HTML]{55A196}\textcolor{white}{\textbf{S}} & \cellcolor[HTML]{55A196}\textcolor{white}{\textbf{S}} & \cellcolor[HTML]{F2F2F2}\textcolor[HTML]{999999}{\textbf{—}} \\
\textit{Dubai (subnational)} & \cellcolor[HTML]{55A196}\textcolor{white}{\textbf{H}} & \cellcolor[HTML]{F2F2F2}\textcolor[HTML]{999999}{\textbf{—}} & \cellcolor[HTML]{55A196}\textcolor{white}{\textbf{H}} & \cellcolor[HTML]{F2F2F2}\textcolor[HTML]{999999}{\textbf{—}} \\
\textit{Dubai Intl. Financial Centre} & \cellcolor[HTML]{F2F2F2}\textcolor[HTML]{999999}{\textbf{—}} & \cellcolor[HTML]{55A196}\textcolor{white}{\textbf{H}} & \cellcolor[HTML]{1F5C8B}\textcolor{white}{\textbf{H}} & \cellcolor[HTML]{F2F2F2}\textcolor[HTML]{999999}{\textbf{—}} \\
Peru & \cellcolor[HTML]{F2F2F2}\textcolor[HTML]{999999}{\textbf{—}} & \cellcolor[HTML]{F2F2F2}\textcolor[HTML]{999999}{\textbf{—}} & \cellcolor[HTML]{1F5C8B}\textcolor{white}{\textbf{H}} & \cellcolor[HTML]{F2F2F2}\textcolor[HTML]{999999}{\textbf{—}} \\
Switzerland & \cellcolor[HTML]{F2F2F2}\textcolor[HTML]{999999}{\textbf{—}} & \cellcolor[HTML]{F2F2F2}\textcolor[HTML]{999999}{\textbf{—}} & \cellcolor[HTML]{55A196}\textcolor{white}{\textbf{H}} & \cellcolor[HTML]{F2F2F2}\textcolor[HTML]{999999}{\textbf{—}} \\
Chile & \cellcolor[HTML]{F2C879}\textbf{H $\triangleright$} & \cellcolor[HTML]{F2C879}\textbf{H $\triangleright$} & \cellcolor[HTML]{F2C879}\textbf{H $\triangleright$} & \cellcolor[HTML]{F2F2F2}\textcolor[HTML]{999999}{\textbf{—}} \\
Nigeria & \cellcolor[HTML]{F2C879}\textbf{H $\triangleright$} & \cellcolor[HTML]{F2C879}\textbf{H $\triangleright$} & \cellcolor[HTML]{F2F2F2}\textcolor[HTML]{999999}{\textbf{—}} & \cellcolor[HTML]{F2F2F2}\textcolor[HTML]{999999}{\textbf{—}} \\
South Africa & \cellcolor[HTML]{F2F2F2}\textcolor[HTML]{999999}{\textbf{—}} & \cellcolor[HTML]{F2F2F2}\textcolor[HTML]{999999}{\textbf{—}} & \cellcolor[HTML]{F2F2F2}\textcolor[HTML]{999999}{\textbf{—}} & \cellcolor[HTML]{F2F2F2}\textcolor[HTML]{999999}{\textbf{—}} \\
\end{tabularx}}

\textbf{Key:} H = a horizontal instrument provides the coverage at the strongest tier. \\
S = coverage at the strongest tier is sectoral only.

\begin{table}[!ht]
\centering
{\small
\hyphenpenalty=50\exhyphenpenalty=50\emergencystretch=1em
\setlength{\tabcolsep}{4pt}
\renewcommand{\arraystretch}{1.15}
\begin{tabularx}{\textwidth}{C{0.035\textwidth}Y}
\toprule
\cellcolor[HTML]{1F5C8B} & Hard law, in force \\
\cellcolor[HTML]{55A196} & Soft law, in force \\
\cellcolor[HTML]{A8D5BA} & Strategy scaffold \\
\cellcolor[HTML]{DDE6C4} & Non-binding (published) \\
\cellcolor[HTML]{F2C879}$\triangleright$ & Proposed / in draft or committee ($\triangleright$) \\
\cellcolor[HTML]{F2F2F2}\textcolor[HTML]{999999}{—} & Confirmed absent (—) \\
\bottomrule
\end{tabularx}}
\end{table}
\FloatBarrier

{\footnotesize \textit{Table 5. GPAI governance coverage by jurisdiction and governance area. Rows are jurisdictions; columns are the four governance areas. Each cell describes the most legally forceful in-scope instrument addressing that area in that jurisdiction. Colour gives its tier: hard law in force (darkest), soft law, strategy scaffold, proposed or in draft (lightest). Grey records a confirmed absence. The letter gives the scope of that instrument: H where a horizontal AI instrument provides the coverage, S where coverage is sectoral only. France and Germany appear on two rows each, one for domestic instruments and one for directly applicable EU law.}}

The table above records, for each jurisdiction and each of the four GPAI governance areas, the most legally forceful instrument addressing that area, described in the tier it sits in, and whether it governs AI horizontally or reaches AI only from within a sector. Concerning horizontal law, only the European Union binds across all four governance areas, while Japan and Kenya register horizontal coverage in every area at the same time as holding no hard law at all. South Africa is empty across all four, and Chile and Nigeria carry nothing but proposals, so their coverage exists only if their bills pass. The table says nothing about the volume of governance activity in an area – a single provision or twenty may produce the same cell – because the question here is what kind of instrument a jurisdiction has reached for, not how densely it has legislated.

The findings follow the research questions. Section 5.1 disposes of two threshold matters: the only binding international treaty on AI, which falls outside scope, and one jurisdiction holding no general-purpose AI governance at all. Section 5.2 answers how general-purpose AI is being legally defined, and what scoping criteria those definitions employ. Sections 5.3 to 5.6 take the four key governance areas in the order of the accountability chain – systemic risk assessment, evaluation and verification, prohibitions with monitoring and detection, and serious incident reporting – and together answer how legal architectures distribute obligations across them. Section 5.7 answers where sectoral instruments carry the governance load, and which mechanisms they employ. Section 5.8 answers how instruments interact across tiers, and which architectures recur. The claim that this section lands on is that there is convergence in GPAI governance subject matter and divergence in almost everything else, which §6 Discussion then develops.

\subsection*{5.1 International treaties, and South Africa}

The Council of Europe Framework Convention on Artificial Intelligence, opened for signature at Vilnius in September 2024, is the only binding international treaty on artificial intelligence, and it fell outside our scope. Its Article 2 defines an artificial intelligence system on the OECD formulation, neither GPAI-risk brightline appears in its text, and its Article 16 risk and impact framework responds to AI systems at large rather than to the capabilities of GPAI models. The eight jurisdictions in the sample that negotiated or signed it therefore acquire no GPAI-directed obligation by having done so.

South Africa did not, during the window of this research, have any in-scope provisions. After a search of its statute book, regulatory guidance, strategy documents, and following expert verification, no instrument met the scope test, and the only candidate, its Draft National AI Policy, was withdrawn in April 2026 before adoption. The absence is a finding about South Africa rather than a gap in the mapping.

\subsection*{5.2 Defining general-purpose AI}

This section asks how AI middle-power jurisdictions define the class of general-purpose AI models they are governing, and what follows for the obligations they attach. It lands on a mismatch: definitional effort is heaviest where legal consequence is lightest.

Among AI middle-powers there is little shared understanding or shared commitment to a unified general-purpose AI definition: seventeen of the twenty jurisdictions hold at least one definition, the exceptions being South Africa, Peru and Switzerland, so the gap appears not to be ignorance of the concept, but a decision not to anchor obligations to it. Of the 101 in-scope instruments, 42 carry a GPAI definition or functional equivalent, so 58 percent seek to shape a class of models without stating what that class is.

In fact, definition and obligation are only loosely coupled across the corpus. After the fifty-nine instruments that impose at least one positive GPAI provision while defining nothing, a small group runs the other way, defining the class while governing none of it: the UAE’s 2025 white paper Towards a Future of Responsible AI, the European Commission’s 2026 proposal for a Cloud and AI Development Act, Cyber Security New South Wales’ 2023 end-user guidance on generative AI, and Singapore’s 2026 PDPC Advisory Guidelines on personal data in generative AI: these are all documents that set direction or explain an existing regime, so definitional effort runs ahead of regulatory effort.

Where jurisdictions define GPAI, they converge on how they scope it. Fifteen of the seventeen bring a model into scope by matching a written description, by naming the kind of content it produces, or by both. This technique is the same whether a jurisdiction holds as many as six GPAI definitions – as Australia, the United Kingdom and Japan do – or one. Only the European Union and Korea scope by anything else: both add a capability trigger and a power to designate a model into scope, and Korea adds a use-context test. This is the divergence that matters for a provider deciding whether it is caught, because in fifteen jurisdictions the question is whether a description fits; in only two, it is a threshold or a designation that settles it.

The same pattern appears in which dimensions of generality the definitions engage (Table 5.2). Training on vast unlabeled data is the most frequent, on 22 of 52 definitions. Just over two thirds do not engage generality itself, which is the capacity to perform a variety of tasks, describing breadth instead by listing content types, and eighteen definitions engage none of the six dimensions at all because they define generative AI, which is a GPAI subclass named for what a model produces \citep{lorenz2023initial}. Twelve instruments rest on such a definition alone, and the sensitivity analysis at §4.5 reports what happens to the findings when they are excluded. Only the EU engages all six, and Australia alone reaches five.

Two features of Table 5.2 carry the argument. The first is how few definitional dimensions any jurisdiction engages. The second is that holding a definition and defining generality are different things. Canada, Nigeria and Taiwan each hold a definition that engages no dimension at all, because each defines generative AI by what it produces rather than by what it can do. France and Germany are shown on their domestic definitions alone; read with the five EU definitions that bind them through direct effect, each would register all six dimensions and hard law, and the gap between the two readings measures what each has authored rather than inherited.

The same displacement appears in the technology layer. The OECD distinguishes a model, whose parameters are fixed after training, from a system, which adds the components needed to operate it \citep{OECD_ExplanatoryMemorandum2024}, and the distinction decides who bears a duty: obligations on a model fall on its developer, obligations on a system can reach a deployer or integrator. Half of the 52 definitions describe a model, and only four describe a system in the EU AI Act’s sense (Article 3(66)). The rest do not resolve the technology layer at all: eight use the word “system” while listing model-level properties, five name both, two define a service, and eight refer only to “AI” or “technology”. The EU is the only jurisdiction separating the two concepts in binding law, at Articles 3(63) and 3(66) of the 2024 AI Act. Elsewhere, it is unclear whether a rule reaches the training of a model, the deployment of a system built on it, or both, and the most common term in the corpus – generative AI – is used inconsistently across layers within single jurisdictions.

Definitional practice and legal force move in opposite directions: only five hard-law instruments in the sample define GPAI or similar, which are the 2024 EU AI Act, Korea's 2025 Framework Act and its 2026 Enforcement Decree, Australia's 2024 Online Safety DIS Standard, and Brazil's 2025 Resolution CNJ No. 615. The remaining fifteen hard-law instruments – three quarters of the binding corpus – impose obligations without defining the class they apply to, while nearly half of the soft-law, non-binding and strategy instruments define it. Jurisdictions are most precise about what GPAI is exactly where precision carries the least consequence.

\begin{table}[!ht]
\centering
\caption*{\textbf{Table 5.2. GPAI definitions by jurisdiction: dimensions and technology layer}}
\label{tab:table-5-2-gpai-definitions-by-jurisdicti}
{\scriptsize
\hyphenpenalty=50\exhyphenpenalty=50\emergencystretch=1em
\setlength{\tabcolsep}{2pt}
\renewcommand{\arraystretch}{1.15}
\begin{tabularx}{\textwidth}{L{0.19\textwidth}C{0.050\textwidth}C{0.038\textwidth}C{0.038\textwidth}C{0.038\textwidth}C{0.038\textwidth}C{0.038\textwidth}C{0.038\textwidth}C{0.050\textwidth}C{0.038\textwidth}C{0.038\textwidth}C{0.038\textwidth}Y}
\toprule
\textbf{Jurisdiction} & \textbf{Defs} & \multicolumn{6}{>{\centering\arraybackslash}p{0.2783\textwidth}}{\textbf{Dimension (§3.2)}} & \textbf{Total} & \multicolumn{3}{>{\centering\arraybackslash}p{0.1341\textwidth}}{\textbf{Layer}} & \textbf{Highest tier} \\
 &  & \textbf{1} & \textbf{2} & \textbf{3} & \textbf{4} & \textbf{5} & \textbf{6} &  & \textbf{M} & \textbf{S} & \textbf{Sv} &  \\
\midrule
European Union & 5 & ● & ● & ● & ● & ● & ● & 6 & ● & ● & ○ & Hard law \\
Australia & 6 & ● & ● & ● & ● & ○ & ● & 5 & ● & ● & ● & Hard law \\
United Arab Emirates & 4 & ● & ● & ● & ● & ○ & ○ & 4 & ● & ● & ○ & Soft law \\
Japan & 6 & ○ & ● & ● & ● & ● & ○ & 4 & ● & ● & ● & Soft law \\
Republic of Korea & 4 & ● & ○ & ● & ○ & ● & ● & 4 & ○ & ● & ○ & Hard law \\
India & 3 & ● & ● & ● & ● & ○ & ○ & 4 & ● & ○ & ○ & Strategy \\
United Kingdom & 6 & ○ & ● & ● & ○ & ● & ○ & 3 & ● & ● & ○ & Soft law \\
Singapore & 5 & ○ & ● & ● & ● & ○ & ○ & 3 & ● & ● & ○ & Soft law \\
Brazil & 2 & ○ & ● & ● & ● & ○ & ○ & 3 & ○ & ● & ○ & Hard law \\
Kenya & 3 & ○ & ● & ● & ○ & ● & ○ & 3 & ● & ● & ○ & Strategy \\
Israel & 1 & ○ & ● & ● & ○ & ○ & ○ & 2 & ● & ● & ○ & Soft law \\
France\textsuperscript{a} & 2 & ○ & ○ & ● & ○ & ○ & ○ & 1 & ● & ○ & ○ & Soft law \\
Germany\textsuperscript{a} & 1 & ○ & ○ & ● & ○ & ○ & ○ & 1 & ● & ○ & ○ & Soft law \\
Chile & 1 & ○ & ● & ○ & ○ & ○ & ○ & 1 & ○ & ● & ○ & Non-binding \\
Canada & 1 & ○ & ○ & ○ & ○ & ○ & ○ & 0 & ○ & ○ & ○ & Soft law \\
Nigeria & 1 & ○ & ○ & ○ & ○ & ○ & ○ & 0 & ● & ● & ○ & Non-binding \\
Taiwan & 1 & ○ & ○ & ○ & ○ & ○ & ○ & 0 & ● & ○ & ○ & Soft law \\
Peru & 0 & ○ & ○ & ○ & ○ & ○ & ○ & 0 & ○ & ○ & ○ & — \\
South Africa & 0 & ○ & ○ & ○ & ○ & ○ & ○ & 0 & ○ & ○ & ○ & — \\
Switzerland & 0 & ○ & ○ & ○ & ○ & ○ & ○ & 0 & ○ & ○ & ○ & — \\
\bottomrule
\end{tabularx}}
\end{table}
\FloatBarrier

{\footnotesize \textit{Notes. Rows are jurisdictions; Defs gives the number of definition records held. Columns 1 to 6 are the dimensions of generality set out at §3.2: (1) large scale; (2) performs a variety of tasks; (3) trained on vast unlabeled data; (4) foundations for other AI systems; (5) advanced; (6) high-impact capabilities due to reach and risk. A filled circle means the dimension appears in at least one of that jurisdiction's definitions; an open circle means it appears in none. Total counts the filled circles. The layer columns record whether any definition targets the model (M), the system (S) or a service (Sv); a jurisdiction may register in more than one where its definitions differ, and definitions naming no layer register in none. Highest tier is the most forceful tier in which any of the jurisdiction's definitions sits. France and Germany report domestic definitions only.}}

\subsection*{5.3 Systemic risk assessment}

The first link in the accountability chain asks whether anyone is required to identify the systemic risks a GPAI model presents before those risks materialize as harm. This section establishes who bears that duty, at what point in the lifecycle, and what happens to the risk assessment once it is made. This section finds that most assessments go nowhere: fewer than a quarter of provisions require the output to be shared with a named authority so, in most jurisdictions, the analysis stays inside the organization that produced it.

Sixteen of the jurisdictions carry at least one risk assessment provision, so the function is widely recognized. The duty falls almost equally on public authorities and on providers, but that near-parity conceals two groups: the UK accounts for a third of the authority-borne provisions, and five further jurisdictions count as authority-led on the strength of a single provision each. Among the jurisdictions that have legislated in this area at any depth – the EU, Korea, Kenya, and Japan – the duty falls predominantly on the provider. Only four jurisdictions have put any of this into binding law: Korea holds five hard-law provisions, Brazil three, and the EU and Australia one each, though only two of those reach a developer. Everywhere else, the function is recognized without being required.

Where jurisdictions do act, most act on themselves. Forty-six percent of risk assessment provisions place the duty on a public authority rather than on anyone it regulates, and these are concentrated: the UK alone holds ten provisions – a third of the total – with Australia holding four, Brazil three, and the EU, Kenya, and Nigeria two each. Six further jurisdictions hold a single authority-borne provision. Provider-facing provisions are almost as numerous, but they sit in different places again – eight in the EU, five in Korea, four in Kenya and three each in Brazil and Japan. Only three provisions fall on deployers, and three on sectoral actors.

Read against the development-deployment divide, the same behavior appears as a timing pattern. Nineteen provisions carry an upstream assessment tag and six are upstream only; eighteen are downstream and five downstream only; thirteen carry both. The largest single group (31) runs across the whole lifecycle: nineteen with an authority as the assessing body, and a further twelve in assessments conducted by other parties. Therefore, fewer than one provision in three reaches the model before release, and of the nineteen upstream provisions only fourteen fall on a provider at all.

The upstream group is where drafting outruns enactment. It spans seven jurisdictions (Australia, Brazil, EU, Japan, Kenya, Korea, Singapore) and five of its provisions sit in hard law, while Kenya’s provisions and two of Brazil’s four sit in instruments not yet passed. Everything in the sample that uses mandatory assessment language outside the EU and Korea is attached to a bill: the 2026 Kenyan \citep{KEN_065}, 2024 Chilean \citep{CHL_054}, 2023 Brazilian \citep{BRA_032}, and 2025 Nigerian texts \citep{NGA_069} all place must-grade duties on providers, yet none has been enacted. Upstream assessment is therefore attempted mainly where a jurisdiction is writing new law, and achieved mainly where it already has some.

In-force, economy-wide, provider-facing hard-law duties exist in two places only. The 2024 EU AI Act binds providers of GPAI models with systemic risk, along with the 2025 General-Purpose AI Code of Practice Safety and Security chapter, which clarifies systemic risk with a catalog running from major accidents and critical-sector disruption, to loss of control and CBRN uplift. However, the GPAI Code of Practice cannot discharge a duty. Article 55(2) permits providers to "rely on codes of practice […] to demonstrate compliance", but the presumption of conformity conferred by Article 40(1) only applies to harmonized standards published in the Official Journal \citep{EUFRADEU_144}. The Digital Omnibus has since deleted the Commission's power to give a code general validity, reasoning that codes "have limited legal effect" \citep{EUFRADEU_162}. No harmonized standard for general-purpose AI models has yet been established, so the most detailed model-layer machinery in the sample therefore binds only its signatories \citep{ECstandardisation2026}.

Korea’s 2025 Framework Act, with its 2026 Enforcement Decree, also imposes economy-wide, provider-facing hard-law through risk-management and assessment duties on AI business operators; a category spanning GPAI developers and deployers as well as the provider. Three further instruments carry in-force hard-law provisions, but none reaches a model developer: Australia’s 2024 Online Safety DIS Standard binds the supplier of an internet service rather than of a model, Brazil’s 2025 Resolution CNJ No. 615 places duties on the courts over their own adoption, and Brazil’s 2026 Resolution TSE No. 23.755 reaches the distributor of synthetic campaign material, rather than the developer of the model that produced it. The convergence is that binding assessment duties exist, and that they attach to whoever the parent regime already governed.

Where the state assesses, it watches from outside. None of the authority-borne provisions is tagged upstream, so the observation happens after release, and without a duty on the developer. Most of the authority-borne provisions, across 12 jurisdictions including the EU, attach to no point in the chain at all, being risk registers, coordination centers and future regulators’ monitoring mandates – the bulk carried by the United Kingdom – and only eleven provisions point at the model layer specifically. These provisions are best read as jurisdictions building the capacity to understand what happens at the model layer, without requiring that anything happen there. Two provisions watch the market rather than the models: in its 2024 foundation-model review, the UK Competition and Markets Authority commits to monitoring partnerships in order to understand the structural features of foundation-model markets, which is the only place in the area where the assessing body is a competition authority. \citep{GBR_196} \citep{GBR_197}

The chain then breaks at disclosure. Less than a quarter of risk assessment provisions require the assessment output to be shared with a named authority (in Brazil, the EU, Korea, Australia, India, Kenya, the UK); 65 percent expressly do not (across Japan, Nigeria, Chile, the UAE, Canada, France domestically, Taiwan and Singapore), and the remainder are silent. Brazil and the EU lead on disclosure. Where the assessor is itself an authority the omission means little, since the body that would receive the output is the body that produced it. Elsewhere, it is decisive: most upstream and downstream provisions fall on providers or deployers, and no mechanism exists through which the findings reach anyone able to act on them. This is the same break as the reporting-chain gap at §5.6.

What is being assessed diverges from what the framework anticipates. Cross-cutting risks lead at 29 provisions, followed by fundamental rights and cyber at 21 each, and risks to democracy or societal manipulation at fifteen, with loss of control and CBRN uplift on eleven each, and concentration of power on five. Because a provision can be domain-agnostic in one respect while naming domains in another, these figures overlap. The catastrophic domains appear only where the model layer is either governed upstream or watched by an authority: of the eleven CBRN provisions, six are upstream, five of those in EU instruments and the sixth in Singapore’s 2025 voluntary AI Verify framework, and a further three are authority-borne; one of them Taiwan’s 2026 AI Risk Classification Framework, which places a standing mandate on sectoral authorities to classify risks across fifteen domains. \citep{TWN_219} The loss-of-control provisions divide the same way, five upstream and all of them EU, and four authority-borne.

Elsewhere, the duties that exist are keyed to fundamental-rights impact and general societal harm, language inherited from data protection, rather than to what a GPAI model can do. The UK names risk domains more comprehensively than any jurisdiction outside the EU framework, but does so only in instruments that “recommend”, and only in provisions where the state does its own monitoring. Korea holds the sample’s only self-legislated binding upstream duties, but its 2026 Framework Act names no catastrophic domain at all, coding instead to cross-cutting risk and to critical infrastructure, public health, fundamental rights and financial stability. Only the EU and Korea condition a duty on a property of the model itself, and only the EU chain connects assessment to enforceable follow-through.

\subsection*{5.4 Evaluation and verification}

The second link in the accountability chain, evaluation and verification, asks whether the claims made about a GPAI model are tested, and by whom. This section lands on two findings that recur through the paper: evaluation is widely required but rarely binding, and the party conducting the evaluation is usually the same party being evaluated (self-assessment is the most common form).

This is the governance area through which claims about a model are tested before or after deployment. Without it, systemic risk assessment (§5.3) produces findings that no one checks, and GPAI prohibitions (§5.5) lack the technical basis on which to detect breaches. Our sample returned 136 positive evaluation and verification provisions, spread across sixteen of the twenty jurisdictions.

Out of the analyzed provisions, six sit in binding law. Of those six, four impose an evaluation mechanism on someone, and two of the four reach the entity that built the model, which are Article 55(1)(a) of the 2024 EU AI Act, and Article 30 of the 2025 Republic of Korea Framework Act. The other two binding provisions are the European Commission's pre-market conformity assessment for systems built on general-purpose models, which the AI Office performs itself, and a Brazilian judicial resolution requiring impact assessments of court AI to be audited \citep{BRA_209}. Everything else in the area is guidance, strategy, or a bill not yet passed.

That is the pattern the rest of this section elaborates. Across the sample, evaluation duties are widely imposed but rarely enforceable. Almost every jurisdiction directs someone to test something, however these obligations are thin in consequence.

\subsubsection*{5.4.1 Four convergent behaviors in evaluation and verification}

\textbf{Self-assessment is the default.} Ninety of the 136 provisions record “self-assessment”; thirty admit an independent third party, and the two sets overlap where an instrument offers a choice. Self-checking is the default among the entities deploying models and among those building them alike, and independence – where it appears – enters at the moment of certification or conformity assessment, rather than during development.

More telling is where that independence comes from. The Emirates' strongest evaluation duties sit in the Central Bank's 2026 Guidance Note for licensed financial institutions, which requires annual cybersecurity review of third-party AI providers by suitably qualified external parties; that is ordinary vendor risk management, applied to AI because AI now sits in the vendor stack. The Dubai International Financial Centre's 2026 accreditation and certification framework under Regulation 10 builds the most explicit third-party certification architecture in the sample, and it does so through a data protection commissioner's existing accreditation powers over certification bodies. Nigeria's draft bill and Kenya's Bill would each have a regulator accredit AI auditors, borrowing the assurance model from financial and standards regulation. \citep{KEN_065} \citep{NGA_069}. Read together, these provisions show AI middle-powers importing independent evaluation from adjacent regulatory traditions, rather than constructing it specifically for GPAI, and independence accordingly clusters in financial supervision and data protection rather than in AI instruments proper.

\textbf{The state is building capacity to evaluate faster than it is creating duties to be evaluated.} Thirty provisions place the evaluation duty on a national authority, and twenty-three create evaluation architecture while imposing no mechanism on anyone: safety institutes, sandboxes, standards programs, evaluation toolkits and ministerial reporting cycles. This is the first of three institutional patterns the paper draws together at §6.6.

Nearly one provision in six in this area is institution-building. The United Kingdom, Japan, Canada, Nigeria, Singapore and the European Union all do it, and they do it in the same order – capacity first, obligation deferred. The European Commission has committed to fund an EU evaluation capacity, including cybersecurity, by 2027 \citep{EUFRADEU_171}; Canada's 2026 "AI for All" strategy commits to expanding its safety institute, since renamed the Advanced AI Measurement, Evaluation and Science institute; Nigeria's bill requires the Minister to publish an annual report evaluating compliance and sectoral impacts; Singapore's framework proposes a baseline set of required safety tests and, eventually, an accreditation mechanism. \citep{SGP_178} None of these creates a duty on a provider.

Our Canadian expert interviewee described a division of labor in which one body runs academic research programs and another conducts government-directed research, with the national research council recently advertising for a frontier-model evaluation function. That is new capacity rather than new obligation, and the interviewee did not expect a move to horizontal regulation, attributing the sectoral preference to a policy priority on commercialisation, and noting that the voluntary code carries little weight because major frontier developers have not signed it.

\textbf{Jurisdictions converge on naming mechanisms but diverge from stating thresholds.} The instruments name sixteen distinct mechanisms, and they are not equivalent in what they produce. Safety testing, capability testing, performance testing, red-teaming, controllability testing and human uplift testing generate evidence about what a model does; audit, conformity assessment, certification, attestation and accreditation check a claim already being made, or the person making it, or the assessor; security testing checks integrity; a compute metric is a threshold that brings a model within a duty. Safety testing is the property distinction that matters most: NIST's AI Risk Management Framework, adopting the vocabulary of ISO/IEC TS 5723:2022, treats a system as safe where it does not, under defined conditions, lead to "a state in which human life, health, property, or the environment is endangered" \citep{nist2023airmf}. Security is the maintenance of confidentiality, integrity and availability against unauthorised access to the model, so penetration testing and vulnerability scanning measure a different property; accuracy, robustness and reliability are separately defined and concern how well a system performs, not whether it endangers \citep{nist2023airmf}. Red-teaming is defined by its adversarial method rather than by the property measured, and serves safety and security ends alike.

Measured against that list, the distribution is skewed away from the mechanisms that produce evidence about the model. Safety testing leads at twenty-one provisions, but it is thinly spread, and in the great majority of provisions the instrument names the mechanism and leaves the threshold, the benchmark and the failure condition unspecified. Human uplift testing, the only mechanism asking what a model adds to what a user could already achieve, appears twice in the entire sample – in the EU Code of Practice and in the 2024 UK AI Security Institute's approach to evaluations. Controllability testing, whether a system can be stopped, overridden, or prevented from resisting shutdown, appears four times and in only two jurisdictions, three of them in the 2026 New South Wales AI Assessment Framework and one in India's 2025 AI Governance Guidelines. The convergence here is a silence: AI middle-powers have converged on requiring tests without converging on, or in most cases addressing, what passing one would mean.

\textbf{Duties attach to systems in context, not to models by capability.} The EU AI Act runs two classifications, one for high-risk systems (expounded in greater depth in Recital 52) and one for general-purpose AI models (Recital 100). \citep{EUFRADEU_144} Outside Korea, it is the high-risk classification that has been taken up, and the GPAI model classification has not been adopted by name anywhere in the sample. Evaluation duties therefore attach to systems categorized by application context, rather than to models categorized by capability, which is the same divide §5.2 records in definitional practice, and it is consistent with how much of the material addresses deployers: forty-three provisions fall on a deployer, against forty on a provider.

\subsubsection*{5.4.2 Divergence in evaluation and verification}

Divergence runs along the same three design choices identified for systemic risk assessment at §5.3, and the jurisdictions group into four families.

The first is both provider-facing and binding, with the European Union its overarching member. Article 55(1)(a) of the EU AI Act requires providers of models with systemic risk to perform adversarial testing, and Article 55(2) attaches consequences for failure; Articles 91 to 93 give the AI Office post-market supervision powers and administrative fines. This is the minimum threshold at which a jurisdiction compels a developer to demonstrate that its model has been tested, and even here, the technical implementation stops short of requiring third-party involvement.

The second is the state as evaluator, found in the United Kingdom, Japan and Canada, where the government builds the capacity and carries out the evaluation while providers participate by agreement. The UK holds one of the highest provision counts in the sample and its AI Security Institute runs capability testing, red-teaming, human uplift evaluation and safeguard evaluation of frontier systems, but the obligation runs to the Institute rather than to any provider, the models tested are those companies choose to submit, and the Institute holds no statutory mandate. \citep{GBR_188} Japan reaches the same position through purchasing: its 2026 Guideline for Government Procurements and Utilizations of Generative AI imposes conformity assessment and testing requirements through procurement check sheets, with a verification step before release and re-verification after a major model update. Those requirements are mandatory in form but bind only those selling to the government. Our Japanese interviewee observed that ministry guidance is understood by its addressees to mean "obey", offering the screening of telecommunications suppliers on national-security grounds as the model case: issued as guidance, its target understood without being named. What such instruments produce is administrative and commercial compulsion, rather than legal force.

The third is published methods without an obligation to apply them, and Singapore and India are its clearest cases. Singapore's Model AI Governance Framework for Generative AI, its Guidelines on Securing AI Systems, the AI Verify Testing Framework and the 2026 Starter Kit for Testing LLM-Based Applications together address red-teaming, safety testing, adversarial robustness, prompt injection resistance and data pipeline integrity, and AI Verify is one of very few jurisdiction-produced toolkits naming specific technical attacks. None of it binds, and the only firm commitment is the undertaking in the 2023 National AI Strategy 2.0 that the government itself will test frontier models. India is placed similarly: its 2025 AI Governance Guidelines set out lifecycle compliance, the CERT-In advisory cluster addresses AI-assisted vulnerability exploitation, and Recommendations 20 and 24 of the 2025 FREE-AI framework propose red-teaming and independent audit in financial services, but FREE-AI remains a committee proposal rather than a rule.

The fourth is borrowed authority, and Australia and the UAE share it. In both, the evaluation duties carrying any force come from a body that already held authority over the addressee for reasons unconnected to GPAI – such as a purchaser over its suppliers, or a supervisor over its licensees – while the flagship national AI instrument remains advisory. Australia's material attaches almost entirely to government adoption, with thirteen of its 25 evaluation provisions coming from New South Wales, whose 2026 assessment framework is mandatory for agencies by circular, and asks whether a system can be shut down without resistance and whether it exhibits deception, collusion or self-preservation. \citep{AUS_272} The Emirates' fifteen provisions divide between the Central Bank's supervisory guidance, \citep{ARE_185} advisory ethics principles from the federal ministry, \citep{ARE_266} and the DIFC certification framework, which is mandatory in form, sits in a subnational free zone, and verifies rather than imposes: every operative requirement is a duty to assess, and the systems control-retention condition being tested belongs to the 2023 Data Protection Regulations (§5.5).

Four jurisdictions have drafted bills that would join the first family, and none has passed: Chile carries four mandatory evaluation provisions, binding providers and deployers, but the bill has been before the legislature since 2024. \citep{CHL_054} Two Chilean experts consulted for this study expected it to go no further in its present form: one described the government's substitute text as closely modeled on the EU AI Act and noted that academia, the private sector and civil society all opposed it at the public hearings; both observed that the administration taking office in March 2026 has not supported its advance. Brazil, Nigeria and Kenya are in the same position. \citep{BRA_032} \citep{NGA_069} \citep{KEN_065} Peru, South Africa, Switzerland, and Taiwan carry nothing on GPAI evaluation at all.

What separates the families is not the content of the tests, but what follows them. Where the state performs the evaluation, no enforcement chain exists: the UK Institute publishes findings and the Japanese Chief AI Officer reports to the ministry, but neither holds power over the provider whose GPAI model was tested, and neither can compel a change to a model on the strength of what an evaluation found. That is the gap that matters most across the sample, and outside the EU it is a gap in authority rather than in capability.

That gap is visible in how the evaluating bodies were constituted. Five of the nine bodies coded as dedicated AI regulators hold no enforcement powers: the United Kingdom’s AI Security Institute, the Canadian AAMES, and the German, Indian and Australian safety institutes. Widened to every named safety or evaluation institute in the corpus, adding those of Singapore, Korea, Japan, Switzerland and France, not one jurisdiction holds enforcement powers. Of the eleven jurisdictions holding an evaluation body, the European AI Office is the only one whose body can act on what it finds. These institutes were built to measure, but constituted without the power to act on what they find.

\subsection*{5.5 Prohibitions, serious-incident monitoring and detection}

This third link in the accountability chain asks what AI-middle powers have forbidden outright where GPAI is concerned, and whether anyone is charged with noticing a breach. This section also tests whether national practice could supply the raw material for an international agreement on red lines, and lands on the finding that it largely could not: prohibitions concentrate where domestic law already provides a vehicle, and thin out in the domains carrying the highest severity risk.

Calls for prohibitions on artificial intelligence have intensified as the capabilities of frontier systems have advanced. Since 2023, open statements signed by researchers, industry leaders and civil society organizations have argued that some AI capabilities are too dangerous to build \citep{caisStatementAIRisk2023}; \citep{idaisBeijingConsensusRedLines2024}; \citep{pacingFrontierOpenLetter2026}. Where those statements described the risks, the September 2025 Global Call for AI Red Lines proposed a remedy: an international political agreement on verifiable prohibitions by the end of 2026 \citep{millernguyen2025globalcall}. This governance area holds 183 positive provisions, of which 127 prohibit a use, a capability or an outcome. The remaining impose serious incident monitoring or detection duties without prohibiting.

This section tests whether national practice across AI middle-powers can supply the raw materials for a future binding international agreement on AI red lines. We find that it largely cannot. Mapped prohibitions rarely address the harms of the highest potential severity: CBRN uplift and loss of control, which are described in §4.2, have been identified as the areas where harm is least reversible and least containable \citep{cltcIntolerableRiskThreshold2025}. GPAI prohibitions are also rarely accompanied by the detection mechanisms that would make a prohibition meaningful. Lethal autonomous weapons and military escalation are both excluded for reasons covered in §4.2.3.

The jurisdictions can be roughly divided into three groups according to the obligation language in which their prohibitions are cast. First, prohibitions in the EU (28 of 29) and Brazil (12 of 14) use language directly indicating prohibition, such as "shall not", to characterize bans that are either absolute or conditional. Second, those in Japan and Taiwan focus mainly on "must" prohibitions, where the actions required to mitigate a harm are prescribed and the obligation is fulfilled even if the harm then materializes. Third, most of the remaining jurisdictions either lack prohibitions at all or, in the case of Australia, India, the UAE and Israel, rely on "should" language that falls short of a genuine obligation. The United Kingdom sits between the first and second groups, its seven prohibitions splitting evenly between prohibition and mandatory language.

\textbf{One risk, four formulations.} Jurisdictions also vary in how they express prohibitions directed at the same risk. For example, a single risk is described four ways across the dataset: loss of control appears categorically in Emirati ethics instruments (autonomously hurting or deceiving humans), operationally in the New South Wales AI Assessment Framework (self-replication limits, real-time kill switches), procedurally in India's 2025 AI Governance Guidelines (autonomous critical actions require human approval), and as a process obligation not to proceed in the EU’s GPAI Code of Practice. None of these shares a threshold or a trigger with any other.

AI middle-power jurisdictions tend to rely on either conditional or on absolute bans, rarely using both – Brazil is the exception, splitting evenly between the two. Which type predominates typically reflects how much a jurisdiction prohibits rather than the legal vehicle it prohibits through. Where a jurisdiction holds only two or three prohibitions, they are almost always absolute: this is the case in Peru, Chile, France, Germany and Singapore, all of which rely exclusively on absolute bans, whether carried in a criminal statute, a soft-law security guide, or a bill not yet passed. Where a jurisdiction says little about prohibition, it says it categorically. By contrast, where administrative-regulatory regimes are the basis for prohibitions, those prohibitions tend to exist in conditional form. This is true of the EU, where prohibitions sit within risk-tiered product safety regulation and 90 percent of them are conditional. In the case of online-safety and guidance regimes, restrictions rather than bans are the norm, as in Australia, India and Korea.

\textbf{Strict prohibitions are the least enforced.} The rough jurisdictional divide between absolute and conditional prohibitions is mirrored in how detection is treated. Among the absolute prohibitions mapped in these jurisdictions, 27 of 39 name no detection mechanisms. This is a markedly higher proportion than among conditional bans or among the provisions that limit an activity without banning it. It reflects the fact that absolute prohibitions are typically found in criminal statutes and ethics charters, which rely on complaints, ordinary policing, or reputational pressure to surface a violation. They do not impose a continuing obligation on any party to monitor for breach. By contrast, conditional prohibitions sit predominantly inside administrative regimes that already carry a supervisory apparatus. That apparatus can be applied to general-purpose AI prohibitions directly. This means that the prohibitions that are strongest in form (absolute) and most directly relevant to an international agreement on red lines, are the least likely to be detected in breach.

\textbf{Use rather than capability.} AI middle-powers that engage in GPAI prohibitions do so mostly with respect to how GPAI is used, rather than what it can do. Across the dataset, 90 prohibitions address a use, 24 address a capability, and thirteen address an outcome. The ratio describes the instrument corpus: the EU, Brazil, and Taiwan hold more than half the use-based prohibitions between them. Capability targeting is concentrated in the UAE (10 of 19), Australia (6), and India (3). Use-based prohibitions are both more numerous and more strictly enforceable across the AI middle-powers mapped for this study: about half sit in binding law, and also about half of use-based prohibitions name a detection mechanism. Among the capability-based prohibitions, a fifth are binding, and two fifths also name a detection mechanism.

The focus on use has implications for enforcement, as a capability can be tested before release, whereas uses and outcomes can only be enforced once misconduct has occurred or harm has materialized. Here again, AI middle-power governance does not match well with the international debate on red lines, which is largely focused on capabilities, placing the burden on developers to demonstrate, before a model reaches the market, that it does not possess a specified dangerous capability. For example, the Global Call, while extending to uses as well, is built around capability categories \citep{millernguyen2025globalcall}.

Coverage of the international red lines domains. Coverage of the domains identified in the international debate on AI red lines is uneven. Participants at the Seventh Athens Roundtable judged red lines to be necessary in seven domains: large-scale manipulation, psychological exploitation and children's safety; loss of control; CBRN threats; cyber-offensive capabilities; autonomous control of critical infrastructure; lethal autonomous weapons and military escalation; and human rights and justice \citep{tfsAthensRoundtableRecap2025}. Six fall within scope, lethal autonomous weapons and military escalation having been excluded at §4.2.3.

The Global Call for AI Red Lines, co-led by TFS, Centre pour la Sécurité de l'IA, and the Center for Human-Compatible Artificial Intelligence, reports convergence across its workshop series on six domains: mass manipulation, children's safety, loss of control, CBRN uplift, offensive cyber capabilities, and lethal autonomous weapons \citep{redLinesCampaignFAQ2026}. Read strictly against those six, coverage is thinner rather than broader – they capture 43 percent of mapped prohibitions against 54 percent under the domains defined at the Seventh Athens Roundtable \citep{tfsUnacceptableRisks2026}, mainly because the Athens human rights and justice domain has no counterpart among the six. The fifteen provisions mapped to that domain divide into twelve on surveillance and three on judicial displacement, and nine of the twelve have no home under the six, among them eight EU AI Act Article 5 prohibitions covering the scraping of facial images, biometric categorization inferring political opinions, and real-time remote biometric identification. That is the largest body of binding AI prohibition law in the sample, and on a strict reading of the six convergence domains, it sits outside the taxonomy. The campaign does name mass surveillance among its illustrative red lines, however, citing the UNESCO Recommendation on the Ethics of AI, so the omission sits at the level of the convergence domains rather than of the campaign's position \citep{redLinesCampaignFAQ2026}.

Where the two frameworks agree is at the catastrophic end: a single CBRN prohibition, but none on offensive cyber capability. The Global Call folds critical infrastructure into offensive cyber, and since no provision in the subset engages autonomous control of infrastructure under either reading, this difference does not affect the count. Human impersonation, which the campaign names separately, would add one provision – through a UAE guideline requiring notification where a system can convincingly impersonate a human. \citep{ARE_266}

Across the EU, Brazil, the UK and Taiwan, all but one prohibition sits in manipulation, children's safety, or human rights and justice. Only a few AI middle-powers apply prohibitions to the other domains (UAE (and Dubai), Australia, France, India, Taiwan, and the EU), which carry the highest potential severity: more than half of these provisions are concentrated in loss of control, and nearly three quarters of those are found in Emirati instruments. To be sure, some provisions engaging these domains may lie within the national security and defense carve-outs that are beyond our scope, however.

If, today, an international agreement were to be built on the Athens domains, negotiators would struggle finding existing national practice to draw on in the domains carrying the highest potential severity of harm.

\textbf{Where detection is addressed.} What the mapping records is whether a provision names a mechanism for detecting a breach, and which kind; it does not record whether detection is carried out, or how well. The differences between jurisdictions are therefore differences in what the governance instruments specify. On that measure, the EU, Brazil, Korea, and India name a detection mechanism on most of their prohibitions, while Israel and Taiwan name one on a minority. Where a jurisdiction holds only one or two prohibitions the proportion carries little weight, so Switzerland, Canada and the UAE are better read individually than as rates.

\textbf{Detection is thinly specified.} Of the 127 prohibitions records, 56 percent say nothing at all about who would notice a breach. Of the those that do, over half point to the regulated actor monitoring itself, while 30 percent do not explicitly name the monitoring body bearing the obligation, leaving just 14 percent that identify a monitor other than the party that is being regulated – such as an accredited certification body in the DIFC, a procuring ministry in Japan, sectoral competent authorities and police in Taiwan, and a returning officer in Singapore.

Some of the detection mechanisms proposed in the policy literature are absent in current AI middle-power governance, including the Seveso-style major accident prevention policy, with standing safety reports and an early warning system, which The Future Society proposed for the most capable models in 2023 \citep{tfs2023HeavyHead}. An expert workshop at the February 2026 IASEAI conference identified three requirements for enforceable international red lines: agreement on thresholds, verification infrastructure, and diplomatic coalition-building \citep{zoumpalovaWhereWeDraw2026}. Yet no prohibition in the dataset defines a capability threshold at which it takes effect. On verification, the Global Call is precise, calling for evaluation before deployment and for this to be done by third parties operating independently of the developer, supported by an international body empowered to audit and certify compliance. This, too, is not reflected in AI middle-power provisions.

However, that absence is in law, rather than in the coordinated governance field. A third-party evaluation ecosystem has formed over the past three years \citep{metrFrontierRisk2026}: organizations including METR and Apollo Research now run pre-deployment evaluations of frontier models under voluntary access arrangements with developers \citep{metrEvaluations2026}, \citep{apolloResearchEvaluations2026}, \citep{apolloWhiteBoxAccess2026}, though evaluators report that access is inconsistent, and the time allowed is measured in days rather than weeks \citep{externalAccessDangerousCapability2026}. METR additionally holds a technical assistance contract with the EU AI Office supporting its methods for assessing loss-of-control risk \citep{metrAbout2026}, one of five systemic-risk lots in the Office's third-party assistance procurement under Articles 89, 92 and 93 \citep{aiActNewsletterAIOfficeTender2025}, \citep{EUFRADEU_144}. Parallel technical work addresses the measurement problem directly, developing hardware-enabled and cryptographic methods for verifying properties such as training compute without exposing model weights \citep{hardwareEnabledMechanisms2025}, \citep{verifyInternationalAgreements2024}.

The ecosystem the Global Call anticipates is therefore under construction; what has not followed is the legal step, since no instrument in our sample requires that any of it be done, or attaches a consequence to its absence. Where a body is named, it is usually an existing regulator taking on AI oversight within its established remit, rather than a new authority: most governance actors in the corpus pre-date the AI mandate they exercise, and only nine are dedicated AI regulators (§5.8). None of them, including the EU AI Office, is named as carrying a monitoring duty for a prohibition.

The domains where prohibitions are densest are those where a domestic legal tradition already provided a vehicle, such as criminal law or online safety regulation, and the domains where coverage is thinnest are those with no such antecedent. This underscores the urgency of international consensus-building on how these gaps might be filled consistently across jurisdictions.

\subsection*{5.6 Serious incident reporting}

The fourth link in the chain asks what happens once a harm has occurred. This section treats a serious incident reporting duty as a chain of four elements – an addressee, a trigger, a period and a consequence – and lands on the finding that half the sample carries no such duty at all, and that among those that do, the four elements are incomplete.

GPAI instruments of the United States and China fall outside this study, and that exclusion extends to sub-national instruments, including California’s Senate Bill 53 of 2025 enacting the Transparency in Frontier Artificial Intelligence Act. The Bill is a reference point many readers will bring to a discussion of serious incident reporting; however this paper takes the OECD definition of a serious AI incident as its trigger test, whereas SB 53 works from “catastrophic risk” and “critical safety incident” at a frontier-model compute threshold that has no counterpart in the instruments mapped here.

\textbf{What was looked for.} Of the four governance areas mapped in this paper, serious incident reporting is the one where the absence of a rule is the most common finding. What was looked for is defined in §3.3.4, which adopts the OECD definition of a serious incident as the trigger test. A reporting duty built on that definition is a chain of four elements: an addressee, meaning the body to which the incident must be reported; a serious incident trigger, meaning the harm event that activates the duty to report; a timeframe within which the report must be made; and a consequence, meaning what follows if the report is not made. Where any one of the four is missing, the duty cannot be enforced in practice.

Two bodies of work support this. Gomez et al. find that these four elements are a dependency chain: what counts as a reportable event has to be settled before an operational trigger can be drafted, so definitional work is a precondition, rather than a refinement \citep{gomez2026}. Most of the operative provisions in this area name a body to report to but leave the trigger, the period, or both, undefined, so the volume of reporting they generate cannot be treated as a measure of how many serious incidents have occurred.

The Center for Security and Emerging Technology’s work on mandatory AI incident reporting is used more narrowly, for its component list and its recommendation that reporting formats be mandated \citep{csetAddingStructureAIHarm}. Mature reporting regimes outside AI pair these elements as a matter of course: the GDPR sets a seventy-two hour deadline, a named supervisory authority and fines for failing to notify. A deadline without an addressee, or an addressee without a consequence, is unusual.

Confirmed absences (§4.1) are positive findings about a jurisdiction, not gaps in research. Across the 105 provision records tagged to serious incident reporting, 27 are operative provisions imposing or recommending some form of reporting duty and 78 are confirmed absences. The distribution of those 27 positive provisions does not follow the distribution of governance activity. India holds twenty positive provisions across the four areas and five of them are reporting provisions, more than any jurisdiction outside the European Union. Australia and the United Kingdom hold 45 and 44 positive provisions respectively, but only two each in serious incident reporting. The United Arab Emirates holds 32, versus none in this area. Israel, with eight positive provisions in total, holds three serious incident reporting provisions. Eight jurisdictions record no reporting duty of any kind, and are identified below; France and Germany hold none domestically, their duties arriving through direct effect of EU law, so ten of the twenty jurisdictions carry no reporting duty of their own. Where these duties exist outside the EU, they are also frequently lodged in instruments that cannot carry them: three of Brazil’s four sit in a bill that has been before the legislature for more than three years, and Korea’s sits outside its AI legislation altogether, as set out below.

\textbf{The European Union chain.} Only three provisions state both a reporting period and a consequence for failing to report, and two of the three originate in the EU. Article 73 of the 2024 EU AI Act sets a graduated schedule backed by administrative fines, and Article 75(1a), as inserted by the 2026 Digital Omnibus on AI, carries both across. Article 55(1)(c) of the same Act requires providers of GPAI models with systemic risk to report serious incidents to the European AI Office and is backed by the same fines, but it states no period – the duty arises “without undue delay” – so it does not complete the chain. The third is a Brazilian judicial-administration resolution requiring adverse events involving GenAI in the judiciary to be reported within 72 hours, backed by corrective actions. The Digital Omnibus changes the addressee element of the chain for a defined class of providers, who now report directly to the AI Office rather than through member states; the periods carry over unchanged; the 2025 General-Purpose Code of Practice operationalizes the Article 55 duty for signatories.

\textbf{Who bears the duty.} Provider-facing reporting is almost entirely an EU phenomenon. All five of the EU's reporting provisions place the duty on the provider. Three other jurisdictions also reach a provider: Brazil, in two provisions of a horizontal AI bill that has not passed; the UK, in a non-binding evaluation template; and Australia, in soft-law guidance. Everywhere else, the duty faces someone other than the entity that built the model. India spreads five provisions across an authority, a deployer, a supplier, and an intermediary without touching a provider; Israel's three provisions fall on deployers and on regulated financial bodies; Korea's single provision falls on financial companies; and Japan's two run inside government, one on a public body, and one on its procurement counterparty. Where an authority bears the duty – in Brazil, India, the UK, Singapore, Kenya, and Japan – the state is either reporting to itself, or still building the machinery for it.

Of the nine provider-facing provisions, three are binding and in force, and all three stem from the EU: Articles 55(1)(c) and 73 of the AI Act, and Article 75(1a) as inserted by the Digital Omnibus. The remaining six impose no enforceable duty. Two are EU soft law interpreting the binding provisions above; two ask that reporting procedures exist without stating what activates them; and two sit in Brazil’s horizontal AI Bill, PL 2338/2023, which has not yet been enacted. Outside the EU, no provider is under a binding, in-force duty to report serious AI incidents.

The only binding, in-force duty outside the EU that reaches a private actor does not fall on a GPAI model provider, but on intermediaries. Under India's 2026 intermediary rules, an intermediary must report to the appropriate authority where the material involves an offense that another statute requires to be mandatorily reported, and an intermediary that fails to act loses safe harbour. That reporting duty carries no period; the clocks the amendment shortens are takedown deadlines, not notification deadlines. \citep{IND_164}

The eight authority-facing provisions divide in two ways. Six are commitments to build or empower reporting machinery that does not yet operate, each deferring the operative element to a step not yet taken: a proposed national database, recommendations that an authority design a framework or set a materiality threshold, a committed-to but unestablished channel, a rule-making power in an unenacted bill, and a regulation-making power under which no duty arises until regulations are made, and none have been. \citep{SGP_178} The remaining two impose a duty that already operates, and both run inside the state’s own machinery: Brazil’s judicial resolution and Japan’s government procurement guideline each require public bodies to report AI incidents to another organ of the same state. This distinction matters because a chain in which the state reports to itself generates no external accountability: it produces a record, but no entity outside of the government itself is accountable.

\textbf{Korea: a reporting duty in the wrong instrument.} The Republic of Korea is the only jurisdiction in the sample to have legislated binding obligations on GPAI providers of its own accord, through the 2025 Framework Act and its Enforcement Decree, which together set a training-compute threshold. Neither instrument creates a serious incident reporting channel. The duty sits elsewhere entirely: the 2026 consolidated AI Guidelines for the Financial Sector requires financial companies to report immediately to the supervisory authority where an AI incident risking escalation into systemic risk has occurred or may occur. Reports travel through the pre-existing financial incident reporting system, rather than an AI specific channel. The reporting duty was written by the financial supervisor for its own licensees, and reaches only one sector. The Guidelines are voluntary and attach no consequence to a failure to report. On the four-element test, Korea supplies an addressee and a trigger, but neither a period nor a consequence. A draft notification under the Framework Act, applying to high-performance AI operators above the compute threshold, was out for consultation as this paper was written, and may close part of the gap.

\textbf{Purpose-built versus borrowed channels.} Some jurisdictions build a reporting channel for AI, while others route AI reporting incidents through regimes they already run. The European Union is one that builds: all five of its reporting provisions create an AI-specific channel, as do both Japanese procurement provisions, Brazil's judicial resolution, Kenya's 2025 strategy commitment, and Singapore's generative AI framework. India, Israel, and Korea are jurisdictions that borrow: three of India's five provisions route through CERT-In cyber-incident reporting or intermediary liability, two of Israel's three through anti-money-laundering and terror-financing, and privacy regimes, and Korea's single provision through the financial incident reporting system that predates it. Korea is the sharpest case, holding a horizontal AI statute that creates no reporting channel, so the only route available runs through financial supervision. \citep{KOR_083}

Where a channel is borrowed, it is almost always anchored in soft law. Seven of the eight borrowed routes sit in guidances, with India's 2026 intermediary rules the exception. These eight borrowed routes draw from four donor domains: financial crime, cyber-incidents, data-protection and health breach notifications, and online-safety channels.

Borrowing from sectoral regimes has a clear rationale: the regulator already holds powers over the same entities, the route is known, and the harm it addresses is one the law already recognizes. In jurisdictions without a horizontal AI statute it may be the only route available. Where the donor regime is mature it can do more than any AI-specific instrument here does: financial operational-resilience regimes oblige firms to aggregate incidents in order to identify common root causes, and to set impact tolerances. Neither requirement has an equivalent anywhere in the corpus.\citep{doraDelegatedRegulation2024}, \citep{praOperationalResilience2022} A borrowed channel also inherits the trigger it was built for: an AI incident producing no personal data breach, no service outage, and no suspicious transaction may activate no duty at all, however severe its consequences. This is a defect in trigger logic resting on a definitional foundation built for another domain \citep{gomez2026}, and it is consistent with the pattern documented for Indian telecommunications \citep{agarwalIncorporatingAIIncidentReporting2025} and for financial regulation \citep{mazziniConsiderationsRegulationAI2023}, in which dense adjacent infrastructure nonetheless yields no trigger attributable to a GPAI failure as such. At the sectoral level, what counts as a serious incident caused by a GPAI model remains operationally undefined.

Two jurisdictions show that inheritance is not inevitable. Israel's Money Laundering and Terror Financing Prohibition Authority does not leave AI-related harm to be caught by the existing suspicion trigger; its red-flags document instructs reporting bodies to file where suspicion arises through the use of generative AI or deepfake technology, whether in the body's own dealings or in a customer's conduct: so the channel is borrowed, but the trigger has been adapted and the AI-related subset remains retrievable. India does something similar on a smaller scale, naming AI-assisted exploitation among events activating an existing six-hours notification duty by CERT.

\textbf{Reporting periods.} Reporting periods, where specified at all, are not comparable across the corpus. Six provisions state a defined clock: six hours in each of the two CERT-In guidelines, seventy-two hours in the Brazilian judicial resolution, the graduated schedule of the EU AI Act, carried across by the Digital Omnibus, and the staged schedule of the GPAI Code of Practice. Everything else is either an open formulation – “without undue delay”, “immediately”, “promptly”, “in a reasonable timeframe”, “timely” – or a deferral, as in Article 42 of Brazil’s draft 2023 AI Bill, where the period is left to a sectoral authority that has yet to set it. The distinction that matters most is therefore not the length of the deadline, but whether one exists: a duty with no stated period has no point at which it can be breached, and so fails the chain test set out above.

Figure 5.6 illustrates reporting timeframes across the 27 operative serious incident reporting provisions, grouped by the kind of period the instrument sets rather than by its length. Each provision appears once.

\begin{figure}[!ht]
\noindent \textbf{Figure 5.6 Serious incident reporting by timeframe}\par\vspace{4pt}
{\centering
\includegraphics[width=0.70\textwidth]{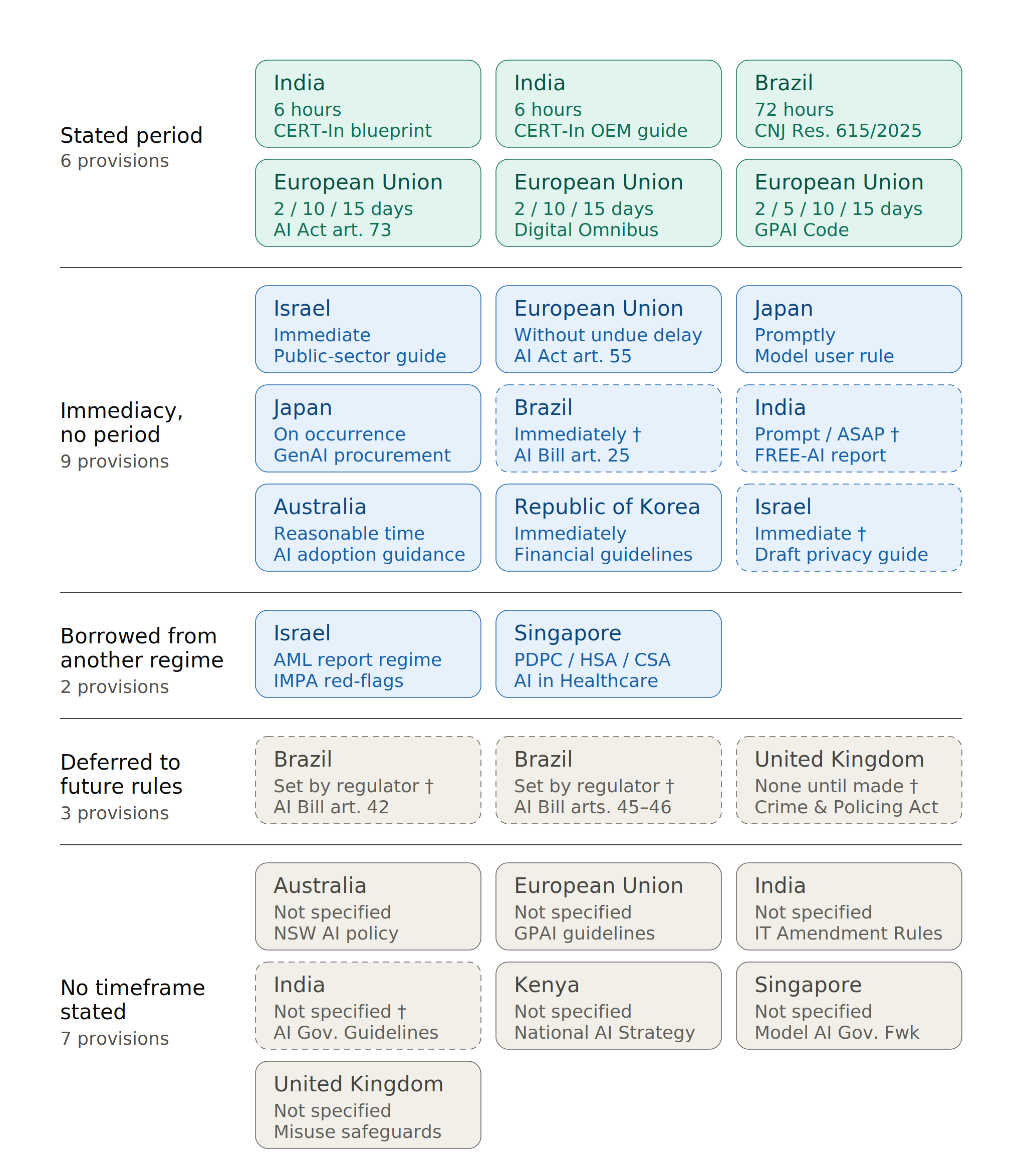}
\par}
\vspace{4pt}
\noindent {\footnotesize
\textit{\textbf{Caption:}} \textit{Stated period – a defined clock, expressed in hours or days. Immediacy, no period – a duty framed as immediate, prompt or timely, with no period fixed. Borrowed from another regime – the timeframe is that of a pre-existing sectoral reporting regime, not one the AI instrument sets. Deferred to future rules – the period is left to rules or a determination not yet made. No timeframe stated – the instrument is silent on timing. † – non-operative, future-intent, or not yet in force.}
}
\label{fig:figure-5-6-serious-incident-repo}
\end{figure}
\FloatBarrier

\textbf{Consequences, and near miss incident reporting.} Consequences for failure to report a serious incident are more often unspecified than not, and 19 of the 27 provisions state no consequence at all. The only administrative fines in the dataset originate in EU law, and they attach to three provisions. Two Brazilian provisions carry a corrective order, and one Indian provision carries loss of safe harbour. However, another two Indian records run the other way, deliberately decoupling reporting from penalty: the national AI incident database, so that reporting carries no threat of penalty, \citep{IND_166} and FREE-AI recommends that the financial-sector reporting framework "adopt a tolerant, good-faith approach to encourage timely disclosure". \citep{IND_167} India is the only jurisdiction in the sample to treat the absence of a consequence as a design choice rather than a gap.

No instrument outside of the EU and its member states ties a defined severity level to a reporting deadline; the severity is left undefined wherever a timeframe is open ended. Nor does any instrument impose a binding duty to report a near miss. Two reach anticipated harm at all, and neither is binding. The EU GPAI Code of Practice defines the near miss and then routes it through a materiality assessment made by the provider, engaged only where the provider has reasonable grounds to believe that the systemic-risk justification for the model has been materially undermined; the quickest route by which a near miss reaches the AI Office is therefore not the incident channel but an update to a Safety and Security Framework or Model Report. Korea’s financial-sector guidance takes the opposite approach, treating a concern that an incident may occur as itself sufficient to trigger a report to the supervisor, without any intervening judgment by the reporting entity. The Korean formulation is the more demanding of the two on this narrow point, and the weaker on every other. Gomez et al. describe the problem the EU formulation creates as detection asymmetry: the developers hold both the visibility of safety problems and the discretion over whether those problems are escalated \citep{gomez2026}.

\textbf{Where no incident reporting duty exists.} Eight jurisdictions record no provision imposing a duty to report serious incidents caused by GPAI. These are Canada, Chile, Nigeria, Peru, South Africa, Switzerland, Taiwan and the UAE. Two of the eight – Canada and the UAE – have the institutional machinery that a reporting duty would use, an evaluation body and a financial supervisor respectively, but attach no duty to it. This is the separation between capacity and authority described at §6.6, appearing here in this area.

\textbf{One actor, one perimeter.} One point runs underneath all of the above. Every reporting obligation in the corpus asks a single actor to report on what lies within its own control: a provider on its own model, a deployer on its own deployment, an intermediary on its own service, a public body on itself. None requires that reports be aggregated across providers, compared across models, or read together with what the platforms carrying a model’s outputs have separately assessed. The EU comes closest to a complete chain, yet still produces a provider-by-provider view. One provision gestures at a composite picture, directing a supervisor to grasp the overall content of the models and systems its licensees develop and use - but that is an aggregate held inside one regulator, for one sector, built from a voluntary report. A harm visible only in the aggregate – the same failure appearing across several GPAI models, or the same population affected through several deployments – has no addressee anywhere in the sample, because no instrument in the corpus creates a body whose function is to hold the composite picture. That gap is not only a drafting omission: detecting AI failure requires machinery to report serious incidents, aggregate them, and identify patterns across them, and the jurisdictions with the least of that machinery are precisely those where a failure is hardest to absorb \citep{malonzaAutomationFirstAdoption2026}.

\subsection*{5.7 Sectoral coverage}

Where a jurisdiction holds no horizontal AI statute, governance arrives through regimes built for something else. This section asks which sectors carry that load, and what those routes buy and cost. It lands on a trade: sectoral instruments reach their addressees through a relationship that predates the GPAI question, but are limited by that same relationship.

Two terms carry the analysis that follows. A governance instrument is horizontal where it is specifically about AI: its provisions attach to AI systems or models as such, and apply wherever those systems appear, whatever the field of activity. An instrument is sectoral where it governs a particular field of activity – e.g., financial services, health, elections, online safety – and carries AI provisions within it. The distinction is one of subject matter, and does not track the identity of the issuer; this is because a body with a sectoral mandate can produce an instrument that is horizontal in subject matter.

Twenty-five of the 101 in-scope instruments govern a single sector rather than AI in general, and between them they carry 97 of the 382 positive provisions. Sectoral instruments therefore account for about a quarter of the instrument collection, and also about a quarter of its coded content. Governing general-purpose AI through an existing sectoral regime is not a marginal route in this dataset; it is one of the two main ones that surfaces.

The sectoral instruments fall into nine sectors. Financial services (21 provisions) and digital (17) are the largest; criminal law follows (12), democracy and elections (11), judicial oversight (10), children's safety (9), health (8), cybersecurity (5), and market competition (4). The digital and children's-safety material is concentrated in the UK, Australia and Taiwan; financial services draws on India, Israel, Korea, the UAE and the UK; judicial oversight is Brazilian, health Singaporean, cybersecurity French, and market competition British. These ten jurisdictions supply the whole sectoral set; the other ten have none, operating through horizontal AI governance instruments alone.

\textbf{Sectors with no recorded coverage.} Two of the coded sectors record nothing at all. The education instruments we tested are guidance on classroom and anti-bias assessment use, which did not meet the severity thresholds for serious incidents or systemic risk under the definitions given in §3.3.1. The environmental sector also returned nothing, but this should not be read as a finding. Environmental harm appears as a risk domain enumerated inside horizontal risk taxonomies: seven provisions across Taiwan, India, Kenya, the UK and Brazil \citep{TWN_219}, \citep{IND_166}, \citep{KEN_065}, \citep{GBR_103}, \citep{BRA_032}, while no jurisdiction in the sample has legislated an AI-specific environmental obligation. Adjacent questions – carbon and resource disclosure for model training, grid stress from concentrated compute, and AI used in emissions monitoring or environmental impact assessment – sit outside our scope boundary. The blank records represent the absence of GPAI-specific environmental instruments, not the absence of environmental risk or of law touching it.

\textbf{Financial services: soft law that binds.} The financial-services group matters for more than its size. All five instruments sit in soft law and none creates a statutory duty, yet fifteen of their 21 provisions are coded as enforceable soft law, just over a quarter of the 55 such provisions in the dataset. That enforceable weight is spread across four of the five instruments, issued in the UAE, Korea, the UK, and Israel \citep{ARE_185}, \citep{KOR_083}, \citep{GBR_105}, \citep{ISR_160}; the fifth, India's, contributes none, because it is the report of an advisory committee rather than an instruction from the supervisor \citep{IND_167}.

What makes the difference is the relationship behind the document rather than anything in its text. A financial regulator writing to its own licensees is not making a suggestion in the way a ministry writing to the public is. The firms are already on its register, a supervisory channel is already in routine use, and the harm the guidance names is one the regulator already has a mandate to prevent. A firm that disregards it can expect the point to be put at the next supervisory contact. That is why the document can begin to bite without waiting for a legislature. If that is right, the body of law a guidance note is issued under should matter less than the relationship it travels along – and across the five instruments that varies considerably: anti-money-laundering and terror-financing in Israel \citep{ISR_160}, consumer protection in the UAE \citep{ARE_185}, operational resilience in the UK \citep{GBR_105}, and ordinary prudential supervision in Korea \citep{KOR_083} and India \citep{IND_167}. No two reach their firms by the same route. What they share is not a common legal basis, but a supervisory relationship that predates the GPAI question.

The mechanism is not confined to financial services. An expert interviewee on Chilean regulation described a consumer-protection recommendation on AI transparency, issued by the national consumer agency and initially resisted by industry, that has since changed observable market behavior: chatbots on Chilean commercial sites now identify themselves as AI. The same interviewee argued that sectoral authorities who know their own perimeter are better placed to keep pace with the technology than a legislature, while accepting that this carries a democratic deficit.

\textbf{Subnational behavior.} Subnational records must be read carefully. The Dubai International Financial Centre appears not through a financial-sector instrument; its entry is the 2023 Data Protection Regulations, coded horizontal, supplying two conditional prohibitions on commercial deployment, with the third-party certification machinery those prohibitions depend on, supplied separately (§5.5) \citep{AREAE-DU_243}. Those two provisions are nevertheless the only binding provisions recorded anywhere in the UAE, national or subnational; the country's other four instruments are coded soft law, non-binding, or strategy documents. This is the reverse of what a financial-hub reading would predict. In the UAE, it is data protection law inside a financial free zone, not financial regulation, that produces binding text.

\textbf{Sectoral-instrument reach.} A sectoral instrument reaches only as far as its parent regime does. Each sector represented here sits in a regulatory relationship that predates the GPAI question: content regulators already governed online platforms, financial regulators already supervised firms, electoral authorities already controlled campaign material, and the criminal statutes already prohibited the same conduct in its non-synthetic form.

Sectoral instruments do reach the model layer, but barely. Twenty-five of the 97 provisions fall on providers, and four of those take the GPAI model itself as the object of evaluation - three Singaporean and one British \citep{SGP_278},\citep{GBR_197}. All four of these provisions are coded non-enforceable, however. The sectoral provisions that reach the developer's own model are therefore the least binding in the set, and everything sectoral that is enforceable attaches instead to a system already running inside a supervised organization. The sectoral route therefore buys enforceability and speed at the cost of reach. A jurisdiction whose only positive provision sits in a sector-specific instrument has coverage of a different kind from one whose provision sits in a horizontal instrument.

\subsection*{5.8 Cross-tier instrument interactions}

This section asks how governance instruments interact across tiers within a jurisdiction, which legal architectures those interactions produce, and which institutions carry them. It lands on six recurring architectures, and on the finding that the bodies exercising GPAI mandates were, in the main, pre-built for something else.

For the scope of this research, cross-tier interactions have been defined as those relationships by which GPAI governance is distributed across instruments sitting in differentiated tiers of law. These tier categories are defined under §4.2, as are the distinctions between hard law, soft law, non-binding instruments and strategy scaffold documents. One point of method: an instrument's position in a legal architecture was not a route to scope. An instrument that failed the scope test was not brought back on the strength of where it sat in relation to other in-scope instruments. This is because our research captures purposeful GPAI governance, rather than incidental GPAI governance. Cross-tier interaction is therefore something we observe between instruments that are already, independently, in scope.

Soft law is not a residue in this sample; it is the hardest working tier. Half the jurisdictions have no binding AI instrument of their own: France and Germany rely on the direct effect of EU law, and Japan, Israel, Switzerland, Kenya, Chile, Nigeria, and South Africa hold no hard-law instrument reaching GPAI at all; for these jurisdictions, guidance, a code of practice, or an unenacted bill is the strongest instruments they have. The ten that do hold binding law are not evenly matched either – the EU and Brazil between them hold more than half of the binding provisions in the sample, and India holds one. What the tiers describe, then, is not a spectrum from weak to strong governance but two groups: jurisdictions that legislate, and jurisdictions that publish. The second group is the larger, and it is where most of the sample's substantive GPAI content sits.

\textbf{Obligation language.} Obligation language points the same way. Provisions were coded by the force of the language they used, alongside the force of the instrument they sit within. The distinction that carries most weight is within the recommendations: a "should" is coded enforceable soft law where the issuing body holds enforcement powers, and non-enforceable where it does not.

Counting the positive provisions, 126 are non-enforceable recommendations (33 percent) and 55 are enforceable soft law (fourteen percent), giving 181 recommendations in all; 94 are requirements (25 percent) and 27 are standing institutional mandates (seven percent), giving 121 requirements; 64 use prohibitory language such as "shall not" (seventeen percent); nine are reporting duties, and seven are permissive discretions.

A second field records what lies behind a recommendation, and it is determined by the issuing body and the instrument type, rather than by the provision's own words. The largest group is advisory, with no named accountability behind it; a second group of 56 are state commitments, made by a government in a strategy or action plan; and 32 are "comply-or-explain" benchmarks, issued by a body that holds enforcement powers. Those 32 are the only recommendations in the corpus with an enforcement route behind them, accounting for eight percent of all positive provisions. The distribution is therefore weighted toward the weakest end of the field: more than half of all recommendations rest on nothing beyond the issuing body's authority to publish them.

Mandatory language is not confined to binding instruments. Of the 94 "must" requirements, most sit outside hard law. An instrument's tier records whether it creates duties enforceable against those it binds. The obligation language of its provisions records how firmly each one is put. The two vary independently, and the Japanese case discussed below is the clearest illustration of the distance between them.

\textbf{A thin implementation tier.} The instrument tier that would carry implementation is thin: only six of the 101 scoped governance instruments are secondary hard law (regulations, statutory instruments or agency rules made under the authority of a parent statute, rather than enacted directly by a legislature). They sit in Korea, India, Australia, Brazil and the Emirate of Dubai, and only one of the six sits beneath an horizontal AI statute. The other five implement online-safety, data-protection, judicial-administration or electoral law, reaching general-purpose AI through those subject matters, rather than through a specialized AI regime. Therefore, where a jurisdiction has set up a binding horizontal regime, the instruments that would make that regime operational have, with the single Korean exception, still not yet been made.

Korea's Decree is the clearest instance of cross-tier interaction in the corpus, as it bounds, rather than implements. The Framework Act disapplies itself to AI developed and used solely for national defense or national security "as prescribed by Presidential Decree", and the Enforcement Decree supplies that content. The 2024 Council of Europe Framework Convention makes the same reservation. The reach of a hard-law statute is therefore not fixed by the statute itself; it is delegated downwards, to the tier beneath. This is the only place in the dataset where a secondary instrument does work of that consequence.

\textbf{Instrument cross-referencing.} Beyond that, recorded interaction across tiers is sparse, and almost all of it runs upward: guidance, codes of practice, and strategies name the statute they are built around. Thirty-six instruments carry a recorded cross-reference to another in-scope instrument, and a minority of those references cross a tier boundary. Clusters of these form in the EU, UK, and Singapore.

Eight instruments point upward at the EU's AI Act or the Digital Omnibus amending it and five at the UK's Online Safety Act. Three point at something other than a statute – a strategy scaffold, a set of soft-law guidelines, and a non-binding “Starter Kit” – and the last of these is the only reference in the corpus that runs downward.

A cross-reference is recorded only where an instrument names another in-scope instrument in its operative text; an instrument that operates on another without naming it leaves no trace in the mapping. The count above is therefore a floor on cross-tier interaction.

\textbf{Six architectures.} Six cross-tier architectures appear in the sample. The first is an example of anchoring – a single horizontal statute with subordinate instruments made beneath it. Korea is a clear case. It holds five mapped instruments, of which the 2025 Framework Act and its Enforcement Decree are the only two at the hard-law tier. The second is inherited from the EU: binding obligations arrive through the direct applicability of supranational law, while France and Germany hold no domestic hard-law AI instrument imposing GPAI obligations. Their entire binding layer is the EU's 2024 AI Act and the 2026 Digital Omnibus amending it, and what each state adds domestically is soft law and strategy. The third is a distributed architecture: governance spread across a large number of lower-tier instruments, with no horizontal statute. The UK is the clearest case here: it holds sixteen instruments, twelve of them soft law or non-binding, and two more strategy scaffolds; its only two hard-law instruments are both sectoral. Australia, with thirteen instruments of which ten are soft law, sits in the same pattern. Singapore is close but not identical: of its thirteen instruments three are hard law, and while two are sectoral, the third is horizontal.

The fourth is an example of "announced" architecture: a national strategy sets the governance direction, while binding horizontal legislation is still pending. Kenya is the cleanest instance, with national AI strategy, a bill still in committee, a code of practice, and no hard law of any kind. Brazil is a partial instance only: its horizontal AI bill remains in committee over three years later, but three binding instruments are already in force, all of them criminal, judicial or electoral. The fifth architecture is sectoral-binding: no horizontal hard law, with the binding rules that do exist confined to particular sectors, and horizontal governance carried in soft law. Taiwan is the clearest case, with two sectoral statutes in force alongside a horizontal AI Risk Classification Framework issued as soft law.

The UAE belongs here too, but by a narrower route: its only binding in-scope instrument is subnational, with everything at federal level sitting in soft law, strategy or non-binding guidance. The sixth is proposal-only: a single pre-legislative proposal, with no strategy scaffold and nothing yet in force. Chile and Nigeria each hold exactly one mapped instrument, both horizontal bills still in committee. Beyond those six architecture types, seven jurisdictions hold no hard-law instrument at all: Japan, Israel, Switzerland, Kenya, Chile, Nigeria, and South Africa. Switzerland is the limiting case, with a single mapped instrument; expert review attributes the absence to reliance on extra-territorial reach of the EU AI Act - the inherited architecture above, arrived at without a membership.

The same retrofitting appears at the level of institutions. The corpus names 144 governance actors. Thirty-one are AI-specific, created for the purpose; 84 pre-date AI governance and have since acquired a formal AI mandate; and a further 29 pre-date the AI question while exercising functions that reach AI with no formal mandate at all. Close to four in five are therefore institutions that existed before the mandate they now exercise. By type, the largest groups are ministries and departments (53) and sectoral regulators (33), followed by advisory bodies (21) and data protection authorities (15); only nine are dedicated AI regulators, and two are standards bodies. These jurisdictions have routed GPAI oversight through ministries and existing sectoral supervisors rather than by building new authorities, which is the instrument-level pattern of this section repeated at the institutional level: duties are built onto borrowed regimes, and the bodies carrying them have been borrowed too.

What that inheritance does not carry is enforcement reach. Forty-eight percent of the 144 bodies hold enforcement powers, 37 percent hold none, and 14 percent hold powers that are constrained, and the remaining two are national legislatures recorded against bills with no sponsoring department. The powers and the AI mandates sit in different institutions. Of the 31 AI-specific bodies, only four hold enforcement powers: the EU AI Office; the UAE’s Artificial Intelligence and Advanced Technology Council; the UAE’s Federal Authority for Artificial Intelligence and Data, established in June 2026; and Kenya’s Office of the Artificial Intelligence Commissioner, which exists only in a Bill still before committee. Outside the direct effect of EU law, the set of purpose-built AI bodies with power to enforce anything is two Emirati councils, and one Kenyan office that has not yet been created.

\textbf{Japan: a case of form versus force.} One case cuts against reading the instrument tier as a proxy for force. Japan holds five mapped instruments, and no hard law at all: four soft-law instruments and one strategy scaffold. Yet the coding does not read like advice: every positive provision on its 2026 government procurement guideline is coded as a requirement or an enforcement law, on an instrument that creates no legally enforceable duty. Our Japanese interviewee's account explains the pattern. Guidance issued by a ministry is understood by its addressees to mean "obey", and the compliance it produces runs through sourcing, bidding and procurement rather than through enforcement (§5.4).

Two things follow. First, though tier is a reliable statement of an instrument's legal form, it is an unreliable proxy for the behavior it produces: a jurisdiction with no hard law is not necessarily a jurisdiction without compulsion. Second, our method could not have caught this on its own. We record legal form because legal form is what instruments state; where practice diverges from form, only interview evidence has recovered it. The Chilean consumer-protection case at §5.7 is the same phenomenon in a different legal culture, which suggests the divergence is not peculiar to Japan. Expert review of the UAE supplies a third instance: the Central Bank's 2026 Guidance Note is coded soft law, as its legal form requires, but operates as hard law in practice, because a licensed institution that ignores it risks its license. Japan, Chile and the UAE thus show the same divergence in three legal cultures.

\section*{6. Discussion}

The discussion proceeds at two levels. The first asks where the twenty jurisdictions converge and where they diverge, such as in the architectures they have built, in which governance functions they have filled, in how they scope GPAI, and in the route by which obligations reach their addressees. This offers the answer to the study's comparative question, and occupies §6.1-6.5. The second asks what cuts across those differences, in §6.6.

Three findings, in particular, carry this discussion. First, the twenty jurisdictions have made structurally different choices about how to govern GPAI: six legal architectures recur, and where one jurisdiction covers a governance area another leaves it empty, the reverse also holds: Peru and Switzerland hold prohibitions and nothing else, while Nigeria holds assessment and evaluation provisions, and no prohibitions. Second, almost none of what these AI middle-powers have produced is binding: ten jurisdictions hold no binding provision at all, and the European Union and Brazil together hold more than half of those that exist. Third, one structure repeats across those differences – the institutions carrying GPAI governance were built for something else, reaching the actors they already could, while the institutional bodies with the technical capacity to evaluate are not the bodies with the authority to act on what they find.

\subsection*{6.1 What twenty AI middle-powers have built}

The twenty AI middle-power jurisdictions have not converged on a single form of GPAI governance. They have produced several, and the differences between them are structural.

Six architectures recur: Korea anchors governance in a horizontal statute with subordinate instruments beneath it; France and Germany inherit their binding layer from the European Union and add only soft law domestically; the UK and Australia distribute governance across many lower-tier instruments with no horizontal statute at all; Kenya announces governance through a national strategy while binding legislation remains pending, with Brazil a partial variant, in which three binding instruments are already in force; Taiwan and the Emirates bind sectorally, while governing horizontally in soft law; Chile and Nigeria hold a single pre-legislative proposal each, and nothing in force (§5.8).

Within those architectures, jurisdictions differ in which governance functions they have built, and the differences do not run along an axis of maturity: Peru and Switzerland hold GPAI prohibitions, and nothing else; Nigeria holds assessment and evaluation provisions, but no prohibition and no reporting duty; Taiwan is prohibition-heavy; the Emirates hold extensive evaluation and prohibition material, but no serious incident reporting at all; India, Korea, Kenya and Singapore cover all four areas nominally, but each holds a single provision in at least one of them. The mapping shows a set of different governance shapes, not a ranking from weak to strong (§6.3).

AI middle-powers also diverge in how they decide what falls within GPAI governance. Fifteen of the seventeen jurisdictions that define the class chose to bring a model into scope by matching a written description or by naming the kind of content it produces. Only the European Union and Korea scope by a property of the model itself, and only they pair a definition with a quantitative trigger, while Singapore scopes by output modality alone. For a provider operating across these markets there is no single scoping question to answer. Where scope turns on a threshold or a designation, there is an object jurisdictions could align – a number, a power, a register – but where it turns on whether a description fits, alignment has no target (§6.4).

Finally, the route by which GPAI governance arrives differs: a quarter of the instruments in the sample govern a single sector rather than AI in general, and in jurisdictions without horizontal legislation these carry the governance that exists – financial supervision, judicial administration, online safety, procurement. That route works because the regulator already holds powers over the firms in question, so a duty can be imposed without new legislation, and lands on the firms accustomed to answering to it, but the very same fact limits it: a financial supervisor reaches licensed institutions and no one else, and the duty fires on the parent regime's trigger, so a GPAI failure that produces no suspicious transaction, no data breach, and no service outage, may activate nothing at all (§6.5).

Taken together, these answer the study’s main comparative question. Among AI middle-powers governing GPAI, there is existing convergence in subject matter – almost every jurisdiction addresses the same four functions, in similar terms, drawn from the same small set of international reference points – while there is divergence in nearly everything else – in legal architecture, in which functions are built, in scoping technique, in legal force, and in the sectoral route by which obligations reach their addressees.

\subsection*{6.2 Governance activity as poor proxy for force}

The number of instruments produced by an AI middle-power jurisdiction seeking to govern general-purpose AI may be a measure of how much governance activity it has generated, but this says little about how far that governance reaches. A better measure is how many obligations sit in binding law that is in force, and whether those obligations are complete enough to be complied with. On that measure, the sample divides sharply. Ten of the twenty jurisdictions hold no binding, in-force GPAI provision at all. Of the 85 binding provisions, the European Union holds 26 and Brazil 18, so those two account for more than half between them, with the United Kingdom (12), Korea (8) behind them, then followed by Taiwan, Australia, Singapore, Peru, Canada, the Dubai International Financial Centre, and India. Binding GPAI law is therefore not only thinly spread across the sample but concentrated in a few jurisdictions, and absent from half of them.

Read against that measure, the composite ranking set out at §3.1 is a poor guide to GPAI governance. France and the UK sit at the top of the middle-power distribution, and Japan and Korea immediately below them, but their positions at the model layer bear no relation to that order. France and Germany hold fifteen positive provisions between them and no domestic hard-law AI instrument imposing GPAI obligations, their entire binding layer arriving through the direct effect of EU law. Japan ranks third and holds twenty-five positive provisions, none of which sits in hard law. Korea ranks fourth, holds sixteen provisions across five instruments, and is the only jurisdiction in the sample to have legislated a binding chain of its own accord, running from a definition through a compute threshold to assessment and evaluation duties \citep{KOR_085}, \citep{KOR_084}. Lower down the distribution, Brazil holds the second-largest body of binding GPAI law in the sample, eighteen provisions across three instruments already in force, and Peru, second from last on the composite score, carries a binding criminal statute.

The UK is the sharpest case, because its position changes character rather than rank. It holds sixteen governance instruments and 44 positive provisions, more than any jurisdiction outside the European Union, and twelve of those provisions are binding. All twelve sit in two sectoral statutes, the Online Safety Act 2023 \citep{GBR_153} and the Crime and Policing Act 2026 \citep{GBR_159}, and none imposes a duty on the provider of a GPAI model. Its single operative reporting provision asks that firms have reporting procedures, rather than that they report anything \citep{GBR_022}. The binding provisions exist, but they are pointed elsewhere. Australia shows the same shape at a different scale, with thirteen instruments carrying 45 positive provisions, the largest count of any non-EU jurisdiction, and binding provisions that are exclusively sectoral.

The same relationship holds at the smaller end of the sample: Israel holds three instruments carrying eight positive provisions; Singapore holds thirteen instruments carrying 27. Yet Singapore's generative AI framework recommends notification above a materiality threshold and leaves it undetermined \citep{SGP_178}, while Israel's anti-money-laundering and terror-financing guidance attaches an expectation to a statutory regime already in operation \citep{ISR_160}. The difference is not the volume of the documents, but whether they connect to working machinery.

India complicates the picture, albeit in a useful way. Nine instruments produce twenty positive provisions spread across intermediary liability, computer emergency response, banking and national AI policy, and only one of those instruments is binding hard law in force \citep{IND_164}. Volume can therefore extend the breadth of coverage while doing nothing for enforceability. Coverage and enforceability sit on separate axes, and a jurisdiction may be rich on one and poor on the other. Any claim about convergence between AI middle-powers has to be weighted by tier, and not mistake activity for force.

\subsection*{6.3 Absence as a finding}

The 205 confirmed absences in the dataset are not gaps in the research. Each records that the search protocol at §4.1 was completed for that jurisdiction and that governance area, and returned no positive provision. Treated that way, absence becomes evidence: it makes the negative space in AI middle-power governance of GPAI visible at provision level, and it allows the claim that a jurisdiction does not govern something to be stated as a finding. The confidence attaching to it is not the same as that attaching to a positive provision, which can be read directly from an instrument. A negative finding is therefore only as good as the search behind it, and it should be read as documented absence under the specific procedure described; a better-resourced or native-language search might, in principle, overturn it.

The distribution of those absences is itself a result, and it shows where the four-area accountability chain breaks. The United Arab Emirates is the clearest case of a chain broken at both ends: thirty-two positive provisions, almost all of them in evaluation and prohibitions, two in systemic risk assessment, and none in serious incident reporting. Volume does not predict which links a jurisdiction has built: India covers all four areas on twenty provisions, while the Emirates covers just three using more provisions. Only five jurisdictions – the European Union, the United Kingdom, Australia, Japan and Brazil – hold more than one provision in every area. Kenya and Korea cover all four governance areas, but only hold a single reporting provision each; India and Singapore cover all, but only hold a single systemic risk assessment provision each. Nominal coverage of the full chain is therefore eleven jurisdictions (including France and Germany by direct effect), but substantive coverage is made by just five jurisdictions.

Serious incident reporting is the area where absence is the most common finding, with 78 confirmed absences against 27 operative provisions, and ten jurisdictions record no reporting duty of their own.

\subsection*{6.4 Definitions describe outputs, not capabilities}

Instruments in the sample govern general-purpose AI with little definitional foundation to stand on. The sample contains 52 GPAI definitions across seventeen jurisdictions; 58 percent of instruments impose provisions on GPAI without defining it at all. The imbalance is sharpest where it matters most: five of the twenty instruments that bind carry a definition, while nearly half of the soft-law, non-binding and strategy instruments do. Where definitions exist, they mostly describe what a model produces, rather than what it is capable of, and few capture the generality dimension, which is the property that distinguishes GPAI from narrow AI.

The terms themselves show an inversion. The parent class is the least used: general-purpose AI appears as the defined term in eight definitions, while its sub-classes account for far more, with generative AI in 27, large language model in thirteen and foundation model in six. Jurisdictions are more willing to define a sub-class than the category that contains it, which is intelligible as drafting practice, since a sub-class is easier to describe by reference to what it produces. The consequence is that the class most international discussion is concerned with is the class domestic instruments are least likely to name.

This has practical consequences for compliance, and the clearest case lies between the two instruments that do pair a definition with a quantitative trigger. The EU AI Act presumes systemic risk where cumulative training compute exceeds 10\textsuperscript{25} floating-point operations, and Article 52 allows a provider to rebut that presumption with substantiated arguments that its model does not, exceptionally, present systemic risks. Korea's Enforcement Decree sets 10\textsuperscript{26} operations, an order of magnitude higher, and treats it not as a presumption, but as one of three cumulative conditions that must all be satisfied, alongside a state-of-the-art test and a severity test \citep{KOR_084}. A model trained between those two figures is therefore a general-purpose AI model with systemic risk in the European Union and falls outside Korea's class altogether. Singapore's framework diverges in a different direction, scoping by output modality alone, so a model displaying generality without generating content sits outside it while a small generative tool sits within \citep{SGP_178}. For a provider operating across these markets, there is no single scoping question to answer, and no shared mechanism into which these jurisdictions could pool their influence.

Whether the pattern reflects the speed at which these instruments were produced is a hypothesis for drafting-history or more interview-based work, rather than a finding of this mapping. A large part of the corpus was written between 2023 and 2026, before any settled account of what was being governed had emerged, and against a generation of models several capability steps behind those now in deployment. On our reading, establishing tighter and internationally usable definitions – of general-purpose AI, of the model and system layers, and of the capabilities that red lines would prohibit – would do more for convergence than any single further instrument. That is a judgment about sequencing rather than a finding of the mapping.

\subsection*{6.5 Sectoral routes work, but they inherit their limits}

One route into GPAI governance is more productive than its size suggests. The five financial-sector instruments in the corpus carry a disproportionate share of the provisions a regulator could act on: fifteen of the sample's 55 enforceable soft-law provisions sit in four of them \citep{ISR_160}, \citep{KOR_083}, \citep{GBR_105}, \citep{ARE_185}. The reason for this is structural – the regulator already holds powers over the same firms, the sector already runs a working serious incident reporting route with a named recipient and a timer, and the harm the guidance addresses is one the law already recognizes. GPAI risk arrives as a new path to a familiar harm, rather than as a new kind of harm requiring new law.

Other sectors show the same pattern. Brazil's National Council of Justice, a constitutional body that governs its courts, produced the only sectoral instrument in the sample covering all four governance areas, and one of only two anywhere in the dataset that sets out a complete reporting chain \citep{BRA_209}.

However, limits are inherited. A sectoral route appears to keep the trigger it was built for, so a duty attached to fraud reporting fires when there is fraud, not when a model fails. A GPAI incident that produces no personal data breach, no service outage and no suspicious transaction may activate no duty at all, whatever its consequences. The route also tends to produce guidance rather than legislation: none of the five financial instruments is a statute (§5.7). As Okolo and colleagues observe, a jurisdiction taking this route gains speed and real compliance pressure at the expense of permanence \citep{okolo2026}. Even so, the repurposing of existing reporting infrastructure remains a plausible foundation for international serious AI incident reporting machinery \citep{gomez2026}.

\subsection*{6.6 Governance built on borrowed institutions}

The following three parts describe one structure: AI middle-power jurisdictions are constructing GPAI governance through different legal architectures, but across those architectures they rely on pre-existing institutions and the regulatory relationships those institutions already hold. That reliance is consistent with the prevalence of sectoral and application-layer governance, and with the separation between evaluation capacity and enforcement authority. We offer this as a hypothesis for comparative or fieldwork-based research, rather than as a finding.

\textbf{The institutions were built for something else.} Most of the institutions carrying GPAI governance were built for something else: four in five pre-date the AI mandate they now exercise, and only nine of the 144 governance actors recorded are dedicated AI regulators.

\textbf{Obligations reach the actors those institutions already hold.} The sequence audit at §5.3 found that most systemic risk assessment duties attach to parties assessing systems they did not build, or are performed by governments from outside the development chain altogether. Of the positive provisions in that area, 46 percent fall on a public authority. Outside the direct effect of EU law and Korea’s 2025 Framework Act, almost nothing reaches the model before it is released by the party that built it.

This should not be read as a failed attempt at GPAI model governance. AI middle-powers may be governing a different layer for different reasons \citep{leicht2026}, \citep{okolo2026}, and the application-layer and state-performed provisions are better understood as distinct design, rather than as a shortfall. These jurisdictions have, in fact, already made the pivot that Leicht recommends: the great majority of what they have legislated attaches to the application layer of GPAI, to procurement, to deployment in public administration, to what platforms carry, and to what financial institutions rely on. This pattern is not an artefact of how instruments enter the corpus, and excluding every instrument whose only GPAI anchor is an output-defined sub-class leaves it intact (§4.5).

The disconnect Heim and Koessler identify runs underneath: compute thresholds appear in the sample, but they rarely trigger anything \citep{heim2024}. Only the European Union and Korea condition a duty on a property of the model itself, and they use different thresholds. Everywhere else, the duty attaches to the actor’s role or the deployment context, which means the instrument never has to say which models carry systemic risk.

\textbf{Evaluation capacity was built without the authority to act on it.} Evaluation and verification is not a neglected area: it holds more positive provisions than any governance area except prohibitions and monitoring, and sixteen of the twenty jurisdictions carry at least one (§5.4). What is scarce is legal force: only six provisions sit in binding law, and only four of the six impose an evaluation mechanism on anyone; the other two establish evaluation institutions without requiring that anything be evaluated. Roughly nineteen in every twenty provisions impose no legally enforceable duty on anyone. In most provisions, the party required to evaluate is the party whose model or system is being evaluated – a developer testing what it built, a deployer testing what it adopted, an agency testing what it uses: self-assessment is the majority evaluation model in eleven of the seventeen jurisdictions (including UAE DIFC) holding evaluation provisions.

The picture in prohibitions, serious incident monitoring and detection is similar: six jurisdictions name a monitor independent of the party being regulated – the European Union, Taiwan, Japan, the United Kingdom, Singapore and the Emirates (§5.5). In the rest, a prohibition or a monitoring duty either points at the regulated actor to watch itself, or names no monitor at all, and the second is far more common than the first. Where an independent monitor is named, it is always a body that existed for another purpose – a court, a police force, a market surveillance authority, a procurement ministry, a certification body – and never an AI institute. Eleven jurisdictions have established an AI safety institute or an equivalent evaluation body, but not one of them, including the European AI Office, is named as carrying a monitoring duty for prohibitions breaches. The institutional capacity exists, but the legal connection to it does not. This disconnection is structural: no named safety or evaluation institute in the corpus holds enforcement powers, and of the 31 AI-specific governance actors, only four do, and one of which exists solely in a Bill (§5.4, §5.8). \citep{KEN_065} These institutes were constituted to measure and to advise; on that reading, attaching a consequence to what they find would be a question of mandate, rather than of the governance instruments they publish.

\subsection*{6.7 Regional patterns, and the inverse relation between ambition and industry}

Two groupings are worth drawing out. The first is a cluster of Asia-Pacific jurisdictions that govern GPAI principally through soft law and voluntary tooling. Japan, Singapore and India each rely on government-conducted, or government-published, evaluation rather than mandatory provider obligations, and none carries a binding provider-facing evaluation duty or a binding model assessment duty. Korea is the regional outlier. Taiwan governs through sectoral hard law adapted for GPAI rather than through horizontal legislation, and Australia combines the largest body of non-EU soft-law evaluation material with binding provisions that are exclusively sectoral. These jurisdictions are often discussed as a bloc in international fora, but they are not converging on a shared regulatory model. The distance between Korea's framework-act approach and the voluntary tooling of Japan and Singapore is as wide as the distance between either and the European Union \citep{csisJapanAgileAIGovernance}, \citep{itifOneLawSetsSouthKoreaAIPolicy2025}, \citep{cetasAPacAIGovernanceMiddlePower}.

The second grouping runs the other way: Kenya, Nigeria, and Brazil have all drafted comprehensive horizontal bills carrying mandatory obligations and naming specific governance mechanisms, and Chile has done the same from a different legal tradition \citep{KEN_065}, \citep{NGA_069}, \citep{BRA_032}, \citep{CHL_054}. These are the only instruments outside the EU that would, if enacted, impose binding provider-facing evaluation, systemic risk assessment and incident reporting duties together. All four remain unenacted, however. The pattern nonetheless illustrates the Global Majority framing set out by Okolo and colleagues \citep{okolo2026}: three of these four jurisdictions sit in the lower half of the composite distribution at §3.1, and they are drafting the most demanding obligations in the sample, while jurisdictions with more developed AI industries opt for voluntary approaches. Whether regulatory ambition and AI industry presence are inversely related is a question this mapping raises but cannot answer, and we record it as a line for comparative work (§7.3).

\subsection*{6.8 What this study contributes}

The first contribution is empirical. To our knowledge, this is the first provision-level mapping of general-purpose AI governance across a defined set of AI middle-power jurisdictions, coded to a common scheme that separates the legal form of an instrument from the force of the language inside it. That separation is what allows the paper to show that tier is a reliable statement of legal form but an unreliable proxy for the behavior an instrument produces, a point the Japanese and Chilean cases make from opposite ends of the sample. The second is methodological: recording confirmed absence as a coded finding gives comparative GPAI governance research a way to record negative results under a stated procedure, rather than leaving them as silence.

The third contribution concerns sequencing. The debate about international GPAI governance has concentrated on what should be prohibited, but our findings suggest that, for AI middle-powers, the more immediate question may be what can be verified, and that the distance between an evaluation being conducted, and an evaluation carrying consequence, is the shortest distance between governance that these jurisdictions already have, and governance that successfully binds its obligation bearers (§6.6). That is an observation about sequencing, rather than a proposal about substance, and the dataset supports it: the jurisdictions in this sample have built more evaluation capability than they have built the law around which to use it.

A fourth contribution is conceptual. The composite ranking at §3.1 measures national weight in the area of AI governance, but it predicts very little about the governance these jurisdictions produce: France and Germany sit at or near the top and hold no domestic hard-law AI instrument imposing GPAI obligations of their own, while Brazil sits twelfth and holds three already in force. What the twenty have in common is not comparable standing, but a common problem: none hosts a frontier developer, so none can reach the point at which a model's capabilities are set. Each must instead govern models built elsewhere, by companies it cannot license or tax, using instruments addressed to the parties that it can more readily reach: the deployer, the purchaser, the platform, the supervised firm. Leicht describes this as the position of countries without a domestic frontier developer to tax or to regulate \citep{leicht2026periphery}. Our findings suggest it is also the more useful basis for comparison, because it predicts the shape of what these jurisdictions have built where a broader measure of national weight does not.

\section*{7. Conclusion}

The conclusion states where the twenty AI middle-power jurisdictions converge, sets out four axes on which they diverge, and identifies what remains for further research. Section 7.1 reports the convergence the mapping establishes, and its limits. Section 7.2 moves from description to implication, marking each implication as our judgment, rather than as a finding. Section 7.3 sets out five lines of work this study could not undertake.

\subsection*{7.1 Emerging consensus}

Convergence across the twenty AI middle-power jurisdictions is occurring, but it is located in the form that general-purpose AI governance takes, and in what these jurisdictions have so far declined to govern, rather than in binding duties on the parties that build GPAI models. The approach devised in the 2024 EU AI Act and the 2023 G7 Hiroshima Process, which is governing a class of models by what they are capable of rather than a class of applications by what they are used for, has been taken up widely as a subject matter, but so far rarely in the minutiae of regulation.

Eleven of the twenty jurisdictions carry positive provisions in all four governance areas, even if only in specific sectors, and fifteen (excluding subnational jurisdictions) carry provisions in at least three. Every jurisdiction, except South Africa, has produced something addressed to at least one of these functions. Where they diverge is in legal form and carrying force: ten of the twenty hold no binding provision at all, and of the 85 binding provisions that exist across the sample, the European Union and Brazil hold more than half between them. Recommendation, rather than requirement, is the majority register everywhere else.

Read in sequence, as the accountability chain described at §1, each governance area holds machinery that stops short of the connection that would give it effect: in systemic risk assessment, the output of the assessment is not required to reach an authority; in evaluation and verification, instruments name the evaluation mechanism but characteristically leave the standard unstated; in prohibitions, the strictest measures are the least enforceable, with nearly 70 percent of absolute prohibitions naming no detection mechanism; in serious incident reporting, every duty asks a single actor to report on what lies within its own control, so a harm visible only across several models or several deployments has no addressee anywhere in the sample.

Half the sample does not address serious incident reporting. Ten of the twenty jurisdictions – Canada, Chile, France, Germany, Nigeria, Peru, South Africa, Switzerland, Taiwan and the UAE – impose no reporting duty of their own, and in France and Germany the duty arrives solely through the direct effect of EU law. Among the ten that do address it, most name a recipient without stating when a report is due, or what follows if none is made: the OECD's common reporting framework proposes twenty-nine criteria for what such a report should carry \citep{oecdCommonReportingFrameworkAIIncidents2025}, but only three provisions in the corpus state a period and a consequence together. Where these jurisdictions have converged is not in what obligations they require, but in the institutions through which they require them, and that convergence does not reach the parties that build GPAI models (§6.6). Oversight is routed through ministries and existing sectoral supervisors, which is the institutional pattern set out at §6.6. However, the bodies holding enforcement powers are generally not the bodies holding AI mandates, which limits how far that form reaches providers (§5.8). Among AI middle-powers, the GPAI governance mechanisms that do exist diverge in legal form, in trigger, and in who bears the duty. Harmonisation strategies that lean into that divergence, working with what each jurisdiction has actually built, may have more to work with than strategies that wait for the templates to align.

\subsection*{7.2 Axes of divergence}

This paper now moves from description, to implication. The mapping currently shows divergence rather than a common GPAI governance template, and that divergence concentrates on four dimensions. Each is set out in two parts – what the coding shows, and then what we take from it, identified as our judgment, rather than as a finding. Where we draw an implication from a pattern, it is a policy judgment. This paper does not prescribe the substance of what any jurisdiction should enact.

\textbf{Whether a reporting duty carries a consequence.} A reporting duty is a chain of four elements rather than a single obligation: an addressee, a trigger, a period, and a consequence for failing to report (§5.6). Half the sample carries no such duty at all – ten jurisdictions record no reporting provision of their own, and in France and Germany the duty arrives solely through direct effect of EU law. Among the jurisdictions that do address it, the chain comes apart: most name a body to report to but then leave the trigger, the period, or both, undefined; only three provisions in the corpus state a period and a consequence together, against the twenty-nine criteria the OECD's common reporting framework proposes \citep{oecdCommonReportingFrameworkAIIncidents2025}. Two instruments calibrate the period to the severity of the incident, rather than fixing a single deadline: Article 73 of the 2024 EU AI Act, and in principle the Reserve Bank of India's 2025 committee report.

Where no domestic consequence is available, the sample shows consequences being borrowed from existing mechanisms of other sectors and authorities. A jurisdiction maintaining its own database of serious incidents, keyed to the harm categories the 2025 GPAI Code of Practice already uses, can share it with the EU AI Office, which holds powers to request information from providers, conduct its own model evaluations, and require mitigation, restriction of market availability or withdrawal \citep{iradukundaLeveragingEUAICode2025}. Bilateral arrangements with the Office, the International Network for Advanced AI Measurement, Evaluation and Science, and alerts to the EU Scientific Panel of Independent Experts run alongside. Kenya, whose national strategy commits to a reporting channel it has not yet established, is a member of that Network. However, arrangements of this kind reproduce asymmetries, with the better-resourced partner setting the agenda and the methods \citep{iradukundaLeveragingEUAICode2025}.

\textbf{Horizontal statute or borrowed sectoral machinery.} Jurisdictions divide between carrying obligations horizontally and carrying them through regimes built for something else. Sectoral instruments reach their addressees through an existing supervisory relationship, since the regulator already holds powers over the same firms, and the reporting route already runs. The trade is scope: a duty built onto a borrowed regime fires on that regime's trigger, and tends to leave the class of systems it applies to unstated, so coverage follows from who is supervised, rather than from any statement about the technology, which is often only inferred. Of the five financial-sector instruments in the corpus, only the Central Bank of the UAE's 2026 Guidance Note defines the class of systems it covers.

Government purchasing works the same way, with the same borrowing effect originating from the authority the state already holds over the addressee for reasons unconnected to GPAI. This is notable because the most advanced frontier models are developed in two jurisdictions but deployed, purchased and relied upon in all of them, so where several sizeable economies align on what they require of a system entering their markets, that shapes the incentives facing developers, wherever those developers sit \citep{hodesWhoGovernsAI}. Leicht identifies two kinds of procurement condition: one specifying what evidence a supplier must produce, which is neutral as to where a model was built – which we see in Japan's 2026 procurement guideline – \citep{JPN_189} and one favoring domestic suppliers, which he identifies as the first step toward a protectionism that entrenches disadvantage \citep{leicht2026periphery}. No instrument in the sample takes the second form.

\textbf{Where the evaluation duty sits, and whether a finding travels.} Four routes are in use. The European Union places the duty on the provider. The United Kingdom, Japan and Canada build the capacity and perform the evaluation themselves, with providers participating by agreement. Singapore and India publish the methods and leave their application to others. Australia and the Emirates rely on bodies that already held authority over the addressee, while their flagship national instruments stay advisory. The routes meet one common limit: eleven jurisdictions have created a body with an evaluation mandate, and of those bodies only the European AI Office holds powers to act on what an evaluation finds; the rest were constituted as advisory institutes. The capability and the consequence sit in different institutions, and the gap between them is one of mandate design, rather than of technical capacity. AI red lines literature notes that this is the part of that agenda which does not depend on prior international agreement about which capabilities should be off-limits \citep{zoumpalovaWhereWeDraw2026}.

Two dynamics run alongside. Where implementation guidance remains unspecified, firms build interim compliance packages to their own interpretation, and those harden into jurisdiction-specific systems that are costly to unify later \citep{aipolicybulletinWindowClosingBrusselsEffect}; and the EU AI Office is under-resourced for the mandate it holds, with a possible result that providers, rather than the Office, lead implementation of the Code \citep{iradukundaLeveragingEUAICode2025}. In parallel, several jurisdictions hold laws that retrofitting a single named element would bring to life, such as a detection mechanism attached to an existing prohibition, or an accredited verification body attached to an existing evaluation duty, and the DIFC's 2026 certification framework performs exactly that retrofit on its own 2023 prohibitions. \citep{AREAE-DU_284}

\textbf{Whether prohibitions attach to conduct or to capability.} The Athens Roundtable identified the domains in which AI red lines are needed \citep{tfsAthensRoundtableRecap2025}, and the mapping shows where national practice already runs parallel to them. The 127 prohibitions in the sample concentrate in manipulation, children's safety, and human rights and justice, which are the domains in which an existing body of law already prohibits the underlying conduct, already identifies who is bound, and already carries an enforcement route, so that covering the AI-enabled version is an amendment rather than a new regime. Capability-based prohibitions have no such antecedent within domestic AI law, and cluster in a handful of jurisdictions. Two drafting grammars are in use alongside this division: some jurisdictions ban outright while others permit subject to conditions – Peru, Chile, France, Germany and Singapore rely exclusively on absolute bans. An agreement drafted for one grammar does not implement cleanly where the working form is the other.

Jurisdictions seeking to harmonize GPAI governance, or to assess how far their own regime travels, can consider these four axes of divergence: whether a reporting duty names an addressee, a trigger, a period and a consequence; whether obligations are carried horizontally or through sectoral machinery that already reaches the addressee; where the evaluation duty sits and whether a consequence attaches to what it finds; and whether prohibitions attach to conduct with an existing legal antecedent or to capability without one. On each axis the sample shows more than one workable route, and what separates them is institutional starting point rather than disagreement about what matters.

\subsection*{7.3 Future research}

Five lines of work follow from what this study could not do. The first is the law-practice gap: whether named addressees receive reports, whether anything follows when they do, and whether the divergence between legal form and practical force observed in Japan, Chile and the United Arab Emirates is general, or local. That is fieldwork rather than document analysis. The second is repetition, since four AI bills remain in committee in Brazil, Chile, Kenya and Nigeria and the EU AI Act's GPAI provisions are only just beginning to generate implementation experience, repeating this mapping in twelve months’ time would establish whether these patterns are stable, deepening, or dispersing. The third is depth: an interoperability analysis asking whether the obligations that exist are mutually compatible for a provider operating across several markets, which the divergence between the EU and Korean compute thresholds suggests they may not be. The fourth concerns capacity; our mapping records that governance actors exist and what mandates they hold, but not whether they have the specialist staff, budget, or technical capability to exercise them. The fifth is breadth: increasing the number of jurisdictions in our sample, and/or adding further governance areas, may surface stronger patterns of GPAI governance convergence or divergence among AI middle-powers.

These gaps exist in domestic legal connection, rather than in international agreement – eleven jurisdictions with evaluation bodies, but almost no legal connection to what those bodies find, and an incident reporting architecture in which no actor holds the composite picture. These gaps can be closed by jurisdictions acting individually, before any agreement exists, and without prejudice to what an international agreement might eventually compel.

\section*{8. LLM usage statement}

Large language models were used at three stages of this research, and their use in building the dataset is recorded per record.

The first is extraction, as set out at §4.3. Claude Opus (versions 4.6, 4.8 and 5.0) was used to locate candidate provisions within primary texts and to propose values for the coded fields, under the configuration specified in our codebook. No proposed value entered the dataset unverified: the extracting researcher checked each provision against the primary source and confirmed or corrected every field before the record was accepted. Of the 587 provision records, 116 were extracted manually, 373 with model assistance, and 98 with model assistance from translated text; of the 52 GPAI definition records, 31 were created through model assistance, and thirteen with model-assisted translation. Where a model-assisted coding was contradicted by expert review, the expert review governed. Every model-assisted record carries a confidence flag, and records flagged low-confidence required two-researcher validation before the coding was accepted. The initial codebook was drafted with Claude Opus assistance, but has since been revised through fourteen major versions in response to coding decisions taken during extraction, and the scope questions the corpus raised. The version governing the dataset reported here is v14.5, and the revision history is documented.

The second use is translation. For instruments published in Arabic, German, Hebrew and Korean, where the team lacked the proficiency described at §4.5, translation was model-assisted and verified by round-trip back-translation between Claude Opus and Mistral, with material divergences flagged for human review, and resolved through expert review where necessary. Provision text is recorded verbatim in the original language alongside the English, so every coding decision can be audited against its source. The third use is writing. Claude Opus was used as a drafting and editing assistant, including for LaTeX formatting, and in verifying reported figures against the dataset.

The scope test and decision flow (§4.2), the four governance areas and their severity thresholds (§3.3), the selection of jurisdictions (§3.1), the expert verification program (§4.4), and all interpretive judgments and conclusions are the authors' own.

\section*{9. References}

\subsection*{Primary Sources}

\subsubsection*{Legal Acts, Statutes, and Treaties}
\printbibliography[category=statutes,heading=none]

\subsubsection*{Governance Instruments and Policy Frameworks}
\printbibliography[category=intlpolicy,heading=none]

\subsection*{Secondary Sources}

\subsubsection*{Scholarly}
\printbibliography[category=scholarly,heading=none]

\subsubsection*{Non-Scholarly}
\printbibliography[category=nonscholarly,heading=none]

\clearpage

\section*{10. Appendices}

\subsection*{Appendix A. Composite jurisdiction scoring}

\begin{table}[!ht]
\centering
\caption*{\textbf{Table A. Composite jurisdiction scoring}}
\label{tab:table-a-composite-jurisdiction-scoring}
{\scriptsize
\hyphenpenalty=50\exhyphenpenalty=50\emergencystretch=1em
\setlength{\tabcolsep}{3pt}
\renewcommand{\arraystretch}{1.15}
\begin{tabularx}{\textwidth}{YL{0.30\textwidth}C{0.075\textwidth}C{0.048\textwidth}C{0.048\textwidth}C{0.048\textwidth}C{0.048\textwidth}C{0.048\textwidth}C{0.048\textwidth}}
\toprule
\cellcolor[HTML]{EFEFEF} & \cellcolor[HTML]{EFEFEF} & \cellcolor[HTML]{EFEFEF} & \multicolumn{6}{>{\centering\arraybackslash}p{0.3635\textwidth}}{\cellcolor[HTML]{EFEFEF}\textbf{Cluster score (0–10)}} \\
\cellcolor[HTML]{EFEFEF}\textbf{Jurisdiction} & \cellcolor[HTML]{EFEFEF}\textbf{Region} & \cellcolor[HTML]{EFEFEF}\textbf{Composite} & \cellcolor[HTML]{EFEFEF}\textbf{(1)} & \cellcolor[HTML]{EFEFEF}\textbf{(2)} & \cellcolor[HTML]{EFEFEF}\textbf{(3)} & \cellcolor[HTML]{EFEFEF}\textbf{(4)} & \cellcolor[HTML]{EFEFEF}\textbf{(5)} & \cellcolor[HTML]{EFEFEF}\textbf{(6)} \\
\midrule
France & Europe (EU) & \textbf{8.15} & 7.28 & 9.23 & 5.42 & 9.87 & 7.85 & 9.24 \\
United Kingdom & Europe (non-EU) & \textbf{8.08} & 7.72 & 9.58 & 4.78 & 9.92 & 7.18 & 9.29 \\
Japan & Asia (East and South East) & \textbf{7.73} & 7.20 & 9.26 & 6.48 & 6.45 & 8.33 & 8.69 \\
Republic of Korea & Asia (East and South East) & \textbf{7.36} & 6.08 & 9.29 & 6.05 & 6.24 & 7.78 & 8.71 \\
Germany & Europe (EU) & \textbf{7.10} & 6.85 & 9.14 & 5.48 & 4.90 & 7.68 & 8.57 \\
Canada & North America & \textbf{6.74} & 7.31 & 7.92 & 4.50 & 4.82 & 6.94 & 8.93 \\
India & Asia (South) & \textbf{6.67} & 6.16 & 9.68 & 5.17 & 5.70 & 4.64 & 8.69 \\
Singapore & Asia (East and South East) & \textbf{6.19} & 6.99 & 7.12 & 4.25 & 4.48 & 6.62 & 7.70 \\
Australia & Pacific & \textbf{6.18} & 5.71 & 8.58 & 2.81 & 4.61 & 6.77 & 8.57 \\
Israel & MENA & \textbf{5.88} & 5.47 & 8.72 & 4.09 & 4.02 & 5.22 & 7.74 \\
Switzerland & Europe (non-EU) & \textbf{5.81} & 5.50 & 5.38 & 3.64 & 6.34 & 5.51 & 8.51 \\
Brazil & Latin America & \textbf{5.61} & 5.59 & 8.64 & 3.29 & 5.78 & 4.84 & 5.52 \\
Taiwan & Asia (East and South East) & \textbf{5.07} & \textit{ex.} & 8.04 & 6.49 & \textit{ex.} & 5.74 & 0.00 \\
Chile & Latin America & \textbf{4.90} & 4.60 & 5.68 & 2.15 & 3.58 & 5.32 & 8.10 \\
United Arab Emirates & MENA & \textbf{4.74} & 6.51 & 5.57 & 3.91 & 6.27 & 5.08 & 1.07 \\
South Africa & Africa (Sub-Saharan) & \textbf{4.19} & 4.63 & 6.43 & 1.84 & 5.41 & 6.52 & 0.33 \\
Nigeria & Africa (Sub-Saharan) & \textbf{4.12} & 4.57 & 6.16 & 1.53 & 3.19 & 4.20 & 5.08 \\
Peru & Latin America & \textbf{3.76} & 4.47 & 5.35 & 0.17 & 3.03 & 4.59 & 4.97 \\
Kenya & Africa (Sub-Saharan) & \textbf{3.42} & 4.07 & 3.65 & 0.56 & 4.25 & 3.17 & 4.82 \\
\bottomrule
\end{tabularx}}
\end{table}
\FloatBarrier

\textbf{Cluster composition.} Each cluster is scored out of 10 and the composite is the unweighted mean of the six. Rank-based inputs are rescaled linearly as ((N $-$ rank) / (N $-$ 1)) $\times$ 10, so rank 1 scores 10 and the lowest rank scores 0. Full methodology is at §3.1.

\textbf{(1) General economic influence.} Adaptive mean of five sub-scores, counting only non-zero fields: population rank (World Bank 2024, N = 217); GDP per capita PPP rank (World Bank 2024, N = 217); OECD inward FDI stock rank (N = 46); AI investment rank (Stanford AI Index, N = 15, scored ((16 $-$ rank) / 15) $\times$ 10 with a floor of 0.5 for jurisdictions assessed but outside the top 15); and AI diffusion relative to GDP per capita (Stanford AI Index), min–max rescaled within the sample. The adaptive denominator prevents absent investment or diffusion data from depressing the score; an adaptive mean of 0.5 or below returns 0.

\textbf{(2) Defence.} Simple mean of two sub-scores, fixed divisor: military expenditure rank (SIPRI 2025, N = 151) and global military power rank (Military Power Index, N = 80).

\textbf{(3) Resource wealth and compute infrastructure.} Adaptive mean of six sub-scores: critical minerals endowment (USGS Critical Minerals Atlas), a weighted count over a maximum of 30, weighting gallium and germanium at 3, rare earths and speciality metals at 2, and other listed minerals at 1; hyperscaler presence (count of 6 providers); notable AI models (Epoch), scaled logarithmically as log(count) / log(663) $\times$ 10 to accommodate a highly skewed distribution; the CNAS Sovereign AI Index (out of 8, excluded when zero); Global AI Vibrancy rank (Stanford AI Index, N = 36, weighting only the R\&D, talent and infrastructure pillars, excluded when zero); and a semiconductor industry sub-score combining industry market capitalization at 0.7 weight, square-root normalized against the sample maximum, with fabrication plant count at 0.3 weight (World Population Review), linearly normalized against the sample maximum. The semiconductor and critical minerals sub-scores always count in the denominator, since a zero there is a substantive finding rather than missing data.

\textbf{(4) Diplomatic agility.} Mean of the sub-scores present, with a variable divisor of 1 or 2: Global Soft Power Index rank (Brand Finance, N = 193, excluded when zero) and UN Security Council membership, scored permanent = 10, non-permanent within the last five years = 3, none = 0. A recorded 'none' counts as a present component and is averaged in as a zero.

\textbf{(5) Democracy and AI governance infrastructure.} Mean of the sub-scores, with a variable divisor of 5 to 7: AI regulatory body status (founding AISI = 10, established = 8, announced = 5, proposed = 2, none = 0); horizontal AI legislation, taking the maximum of the domestic and supranational scores so that EU member states are not credited twice, on a ladder of full enforcement = 10, partial enforcement = 8, adopted = 6, pending = 4, indicated = 2, none = 0; CAIDP AI and Democratic Values Index (out of 12); IMF AI Preparedness Index (0 to 1, multiplied by 10); Transparency International CPI (out of 100); Oxford Insights AI Government Readiness rank (N = 195); and the CORDA Democratic AI-Readiness Index (out of 100).

\textbf{(6) International AI governance participation.} Simple mean of three sub-scores, fixed divisor: OECD GPAI membership (yes = 10); Friend of the Hiroshima Process (yes = 10); and an overall summits score, itself an adaptive mean across the 2023 UK, 2024 Seoul, 2025 Paris and 2026 India summits. Each summit uses a weighted engagement multi-select divided by that summit's own maximum, so that jurisdictions are not penalized for summits predating their engagement. Declining to sign the 2025 Paris statement scores 3 against 4 for signature, since several jurisdictions withheld signature on stated grounds and non-signature there reflects active engagement.

{\footnotesize \textit{Note.} Jurisdictions are ordered by composite score, descending. \textit{ex.} denotes a cluster excluded from a jurisdiction's composite by a named carve-out. Clusters (1) and (4) are excluded for Taiwan, whose composite is computed over four clusters, because the primary sources for both do not publish separate figures for Taiwan given its status as a non-UN-member entity, leaving those clusters unrepresentative. Taiwan's cluster (6) score is retained at zero, since its limited participation in international governance processes reflects a real constraint on its political standing, rather than a gap in the data. Outside that carve-out, the composite does not use adaptive averaging. The European Union is in the instrument sample as a supranational jurisdiction and carries no composite score.}

\subsection*{Appendix B. Decision flow test}

Every candidate instrument was tested against an eight-step decision flow to assist systematic scope testing. After confirming the jurisdiction, the flow directs in either of two ways, which offers a GPAI-relevance test structured as an OR gate. Under the first branch, an instrument is GPAI-relevant if it carries a GPAI definition in the sense set out in §3.2; under the second branch, an instrument that carries no such definition is nonetheless GPAI-relevant if it contains operative provisions in at least one of the four key governance areas that respond to GPAI-specific risks. An instrument satisfying either of these passes; an instrument failing both is out of scope. The gate is disjunctive by design as requiring both criteria would result in under-scoping. See Section 4.2.1 for a figure of the decision flow.

An instrument that carries no GPAI anchor in its own text does not enter scope because it sits within a wider legal architecture containing GPAI-specific instruments. Implementing circulars, covering notices, and guidance that operationalizes a parent framework were tested on their own operative text: where that text carries GPAI-specific content, or engages a brightline risk class, the instrument passes on its own terms; where it does not, it is out of scope. The rule prevents instruments that carry no relevant GPAI provisions of their own, or whose operative content restates a parent, from inflating the instrument count without adding governance content to the mapping. The GPAI-definition test is satisfied by the instrument's own text, or by the text of the legal instrument that directly establishes its operative scope, not by subordinate legislation made under it: rules made under an Act are a child of that Act, and cannot supply it with a GPAI anchor. The child instrument might therefore pass where the parent does not.

The flow does not resolve borderline cases: where a determination was uncertain, the instrument was flagged for the research team lead and methodological supervisor to resolve, rather than decided at the node. The scoping methodology was then clarified to ensure clarity over the borderline case – therefore, no instrument has been either included or excluded on the basis of judgment call or individual carveout.

\subsubsection*{B.1 Prescriptive gating}

Next, the decision flow assessed whether the instrument is part of the jurisdiction’s strategy scaffold for GPAI governance. National strategies, action plans, and white papers fall into this category. This step exempts such instruments from the prescriptive gate, though they must still satisfy the GPAI-relevance test, and their aspirational content is extracted and qualified in the coding. All other instruments must clear a prescriptive gate: the instrument must contain at least one passage directing an identified actor – a provider, deployer, authority, standards body, or equivalent – to do something, in language that prescribes action. Catalogs of good practice, aspirational or doubly conditional formulations outside the strategy scaffold, and descriptions (as opposed to prescriptions) of what regulated actors already do fail the gate. By contrast, concrete issuing-body commitments, including standing institutional mandates by which a body commits to performing a governance function, pass it – subject to an issuer-legal-standing test requiring that the issuing body hold a legal mandate or statutory basis for the governance function concerned. Documents that only describe capabilities or risks, and pure research outputs, were excluded as non-instruments on this basis. Consultative and preliminary regulator outputs were excluded on the same basis: where a regulator proposes principles or expectations for consultation, or to guide its own future work, and characterizes them as preliminary, or subject to further engagement, the prescriptive content is proposed rather than settled.

\subsubsection*{B.2 Institutional architecture tracing}

A separate rule governed institutional architecture. A provision that establishes, empowers or structures an authority – for example by establishing complaints machinery, penalty proceedings, inter-authority cooperation, coordination centers – is retained within a governance area only where it serves an identifiable systemic risk assessment duty, evaluation mechanism, prohibition, or reporting obligation already coded in the same area and within the same governance instrument. Merely establishing architecture without attaching it to an identifiable governance duty therefore falls outside the area. The rule prevents institutional scaffolding from inflating the provision count, but it does not reach provisions that themselves impose a governance-area function on the body they concern. The distinction is between a provision that creates a body and a provision that tells a body to assess. For example, "a Coordination Centre is established," or "the authority shall have the following powers" are both architecture: each brings a body into being. or gives it capacity without telling it to perform any governance-area function. By contrast, "develop and maintain a cross-economy, society-wide AI risk register", "analyse systemic risks and unforeseen risks emerging from GPAI models" are assessment duties in their own right, so each is coded on its own terms, with sequence position recorded as performed by the state itself.

\subsubsection*{B.3 Currency and legal form}

The remaining three steps address policy currency, legal form, and the scope decision that follows from them. One tests currency: expired, withdrawn and wholly superseded instruments are excluded, while partially superseded instruments remain in scope with their supersession recorded and the revising instrument cross-referenced, so that the relationship between an instrument and the one that amends it is preserved. Another classifies the instrument by tier – hard law primary, hard law secondary, soft law, strategy scaffold, or non-binding – according to its formal legal status rather than its practical effect. Primary hard law covers AI-specific statutes and bills, sectoral statutes, international treaty obligations, and broader frameworks and statutes containing GPAI-specific provisions. Secondary hard law covers statutory instruments, agency rules and regulations made under a primary instrument, together with their associated enforcement provisions; supervisory statements and agency expectations are typically soft law rather than secondary hard law, since only rules carrying the force of law qualify. Non-binding pre-legislative proposals are in scope only where they propose primary legislation. Voluntary codes – opt-in commitment frameworks lacking a governance-function issuer – are out of scope unless adherence to them creates a formal presumption of conformity with a hard-law obligation already in force; an interim commitment framework issued in anticipation of future legislation therefore does not satisfy that test.

\subsection*{Appendix C. Coding vocabularies}

This appendix reproduces the controlled vocabularies and coded fields of the project codebook that are used in the analysis reported in this paper. It is drawn from codebook v14.5. Values separated by vertical bars are single-select unless marked multi-select.

Every coded record additionally carries data quality fields referenced in §4.3–4.5 of the paper: extraction\_\allowbreak{}method (Manual | AI-assisted | AI-assisted (translated)); ai\_\allowbreak{}confidence\_\allowbreak{}flag (High | Low | N/A), where Low triggers two-researcher validation; round\_\allowbreak{}trip\_\allowbreak{}check (Not needed | Passed | Flagged) for translated material.

\textit{One row per governing body (regulator, ministry, standards body, advisory body, etc.) within a jurisdiction. Each governance instrument links to the body that issued or administers it.}

{\footnotesize
\hyphenpenalty=50\exhyphenpenalty=50\emergencystretch=1em
\setlength{\tabcolsep}{4pt}
\renewcommand{\arraystretch}{1.15}
\begin{tabularx}{\textwidth}{L{0.19\textwidth}YL{0.34\textwidth}}
\multicolumn{3}{@{}>{\raggedright\arraybackslash}p{\dimexpr\textwidth-2\tabcolsep}@{}}{\normalsize \textbf{Table C.1 Governance Actors}} \\[5pt]
\toprule
\cellcolor[HTML]{EFEFEF}\textbf{Field} & \cellcolor[HTML]{EFEFEF}\textbf{Description} & \cellcolor[HTML]{EFEFEF}\textbf{Values / recording} \\
\midrule
\endfirsthead
\toprule
\cellcolor[HTML]{EFEFEF}\textbf{Field} & \cellcolor[HTML]{EFEFEF}\textbf{Description} & \cellcolor[HTML]{EFEFEF}\textbf{Values / recording} \\
\midrule
\endhead
\midrule\multicolumn{3}{r}{\textit{continued on next page}}\\
\endfoot
\bottomrule
\endlastfoot
body\_\allowbreak{}type & What kind of governance institution. National legislatures are not recorded; the relevant sponsoring department is recorded instead. & Constitutional body | Dedicated AI regulator | Sectoral regulator | Data protection authority | Standards body | Ministry or department | Advisory body | Other \\
ai\_\allowbreak{}specific & Whether the body was created specifically for AI governance. & AI-specific | Pre-existing with AI mandate | Pre-existing without formal AI mandate \\
mandate\_\allowbreak{}scope & Whether the body’s AI mandate extends across all sectors or is confined to one. Classified by the scope of the AI mandate, not institutional origin. & Horizontal | Sectoral \\
sector\_\allowbreak{}if\_\allowbreak{}sectoral & Which sector, if the mandate is sectoral. Same vocabulary as the Instruments table. & See sector\_\allowbreak{}if\_\allowbreak{}sectoral under Table C.2 \\
enforcement\_\allowbreak{}powers & Whether the body possesses formal enforcement powers as an institution, regardless of whether currently exercised over AI. & Yes | Limited | Advisory only | N/A (originating body) \\
established\_\allowbreak{}date / established\_\allowbreak{}by & When the body was established and the instrument that establishes or empowers it. & Date; instrument reference or description \\
\end{tabularx}}

{\footnotesize \textit{One row per governance instrument. Links to its jurisdiction, its issuing body, and (where applicable) a GPAI definition record.}}

{\footnotesize
\hyphenpenalty=50\exhyphenpenalty=50\emergencystretch=1em
\setlength{\tabcolsep}{4pt}
\renewcommand{\arraystretch}{1.15}
\begin{tabularx}{\textwidth}{L{0.19\textwidth}YL{0.34\textwidth}}
\multicolumn{3}{@{}>{\raggedright\arraybackslash}p{\dimexpr\textwidth-2\tabcolsep}@{}}{\normalsize \textbf{Table C.2 Governance Instruments}} \\[5pt]
\toprule
\cellcolor[HTML]{EFEFEF}\textbf{Field} & \cellcolor[HTML]{EFEFEF}\textbf{Description} & \cellcolor[HTML]{EFEFEF}\textbf{Values / recording} \\
\midrule
\endfirsthead
\toprule
\cellcolor[HTML]{EFEFEF}\textbf{Field} & \cellcolor[HTML]{EFEFEF}\textbf{Description} & \cellcolor[HTML]{EFEFEF}\textbf{Values / recording} \\
\midrule
\endhead
\midrule\multicolumn{3}{r}{\textit{continued on next page}}\\
\endfoot
\bottomrule
\endlastfoot
governance\_\allowbreak{}level & The level at which the instrument operates. International instruments may carry multiple jurisdiction records reflecting differing engagement. & International | Supranational | National | Subnational \\
adoption\_\allowbreak{}status & Stage in the legislative lifecycle; distinguishes binding law from proposals. & In force | Adopted not yet in force | In draft or committee | Proposed | Expired or withdrawn \\
superseded\_\allowbreak{}status & Whether the instrument has been replaced by a later instrument. & Yes | No | Pending | Partially or Added to \\
instrument\_\allowbreak{}tier & Primary structural classification. & Hard law | Soft law | Strategy scaffold | Non-binding \\
instrument\_\allowbreak{}subtype\_\allowbreak{}hard\_\allowbreak{}law & Instrument type within the hard law tier. & Primary instrument | Secondary instrument | N/A \\
primary\_\allowbreak{}instrument\_\allowbreak{}subtype & Type of primary hard law instrument. & International treaty obligations | AI-specific bills and sectoral AI statutes | Broader frameworks and statutes containing AI-specific provisions | N/A \\
secondary\_\allowbreak{}instrument\_\allowbreak{}subtype & Type of secondary hard law instrument. & Statutory instrument or agency rule | Enforcement provision | N/A \\
instrument\_\allowbreak{}subtype\_\allowbreak{}soft\_\allowbreak{}law & Instrument type within the soft law tier. & Executive orders and decrees | Ministerial orders, statements and notices | Regulatory guidance and codes of practice | De facto regulatory tools | N/A \\
instrument\_\allowbreak{}subtype\_\allowbreak{}strategy-scaffold & Instrument type within the strategy scaffold tier. & National strategy | White papers | Other | N/A \\
instrument\_\allowbreak{}subtype\_\allowbreak{}non-binding & Instrument type within the non-binding tier. Pre-legislative proposals are in scope only where they propose primary legislation. & Pre-legislative proposals and instruments in draft or committee processes | Policy papers, reports, and reviews | N/A \\
carries\_\allowbreak{}gpai\_\allowbreak{}definition & Whether the instrument carries a GPAI definition. & Yes | No \\
horizontal\_\allowbreak{}or\_\allowbreak{}sectoral & Whether the instrument applies economy-wide or is sector-specific. & Horizontal | Sectoral \\
sector\_\allowbreak{}if\_\allowbreak{}sectoral & Which sector, if sectoral. & Children’s safety | Consumer protection | Criminal law | Critical infrastructure | Cybersecurity | Democracy | Digital | Education | Employment and welfare | Environment | Financial services | Judicial oversight | Market competition | N/A | Other \\
carries\_\allowbreak{}instrument\_\allowbreak{}cross\_\allowbreak{}reference & Whether the instrument explicitly references another in-scope instrument in its operative text; used especially to bind secondary legislation to its primary instrument. & Yes | No \\
\end{tabularx}}

{\footnotesize \textit{One row per distinct GPAI definition encountered. Different instruments within one jurisdiction may define GPAI differently; each definition receives its own record.}}

{\footnotesize
\hyphenpenalty=50\exhyphenpenalty=50\emergencystretch=1em
\setlength{\tabcolsep}{4pt}
\renewcommand{\arraystretch}{1.15}
\begin{tabularx}{\textwidth}{L{0.19\textwidth}YL{0.34\textwidth}}
\multicolumn{3}{@{}>{\raggedright\arraybackslash}p{\dimexpr\textwidth-2\tabcolsep}@{}}{\normalsize \textbf{Table C.3 GPAI Definitions}} \\[5pt]
\toprule
\cellcolor[HTML]{EFEFEF}\textbf{Field} & \cellcolor[HTML]{EFEFEF}\textbf{Description} & \cellcolor[HTML]{EFEFEF}\textbf{Values / recording} \\
\midrule
\endfirsthead
\toprule
\cellcolor[HTML]{EFEFEF}\textbf{Field} & \cellcolor[HTML]{EFEFEF}\textbf{Description} & \cellcolor[HTML]{EFEFEF}\textbf{Values / recording} \\
\midrule
\endhead
\midrule\multicolumn{3}{r}{\textit{continued on next page}}\\
\endfoot
\bottomrule
\endlastfoot
gpai\_\allowbreak{}definition\_\allowbreak{}text & Verbatim definitional text for GPAI or its functional equivalent. & Verbatim; original language plus English where non-English \\
gpai\_\allowbreak{}term\_\allowbreak{}used & Terms the instrument uses. & Multi-select: Advanced/High-performance AI | Generative AI | GPAI | Foundation model | Frontier AI | LLM | High-impact AI | Other \\
scope\_\allowbreak{}mechanism & How the definition brings systems into scope. & Definition-based | Designation-based | Capability-based | Use-context-based | Output-modality-based | Combination (specified) \\
numeric\_\allowbreak{}threshold & Any stated numeric threshold (e.g. training compute). & Verbatim; no conversion; N/A if none \\
threshold\_\allowbreak{}metric & Type of metric used in the threshold. & FLOPs | Parameters | Compute proxy | Other \\
\end{tabularx}}

{\footnotesize \textit{One row per provision, at a default granularity of one record per section of the instrument, with all applicable governance areas tagged on that record. Provision text is recorded verbatim, in original language and English where the source is non-English.}}

{\footnotesize
\hyphenpenalty=50\exhyphenpenalty=50\emergencystretch=1em
\setlength{\tabcolsep}{4pt}
\renewcommand{\arraystretch}{1.15}
\begin{tabularx}{\textwidth}{L{0.19\textwidth}YL{0.34\textwidth}}
\multicolumn{3}{@{}>{\raggedright\arraybackslash}p{\dimexpr\textwidth-2\tabcolsep}@{}}{\normalsize \textbf{Table C.4 Provisions (core fields)}} \\[5pt]
\toprule
\cellcolor[HTML]{EFEFEF}\textbf{Field} & \cellcolor[HTML]{EFEFEF}\textbf{Description} & \cellcolor[HTML]{EFEFEF}\textbf{Values / recording} \\
\midrule
\endfirsthead
\toprule
\cellcolor[HTML]{EFEFEF}\textbf{Field} & \cellcolor[HTML]{EFEFEF}\textbf{Description} & \cellcolor[HTML]{EFEFEF}\textbf{Values / recording} \\
\midrule
\endhead
\midrule\multicolumn{3}{r}{\textit{continued on next page}}\\
\endfoot
\bottomrule
\endlastfoot
governance\_\allowbreak{}areas & Which governance areas the provision addresses; routes the record to the area-specific fields below. & Multi-select: Systemic risk assessment | Evaluation and verification | Prohibitions and serious incident detection/monitoring | Serious incident reporting \\
obligation\_\allowbreak{}bearer & Who the obligation falls on. Additional bearers receive their own records. & Provider | Deployer | Authority | Standards body | Sectoral actor | Supplier | Other | Not specified \\
obligation\_\allowbreak{}type & Nature of the obligation. & Must | Must (standing institutional mandate) | Should | Should (enforceable soft law) | Should (non-enforceable) | May | Prohibit | Report | Other \\
should\_\allowbreak{}commitment\_\allowbreak{}type & Type of accountability behind a Should provision. & Advisory (no named accountability) | Comply-or-explain (regulatory benchmark) | Industry commitment | State commitment | N/A \\
risk\_\allowbreak{}domains\_\allowbreak{}covered & Which risk domains must be assessed. Coded only where a specific passage of the instrument grounds the code. & Multi-select: CBRN | Loss of control | Concentration of power | Cyber | Critical infrastructure | Cross-cutting | Children’s safety | Consumer rights | Data privacy | Fundamental rights | Democracy/societal manipulation | Financial | Public health | Worker rights | Environmental | Not specified \\
negative\_\allowbreak{}record & Whether the record is a confirmed-absence finding, distinguishing ‘nothing found’ from ‘not yet looked’. & No | Confirmed absent | Coverage uncertainty \\
absence\_\allowbreak{}confidence & Confidence in an absence finding. Confirmed absent: comprehensive search completed, no provision found. Coverage uncertainty: search was limited or coverage unsure. & Confirmed absent | Coverage uncertainty (completed only for negative records) \\
\end{tabularx}}

{\footnotesize \textit{(Table C.5 over)}}

\clearpage
{\footnotesize
\hyphenpenalty=50\exhyphenpenalty=50\emergencystretch=1em
\setlength{\tabcolsep}{4pt}
\renewcommand{\arraystretch}{1.15}
\begin{tabularx}{\textwidth}{L{0.19\textwidth}YL{0.34\textwidth}}
\multicolumn{3}{@{}>{\raggedright\arraybackslash}p{\dimexpr\textwidth-2\tabcolsep}@{}}{\normalsize \textbf{Table C.5 Provisions (systemic risk assessment fields)}} \\[5pt]
\toprule
\cellcolor[HTML]{EFEFEF}\textbf{Field} & \cellcolor[HTML]{EFEFEF}\textbf{Description} & \cellcolor[HTML]{EFEFEF}\textbf{Values / recording} \\
\midrule
\endfirsthead
\toprule
\cellcolor[HTML]{EFEFEF}\textbf{Field} & \cellcolor[HTML]{EFEFEF}\textbf{Description} & \cellcolor[HTML]{EFEFEF}\textbf{Values / recording} \\
\midrule
\endhead
\midrule\multicolumn{3}{r}{\textit{continued on next page}}\\
\endfoot
\bottomrule
\endlastfoot
risk\_\allowbreak{}assessment\_\allowbreak{}trigger & What activates the duty to assess. & Verbatim \\
risk\_\allowbreak{}assessor & Who performs the assessment. & Named actor \\
risk\_\allowbreak{}assessment\_\allowbreak{}timing & When the assessment is performed relative to the development/deployment divide. & Upstream | Downstream | Whole-lifecycle (authority) | Whole-lifecycle (other) | Architecture trace | Not specified \\
output\_\allowbreak{}to\_\allowbreak{}authority & Whether the assessment output must be shared with an authority. & Yes | No | Not specified \\
output\_\allowbreak{}addressee & The authority or body to whom the output must be submitted or disclosed. & Named governing body \\
\end{tabularx}}

{\footnotesize
\hyphenpenalty=50\exhyphenpenalty=50\emergencystretch=1em
\setlength{\tabcolsep}{4pt}
\renewcommand{\arraystretch}{1.15}
\begin{tabularx}{\textwidth}{L{0.19\textwidth}YL{0.34\textwidth}}
\multicolumn{3}{@{}>{\raggedright\arraybackslash}p{\dimexpr\textwidth-2\tabcolsep}@{}}{\normalsize \textbf{Table C.6 Provisions (evaluation and verification fields)}} \\[5pt]
\toprule
\cellcolor[HTML]{EFEFEF}\textbf{Field} & \cellcolor[HTML]{EFEFEF}\textbf{Description} & \cellcolor[HTML]{EFEFEF}\textbf{Values / recording} \\
\midrule
\endfirsthead
\toprule
\cellcolor[HTML]{EFEFEF}\textbf{Field} & \cellcolor[HTML]{EFEFEF}\textbf{Description} & \cellcolor[HTML]{EFEFEF}\textbf{Values / recording} \\
\midrule
\endhead
\midrule\multicolumn{3}{r}{\textit{continued on next page}}\\
\endfoot
\bottomrule
\endlastfoot
eval\_\allowbreak{}type & Type of mechanism. Where evaluation timing does not mirror every selected type, the provision is split into separate records so that timing per type remains unambiguous. & Multi-select: Conformity assessment | Controllability | Capability testing | Compute metric | Bias testing | Benchmarking | Performance testing | Safeguard efficacy | Safety testing | Security testing | Red-teaming | Certification | Audit | Attestation | Accreditation | Human uplift | Other | Not specified | None (institutional architecture) \\
eval\_\allowbreak{}timing & When the evaluation occurs. & Pre-deployment | Post-deployment | Periodic | Triggered | Continuous | Not specified \\
eval\_\allowbreak{}performer & Who performs the evaluation. & Provider | Deployer | Independent third party | National authority | Sectoral regulator | Standards body | Other | Not specified \\
eval\_\allowbreak{}independence & Whether the evaluator is independent of the evaluated party. & Self-assessment | Independent third party | Authority | Other \\
eval\_\allowbreak{}subject & What is evaluated or verified. & Free text \\
\end{tabularx}}

{\footnotesize
\hyphenpenalty=50\exhyphenpenalty=50\emergencystretch=1em
\setlength{\tabcolsep}{4pt}
\renewcommand{\arraystretch}{1.15}
\begin{tabularx}{\textwidth}{L{0.19\textwidth}YL{0.34\textwidth}}
\multicolumn{3}{@{}>{\raggedright\arraybackslash}p{\dimexpr\textwidth-2\tabcolsep}@{}}{\normalsize \textbf{Table C.7 Provisions (prohibitions, serious incident monitoring and detection fields)}} \\[5pt]
\toprule
\cellcolor[HTML]{EFEFEF}\textbf{Field} & \cellcolor[HTML]{EFEFEF}\textbf{Description} & \cellcolor[HTML]{EFEFEF}\textbf{Values / recording} \\
\midrule
\endfirsthead
\toprule
\cellcolor[HTML]{EFEFEF}\textbf{Field} & \cellcolor[HTML]{EFEFEF}\textbf{Description} & \cellcolor[HTML]{EFEFEF}\textbf{Values / recording} \\
\midrule
\endhead
\midrule\multicolumn{3}{r}{\textit{continued on next page}}\\
\endfoot
\bottomrule
\endlastfoot
prohibition\_\allowbreak{}text & Verbatim text of the prohibition. & Verbatim \\
prohibition\_\allowbreak{}target & What is prohibited. ‘Capability’ targets what the system can do regardless of use; ‘Use’ targets a specific application or deployment context regardless of capability; ‘Outcome’ targets a specified harmful result regardless of design or intent. & Use | Capability | Outcome | N/A \\
prohibition\_\allowbreak{}structural\_\allowbreak{}form & Legal structure of the prohibition. & Absolute ban | Conditional ban | Moratorium | Restriction | Other | N/A \\
detection\_\allowbreak{}mechanism & How compliance is monitored or breaches detected. & Mandatory disclosure | Audit rights | Whistleblower | Complaint | Automated content screening | Post-deployment monitoring (provider) | Regulatory supervision | Usage and access monitoring | Other | Not specified \\
monitoring\_\allowbreak{}obligation\_\allowbreak{}bearer & Who is responsible for monitoring. & Named governing body or actor \\
\end{tabularx}}

{\footnotesize
\hyphenpenalty=50\exhyphenpenalty=50\emergencystretch=1em
\setlength{\tabcolsep}{4pt}
\renewcommand{\arraystretch}{1.15}
\begin{tabularx}{\textwidth}{L{0.19\textwidth}YL{0.34\textwidth}}
\multicolumn{3}{@{}>{\raggedright\arraybackslash}p{\dimexpr\textwidth-2\tabcolsep}@{}}{\normalsize \textbf{Table C.8 Provisions (serious incident reporting fields)}} \\[5pt]
\toprule
\cellcolor[HTML]{EFEFEF}\textbf{Field} & \cellcolor[HTML]{EFEFEF}\textbf{Description} & \cellcolor[HTML]{EFEFEF}\textbf{Values / recording} \\
\midrule
\endfirsthead
\toprule
\cellcolor[HTML]{EFEFEF}\textbf{Field} & \cellcolor[HTML]{EFEFEF}\textbf{Description} & \cellcolor[HTML]{EFEFEF}\textbf{Values / recording} \\
\midrule
\endhead
\midrule\multicolumn{3}{r}{\textit{continued on next page}}\\
\endfoot
\bottomrule
\endlastfoot
reporting\_\allowbreak{}addressee & To whom serious incidents must be reported. & Named governing body \\
reporting\_\allowbreak{}trigger & What event activates the reporting duty. & Verbatim; ‘Not specified’ if silent \\
reporting\_\allowbreak{}timeframe & Time allowed for reporting. & Verbatim; ‘Not specified’ if silent \\
reporting\_\allowbreak{}consequence\_\allowbreak{}type & Category of consequence for failure to report. & Administrative fine | Market access restriction | Licence revocation | Criminal penalty | Public disclosure | Corrective order | Other | Not specified \\
reporting\_\allowbreak{}consequence & Consequence of failure to report, verbatim. & Verbatim; ‘Not specified’ if silent \\
reporting\_\allowbreak{}channel\_\allowbreak{}type & Whether the reporting channel is AI-specific, repurposed from a sectoral regime, or general. & AI-specific | Repurposed sectoral | Not specified | Other \\
\end{tabularx}}

\subsection*{Appendix D. Data availability}

The dataset underlying this study is openly available in the Zenodo repository at https://doi.org/10.5281/zenodo.21978946 (Schwab et al., 2026), released under a Creative Commons Attribution 4.0 license. The deposit is versioned. The DOI above resolves to version 1.0.0, the snapshot on which the analysis reported here was performed.

The provision-level records – 587 in total, comprising 382 positive provisions and 205 confirmed absences – are not released with this paper. The dataset is the output of an ongoing research program: further jurisdictions may be added, and the coding of several governance areas are constantly subject to continuing methodological review. Provision-level records are available to editors and peer reviewers on a confidential basis, and to other researchers on reasonable request to the corresponding author, on terms to be agreed. The deposit contains the following files.

\textbf{How to cite the dataset.} Schwab, J., Naidoo, N., Barazzutti, F., Lee, S., and Machado, C. (2026). \textit{Mapping General-Purpose AI Governance in Twenty AI Middle-Power Jurisdictions} [Dataset]. Version 1.0.0. Zenodo. doi:10.5281/zenodo.21978946

\textit{(Table over)}

\begin{table}[!ht]
\centering
{\footnotesize
\hyphenpenalty=50\exhyphenpenalty=50\emergencystretch=1em
\setlength{\tabcolsep}{5pt}
\renewcommand{\arraystretch}{1.15}
\begin{tabularx}{\textwidth}{L{0.20\textwidth}C{0.09\textwidth}Y}
\toprule
\cellcolor[HTML]{EFEFEF}\textbf{File} & \cellcolor[HTML]{EFEFEF}\textbf{Records} & \cellcolor[HTML]{EFEFEF}\textbf{Contents} \\
\midrule
instruments.csv & 101 & One row per in-scope governance instrument. Fields: instrument ID, title, local-language title, issuing body, jurisdiction, level, adoption status, publication date, tier, sub-tier, scope, cross-references, GPAI definition flag, governance areas, source URL. Replaces Appendix F. \\
governing\_\allowbreak{}bodies.csv & 144 & One row per governing body: body ID, name, local-language name, AI status, body type, mandate scope, sector, enforcement powers, linked instruments. Replaces Appendix G. \\
Appendix\_\allowbreak{}E\_\allowbreak{}out\_\allowbreak{}scoped\_\allowbreak{}instruments.pdf & 12 & One row per worked exclusion: instrument name, jurisdiction, superseded and adoption status, publication date, tier, sub-tier, horizontal or sectoral, sector, and the decision-flow step at which the instrument failed. Replaces Appendix E. \\
gpai\_\allowbreak{}definitions.csv & 52 & One row per definitional moment, linked to instruments.csv by instrument ID. \\
README.md & — & Deposit contents, license, version history. \\
\bottomrule
\end{tabularx}}
\end{table}
\FloatBarrier

\subsection*{Appendix E. Sample of out-scoped instruments}

The twelve worked exclusions are deposited as Appendix\_\allowbreak{}E\_\allowbreak{}out\_\allowbreak{}scoped\_\allowbreak{}instruments.pdf. Each row carries the instrument’s details and, in the reason\_\allowbreak{}out\_\allowbreak{}of\_\allowbreak{}scope field, the decision-flow step at which it failed together with the reasoning. The sample is illustrative rather than exhaustive: it shows how the scope test was applied at the margin, and §4.2 explains the selection.

\subsection*{Appendix F. Catalog of GPAI governance instruments}

The record-level catalog of all 101 in-scope instruments is deposited as instruments.csv. The fields are those previously printed as columns of this appendix: instrument ID, title, issuing body, level, adoption status, publication date, tier and sub-tier, scope, cross-references, whether the instrument carries a GPAI definition, and which of the four governance areas it addresses. Instruments are grouped by jurisdiction in the deposited file as they were here, and the source URL for each is included.

\subsection*{Appendix G. Catalog of GPAI governance actors}

The record-level catalog of all 144 governing bodies is deposited as governing\_\allowbreak{}bodies.csv. The fields are those previously printed here: body ID, name with the local or colloquial name where applicable, whether the body is AI-specific or pre-existing, body type, whether its mandate is horizontal or sectoral, sector, enforcement powers, and the instruments linked to it. Bodies are grouped by jurisdiction and ordered as described in §5.8.

\subsection*{Appendix H. Catalog of GPAI definitions}

The record-level catalog of all 52 GPAI definitions is deposited as gpai\_\allowbreak{}definitions.csv: definition ID, the instrument and jurisdiction it belongs to, the verbatim text with an English rendering where the source is non-English, the term the instrument uses, the mechanism by which it brings a model into scope, any numeric threshold and its metric, and which of the six dimensions the definition engages. Definitions are grouped by jurisdiction and ordered as described in §5.2.

\end{document}